\documentclass[useAMS,usenatbib]{mn2e}
\usepackage{graphicx}
\usepackage{epstopdf}
\usepackage{multirow}
\usepackage{subfigure}

\newcommand{\apj} {ApJ}
\newcommand{\aj} {AJ}
\newcommand{\apjl} {ApJL}
\newcommand{\aap} {AAp}
\newcommand{\apjs} {ApJS}
\newcommand{\aaps} {AApS}
\newcommand{\mnras}{MNRAS}

\newcommand{\tab} {Table~}
\newcommand{\fig} {Figure~}
\newcommand{\sect} {Section~}

\title[The evolution 
of early-type galaxies in clusters]{The evolution 
of early-type galaxies in clusters from $z\sim 0.8$ to $z\sim 0$: 
the ellipticity distribution and the morphological mix}

\author[Vulcani et al. ]{\parbox[t]{\textwidth}{Benedetta Vulcani$^{1,2}$\thanks{E-mail:
benedetta.vulcani@oapd.inaf.it; }
\thanks{visiting The
Observatories of Carnegie Institution of Washington, Pasadena, CA, USA},
Bianca M. Poggianti$^{2}$, Alan Dressler$^{3}$, Giovanni Fasano$^{2}$, 
Tiziano Valentinuzzi$^{1}$, Warrick Couch$^{4}$, Alessia Moretti$^{2}$, Luc Simard$^{5}$, Vandana Desai$^{6}$, Daniela Bettoni$^{2}$, Mauro D'Onofrio$^{1}$, Antonio Cava$^{7,8}$ and Jes\'us Varela$^{2}$}\\ 
\\$^{1}$Astronomical Department, Padova University, Italy,\\$^{2}$INAF-Astronomical Observatory 
of Padova, Italy,\\$^{3}$The
Observatories of Carnegie Institution of Washington, Pasadena, CA, USA, \\$^{4}$ School of Physics, University of New South Wales, Sydney, Australia\\$^{5}$National Research Council of Canada, Herzberg Institute of Astrophysics, Victoria, British Columbia, Canada,\\$^{6}$Spitzer Science Center, California Institute of Technology, USA,\\$^{7}$Instituto de Astrofisica de Canarias, Spain,\\$^{8}$Departamento de Astrofisica, Universidad de La Laguna, Spain,}

\begin{document}

\date{Accepted .... Received ..; in original form ...}

\pagerange{\pageref{firstpage}--\pageref{lastpage}} \pubyear{2010}

\maketitle

\label{firstpage}

\begin{abstract}
We present the  ellipticity distribution and its evolution for early-type galaxies
in clusters from $z\sim0.8$ to the
current epoch, based on 
the WIde-field Nearby Galaxy-cluster Survey (WINGS) ($0.04 \leq z \leq
0.07$), and the ESO Distant Cluster Survey
(EDisCS) ($0.4 \leq z \leq 0.8$). 
We first investigate a mass limited sample and we find that, above a fixed mass limit 
($M_{\ast}\geq 10^{10.2}M_{\odot}$),  the ellipticity ($\epsilon$)
distribution of early-types noticeably evolves with redshift.
In the local Universe there are proportionally more galaxies with higher ellipticity, hence flatter, than in  distant clusters.
This  evolution is due partly to the change of the mass distribution and mainly to the change
of the morphological mix with z (among the early types, the fraction of ellipticals goes from $\sim70\%$  at high-z to $\sim40\%$ at low-z). 
 Analyzing separately the ellipticity distribution of the different morphological types, 
we find no evolution both for ellipticals and for S0s. However, for ellipticals a
change with redshift in the median value of the distributions is detected. 
This is due to a larger population of very round ($\epsilon <0.05$) elliptical galaxies at low-z. 
In order to compare our finding to previous studies, we also assemble a 
magnitude-``delimited'' sample that 
consists of early-type galaxies on the red sequence 
with $-19.3 >M_{B} +1.208z >-21$. 
Analyzing this sample, we do not recover exactly the same results of the mass-limited sample.his indicates that the selection criteria are crucial to characterize the galaxy properties: 
the choice of the magnitude-``delimited'' sample
implies the loss of many less massive galaxies and so
it biases the final conclusions. 
 Moreover, although we are adopting the same selection criteria, 
our results in the magnitude-``delimited'' sample are also not  
in agreement with those of \cite{h09}. This is due to the fact 
that our and their low-z
samples have a different magnitude distribution because
the \cite{h09} sample suffers from incompleteness 
at faint magnitudes.
\end{abstract}


\section{Introduction}
Ellipticals and lenticulars (S0s) belong to the class of early-type galaxies. This
means that they have several properties in common: they dominate the
total galaxy population at high masses, they preferentially
inhabit dense regions of the
universe, such as rich clusters \citep{dressler97}, they tend to be
passive, 
   they have red colors and their spectra show strong values of the 
characteristic
$D4000$ feature  (see e.g.  \citealt{kauffmann03,
brinchmann04}); they lack spiral arms and in most cases 
exhibit neither major dust features nor a large interstellar gas content. 
For
these reasons, often they are considered together.

On the other hand, elliptical and S0 galaxies differ in several important ways:
S0s are bulge-dominated systems with an identifiable disk (e.g. \citealt{scorza98,laurikainen07}),
 that is mainly rotationally supported (e.g. \citealt{erwin03, cappellari05}),
 their intrinsic
shape is similar to that of spirals \citep{rood67, sandage70} and their
formation is still not well understood. \cite{hubble36} first
proposed their existence as transitional class between ellipticals and spirals.
Understanding how they form and evolve is essential if we wish 
to have a complete picture of how galaxy morphology is related to 
galaxy formation and the environment. 

Then again, 
 ellipticals show ellipsoidal shapes, not rarely with
significant 
kinematic twists, and kinematically decoupled components in their centres.
Most of them are not characterized by strong rotation \citep{bertola75},
 and their luminosity profiles follow 
a Sersic's law.
In the Local Universe disky ellipticals are probably the high bulge mass end of S0.

Morphologically, \cite{dressler97} showed that, at least for bright galaxies, the 
raising fraction of early-type galaxies since $z\sim  0.5$ corresponds 
mainly to an increase of
lenticular S0 galaxies, with a roughly constant elliptical fraction. S0s are quite
rare in clusters at high redshift ($z>0.3-0.4$); as a consequence, they
 have to acquire their shapes 
with different time-scales and later than ellipticals.
The evolving fraction of S0s in clusters might result from the evolving 
population of newly accreted spiral galaxies from infalling groups and the 
field. 

\cite{fasano00} showed that the cluster S0 to elliptical ratio is, on average, a factor of 
$\sim 5$ higher at $z\sim 0$ than at $z\sim0.5$. At higher redshift, there is no 
evidence for any further evolution of the S0 fraction in clusters 
to $z\sim  1$: most of the evolution occurs since $z \sim 0.4$
(see e.g. \citealt{postman05, desai07, wilman09}).

\cite{dressler97} and \cite{postman05} also investigated
the ellipticity distributions of the S0 and 
elliptical galaxies in their magnitude limited samples. They found 
that the ellipticity distribution of S0 and elliptical galaxies show 
no evolution over the broad redshift ranges in their samples.
Moreover, they 
differ from each other, providing evidence for the 
existence of two distinct classes of galaxies. 

In contrast, 
in their magnitude-``delimited'' sample (with both an upper and a lower 
magnitude limit), 
\cite{h09} found no evolution in neither  the median ellipticity nor the shape of
the ellipticity distribution with redshift
for early-type (ellipticals + S0s)
red-sequence galaxies. This lead them to conclude
that there has been little or no evolution in the overall 
distribution of bulge-to-disk ratio of early-type galaxies
from $z\sim 1$ to $z\sim 0$.  
Assuming that
the intrinsic ellipticity distribution of both elliptical and S0 
galaxies separately remains constant,
they finally concluded 
that the relative fractions of ellipticals and S0s do
not evolve from $z\sim 1$ to $z = 0$ for a red-sequence selected 
sample of galaxies.

All the cited works analyzed samples limited in some ways by magnitude cuts. 
For the first time, in this paper we analyze the evolution of the ellipticity distribution
of early-type galaxies also in a mass-limited sample. 
For the sample in the Local Universe, we analyze the data  of the WIde-field
Nearby Galaxy-cluster Survey (WINGS) \citep{fasano06}, while for that in 
the distant Universe we use the ESO
Distant Cluster Survey (EDisCS) \citep{white05}.
These large cluster samples and their
high quality images (see \S 2) allow us to characterize
properly the cluster environment at the two redshifts and
to subdivide galaxies into the different morphological types and obtain robust
estimates of ellipticity. 

This paper is organized as follows: in \S 2 we present the cluster and
 galaxy samples (WINGS  \citep{fasano06} and EDisCS 
 \citep{white05}), describing the surveys, the data reduction, the
 determination of morphologies, ellipticites and masses. We also
 depict the selection criteria we follow to assemble the mass-limited
 and the magnitude-``delimited'' samples.  In \S 3 we show the results of
 our analysis of the evolution of the ellipticity distribution with
 redshift in our mass-limited samples, while in \S 4 we show the same
 for the magnitude-``delimited'' samples.  In \S 5 we try to
 reconcile the results of the different samples, while in \S 6 we
 compare our results with those found in literature (in particular
 with the results drawn by \citealt{h09}). Finally, in \S 7 
 we discuss and summarize our findings.

Throughout this paper, we assume $H_{0}=70 \, \rm km \, s^{-1} \, Mpc^{-1}$, 
$\Omega_{m}=0.30$, $\Omega_{\Lambda} =0.70$.  The adopted initial 
mass function is a \cite{kr01} in the mass range 0.1-100 $M_{\odot}$.

\section{Cluster and Galaxy samples}

To perform the study of the ellipticity ($\epsilon \equiv 1-b/a$, b
$\equiv$ semi-minor axis, a $\equiv$ semi-major axis) distribution and
its evolution from $z\sim 0.8$ to $z\sim 0$ for early-type galaxies
and for ellipticals and S0s separately, we assemble
two different galaxy cluster samples in two redshift intervals: we
draw the samples at low-z from the WIde-field Nearby Galaxy-cluster
Survey (WINGS) \citep{fasano06} and those at high-z from the ESO
Distant Cluster Survey (EDisCS) \citep{white05}.

First of all, we use a mass-limited sample, that ensures
completeness, i.e. includes all galaxies more massive than the limit
regardless of their color or morphological type. We think that this
is the best choice to characterize properly galaxy properties. 

Then, 
since  \cite{h09} have already 
analyzed the ellipticity
distribution using a sample delimited in magnitude both at faint
and bright magnitudes, in
order to compare our results with theirs, 
we also assemble a magnitude-``delimited'' sample,
following their selection criteria.

\subsection{Low-z sample: WINGS}\label{samplew}

The main goal of WINGS\footnote{http://web.oapd.inaf.it/wings}
\citep{fasano06}, a multiwavelength survey of clusters at \mbox{$0.04
< z < 0.07$}, is to characterize the photometric and spectroscopic
properties of galaxies in nearby clusters and to describe the changes
of these properties depending on galaxy mass and environment.  The
project was based on deep optical (B, V) wide field images 
of 77 fields \citep{varela09} centered on nearby clusters of galaxies
selected from three X-ray flux limited samples compiled from ROSAT
All- Sky Survey data \citep{ebeling98, ebeling00} and the X-ray
Brightest Abell-type Cluster sample \citep{ebeling96}.

 WINGS clusters cover a wide range of velocity dispersion
 $\sigma_{clus}$ (typically $500-1100 \, km \, s^{-1}$) and a wide
 range of X-ray luminosity $L_{X}$ (typically $0.2-5 \times10^{44} erg
 \, s^{-1}$).

The survey has been complemented by a
near-infrared (J, K) survey of a subsample of 28 clusters obtained
with WFCAM@UKIRT \citep{valentinuzzi09}, by a
spectroscopic survey of a subsample of 48 clusters, obtained with the
spectrographs WYFFOS@WHT and 2dF@AAT \citep{cava09},  and by U broad-band and
$H_{\alpha}$ narrow-band imaging of a subset of WINGS clusters,
obtained with wide-field cameras at different telescopes (INT, LBT,
Bok) (\citealt{omizzolo10}).

The spectroscopic target selection was based on the WINGS 
B, V photometry. The aim of the target
selection strategy was to maximize the chances of observing
galaxies at the cluster redshift without biasing the cluster sample.
Galaxies with a total $V \leq 20$ magnitude, a V magnitude
within the fiber aperture of V $<$ 21.5 and with a color within a 
5 kpc aperture of $(B-V)_{5 kpc} \leq1.4$ were selected, to reject background
galaxies. The exact cut in color was varied slightly from cluster
to cluster in order to account for the redshift variation and
to optimize the observational setup. 
These very loose selection limits were applied so as to avoid any bias
in the colors of selected galaxies.

Our optical imaging covers a
 $34^{\prime} \times 34^{\prime}$ field.
This imaging corresponds to about 0.6
 $R_{200}$ or more, for most clusters, although in a few cases only  
$\sim 0.5 R_{200}$ is covered. 
$R_{200}$  is defined as the radius delimiting a 
sphere with interior mean density 200 times the critical 
density of the Universe at that redshift,
and is commonly used as an approximation for the cluster 
virial radius. The $R_{200}$  values for our structures are computed 
from the velocity dispersions by \cite{cava09}.

\subsubsection{Morphologies}

Morphological types are derived from V-band images using MORPHOT, an
automatic tool for galaxy morphology, purposely devised in the
framework of the WINGS project. MORPHOT was designed with 
the aim to reproduce as closely as possible
visual morphological classifications.

\begin{figure}
\includegraphics[scale=0.4]{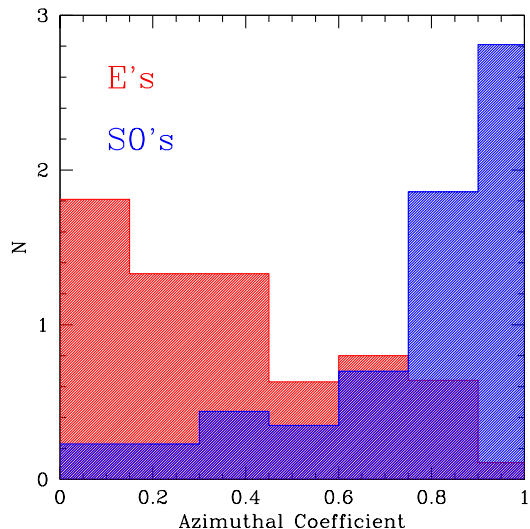}
\caption{
Normalized distributions of the MORPHOT Azimuthal Coefficient for the
visually classified ellipticals (366 objects, red histogram) and S0
galaxies (267 objects, blue histogram) of the MORPHOT calibration
sample. The Azimuthal Coefficient measures the correlation between
azimuth and pixel flux relative to the average flux value of the
elliptical isophote passing through the pixel itself (from \citealt{morph}).
\label{morph}}
\end{figure}

 MORPHOT extends the classical CAS
(Concentration/Asymmetry/clumpinesS) parameter set \citep{conselice03}, by
using 20 image-based morphological diagnostics. Fourteen of them have
never been used, while the remaining six [the CAS parameters,
the Sersic index, the Gini and M20 coefficients \citep{lotz04}] are
already present in the literature, although in slightly
different forms. 
An exhaustive description of MORPHOT
will be given in a forthcoming paper (Fasano et al. 2010b), where also
the morphological catalogs of the WINGS clusters will be presented and
discussed. Provisionally, we refer the reader to \citet{fasa07} and
\citet[][Appendix~A therein]{fasa10} for an outlining of the
logical sequence and the basic procedures of MORPHOT.
 Here we just mention that, among the 14 newly
devised diagnostics, the most effective one in order to disentangle ellipticals
from S0 galaxies turned out to be an Azimuthal coefficient, measuring the
correlation between azimuth and pixel flux relative to the average
flux value of the elliptical isophote passing through the pixel
itself. From
\cite{morph}, we report here in Figure~\ref{morph} a plot
illustrating the capability of 
the distributions of the
Azimuthal coefficient in disentangling elliptical from S0
galaxies, a crucial point in the present analysis.

\begin{figure}
\includegraphics[scale=0.4]{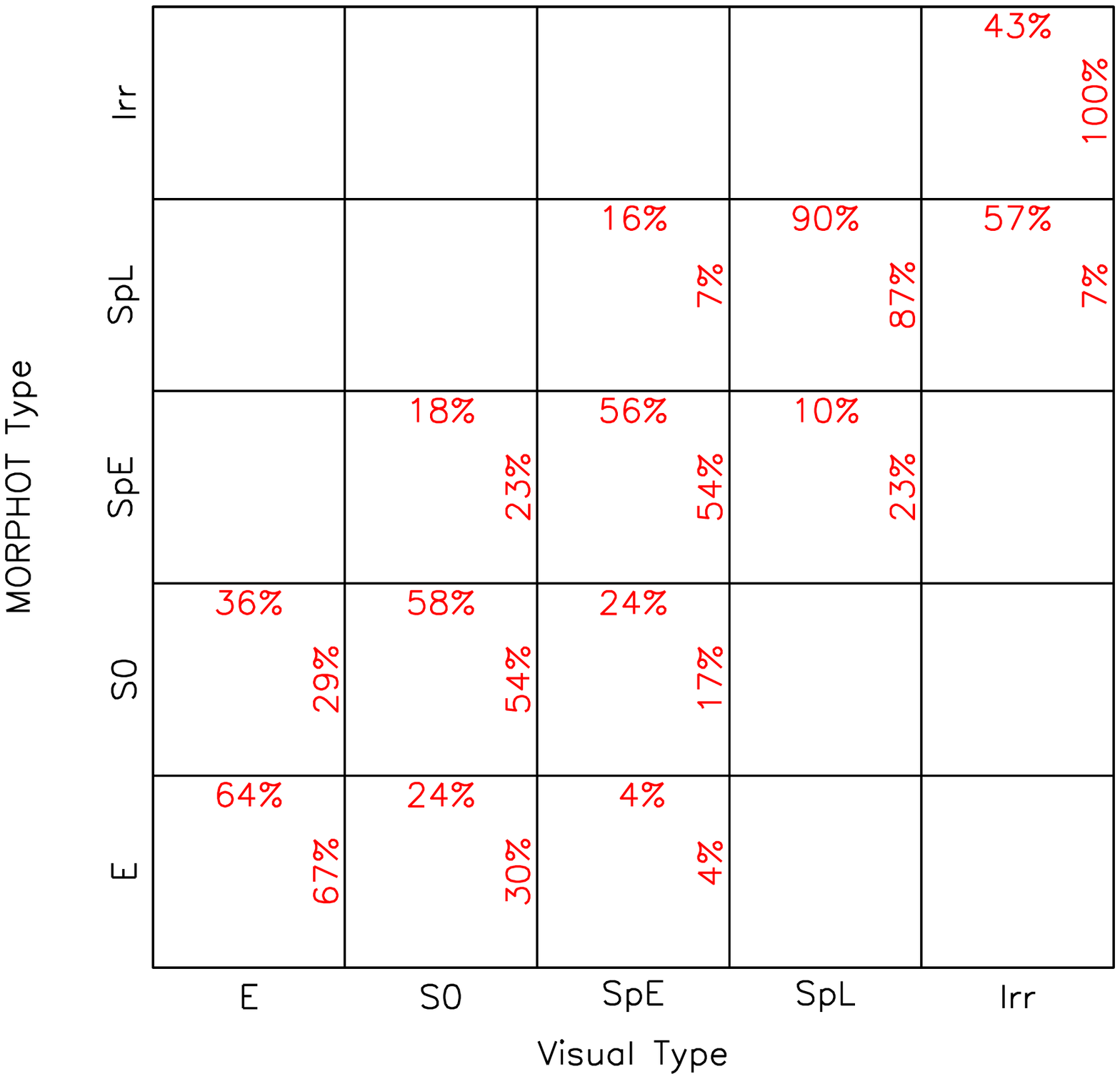}
\caption{Comparison between visual and MORPHOT broad morphological
types for the 931 galaxies of the MORPHOT calibration sample. In each
one of the 2D bins of the plot the percentages of the visual broad
types (Es, S0s, early spirals [SpE], late spirals [SpL] and
irregulars) falling in different bins of the MORPHOT (broad)
classification are reported on the top. Similarly, on the right side
of each bin the percentages of MORPHOT types falling in different bins
of the visual types are reported.
\label{morph2}}
\end{figure}

More importantly for our purposes, the quantitative discrepancy
between automatic (MORPHOT) and visual classifications turns out to be
similar to the typical discrepancy among visual classifications given
by experienced, independent human classifiers (r.m.s.$\sim$1.3-2.3 T
types). The last one has been estimated from a sample of 233 SDSS
galaxies included in the Third Reference Catalog of Bright Galaxies
(de~Vaucouleurs et al. 1991, RC3), whose visual classification was
carried out independently by GF and AD and also compared with that
given in the RC3. The comparison between visual
(GF) and automatic (MORPHOT) classification is illustrated in
Figure~\ref{morph2} for the MORPHOT calibration sample (931
galaxies). In this figure (from \citealt{morph}) the automatic classification is also shown
to be bias-free in the overall range of morphological types, perhaps
apart from the last bin, i.e. that relative to the very late and
irregular galaxies.

For now, we can apply MORPHOT just to the WINGS imaging,
because the tool is calibrated on the WINGS imaging characteristics,
and we defer to a later time a more generally usable version of the
tool. In the following, for EDisCS we will use visual morphological
classifications.
To verify directly that the two methods adopted at different
redshifts (see \sect\ref{dataediscs})
are consistent, we can apply
the same ``method'' (visual classification and persons)
that was used at high-z on the low-z images.

To this aim, 3 of the classifiers that in 2007 visually classified all
the EDisCS galaxies (BMP, AAS, VD) now performed a visual
classification of WINGS galaxies. This was done on the subset of
WINGS galaxies that was used to calibrate MORPHOT on the visual WINGS
morphologies, including only galaxies that enter
the sample we analyze in this paper (173 galaxies).

\begin{figure*}
\begin{minipage}[c]{175pt}
\centering
\includegraphics[scale=0.35,clip = false, trim = 0pt 150pt 0pt 0pt]
{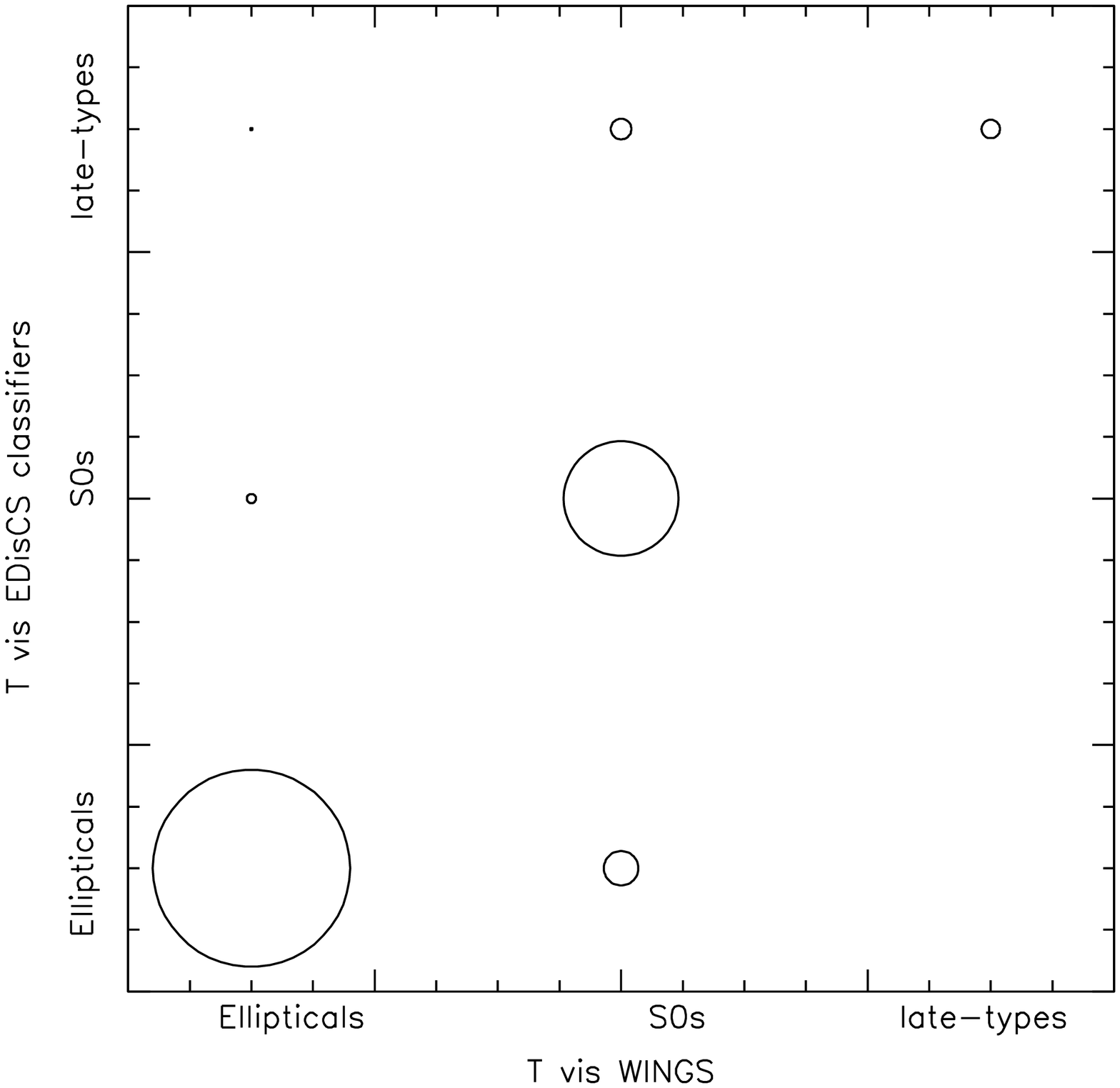}
\vspace*{2.2cm}
\end{minipage}
\hspace{2.5cm}
\begin{minipage}[c]{175pt}
\centering
\includegraphics[scale=0.35,clip = false, trim = 0pt 150pt 0pt 0pt]
{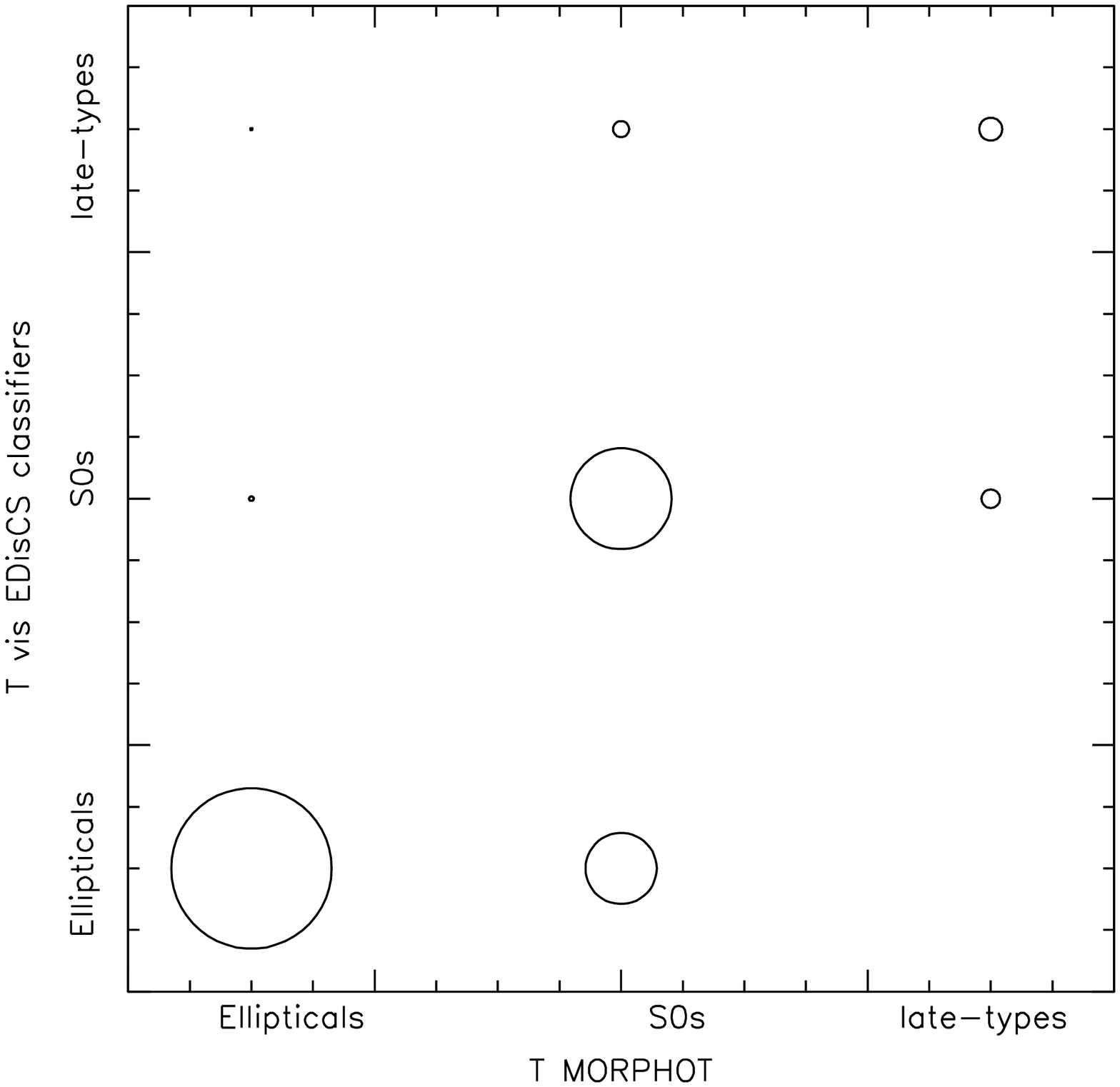}
\vspace*{2.2cm}
\end{minipage}
\hspace{1cm}
\caption{Comparison between the visual classification performed by the EDisCS classifiers
and the WINGS visual classification (left panel)  and the automatic classification performed by MORPHOT (right panel). The circle radius is proportional to the number of galaxies.\label{mor}}
\end{figure*}

The results (see Figure \ref{mor} from \citealt{morph}) show agreement between the three broad
morphological classes assigned by the EDisCS classifiers with the
WINGS visual classification in $\sim$83\% of the
cases, and with MORPHOT in $\sim$75\% of the cases.
Again, these discrepancies turn out to be similar
to the typical discrepancy among visual classifications given by
experienced, independent human classifiers, so we conclude that the
different methods adopted provide a comparable classification.

\subsubsection{Ellipticity measurements}

Galaxy ellipticities have 
been computed using the tool GASPHOT \citep{pignatelli06}.
This is heavily based on SExtractor  (``Source 
Extractor'') galaxy photometry package \citep{bertin96} and  provides, among other quantities,  ellipticity profiles of galaxies 
extracted from CCD frames. It fits simultaneously the major and minor axis light growth curves
of galaxies with a 2D flattened Sersic-law, convolved 
by the appropriate, space-varying point-spread-function (PSF), which was previously evaluated by the tool itself
 using the stars present in the frame. This approach exploits the robustness of the 1D 
 fitting technique, saving at the same time the capability, typical 
of 2D approaches, of dealing with PSF convolution of flattened 
galaxies. The tool was previously tested for non-Sersic profiles 
and blended objects and its results compared with other tools, such as GALFIT \citep{peng02}
and GIM2D \citep{marleau98}, as shown in \cite{pignatelli06}.

\begin{figure}
\centering
\includegraphics[scale=0.4]{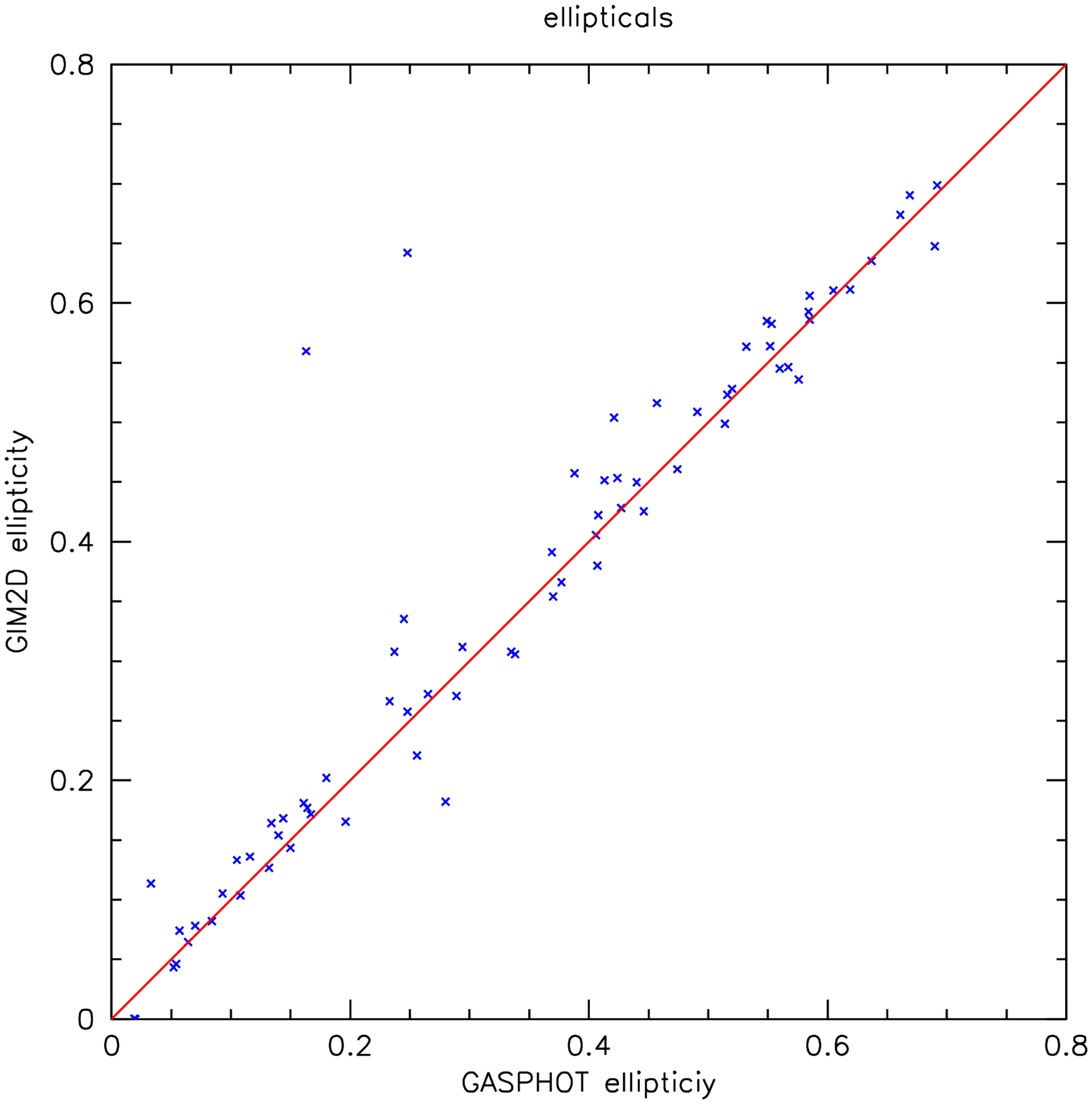}
\caption{Comparison between the ellitpicity estimation from GASPHOT and GIM2D for early-type galaxies of the cluster A119. \label{A119}}
\end{figure}

Since in our analysis we are comparing the ellipticities of WINGS
galaxies to those of EDisCS galaxies, that have been determined using
the tool GIM2D (see \sect\ref{dataediscs}), we also performed a comparison between the
values estimated by the two different tools, for early-types galaxies
in the WINGS cluster A119.  As shown in \fig\ref{A119}, the estimates are in
good agreement (r.m.s.$\sim$0.07).
 
\subsubsection{Galaxy stellar masses}

Stellar masses have been determined using the relation between
 $M/L_{B}$ and rest-frame $(B-V)$ color, following \cite{bj01}, who
 used a spectrophotometric model finding a strong correlation between
 stellar mass-to-light $M/L$ ratio and optical colors of the
 integrated stellar populations for a wide range of star formation
 histories. This method was chosen to be consistent with that adopted
 for galaxy masses at high-z, and because it can be used also for
 galaxies with no spectroscopy in the magnitude-``delimited'' sample.

We use the equation that \cite{bj01} give for
the Bruzual \& Charlot model with a \cite{salpeter55} IMF 
(0.1-125 $M_{\odot}$) and
solar metallicity,
\begin{equation}
\log_{10}(M/L_{B})=-0.51+1.45 \times (B-V)
\end{equation}

The total
luminosity $L_{B}$ has been derived from the total (SExtractor AUTO)
observed B magnitude \citep{varela09}, corrected for distance modulus
and foreground Galaxy extinction, and k-corrected using tabulated
values from \citep{poggianti97}.
The $(B-V)$ color used to calculate masses was
derived from observed B and V aperture magnitudes
measured within a diameter of 10 kpc around each galaxy baricenter, 
corrected as the total magnitude.
 
Then,
we scaled our masses to the more used \cite{kr01} IMF adding -0.19 dex to
the logarithmic value of the masses.

Stellar masses for WINGS galaxies observed spectroscopically
had been previously determined
by fitting the optical spectrum (in the range \mbox{$\sim 3600 - \sim 7000$} 
\AA{}) (\citealt{fritz10}), with the spectro-photometric 
model fully described in 
\cite{fritz07}, and they are in good agreement with the masses
used in this paper.
For a detailed description 
of the determination of masses and a comparison of different
methods see \cite{fritz10}, Fig.~1 in \cite{morph} and \cite{valentinuzzi10}.

\subsubsection{Mass-limited sample}
For the mass-limited sample, we rely on spectroscopy to 
be sure that we are using only cluster members. In WINGS, photo-z techniques
cannot be used to assess cluster membership due to the low redshift
and to the fact the we have at our disposal few photometric bands. 
Galaxies are considered members of a cluster if their spectroscopic
redshift lies within $\pm3 \sigma$ from the cluster mean redshift,
where $\sigma$ is the cluster velocity dispersion \citep{cava09}.
We use only spectroscopically confirmed members
of 21 of the 48 clusters with spectroscopy. 
The clusters used in this analysis are listed in \tab\ref{tab:wi.cl}.  
This is the subset of clusters that
have a spectroscopic completeness (the ratio of the number of spectra
yielding a redshift to the total number of galaxies in the photometric
catalog) larger than 50\%.  
We apply a statistical correction to
correct for incompleteness, weighting each galaxy by
the inverse of the ratio of the number of spectra yielding a redshift
to the total number of galaxies in the V-band photometric catalog, in bins of
1 mag \citep{cava09}.

In each cluster, we exclude the brightest cluster galaxy (BCG),
defined as the most luminous galaxy of each cluster, 
that has peculiar properties and could alter the distributions \citep{fasa10}.

Only galaxies lying within 0.6$R_{200}$ are considered, because this is
the largest radius covered in the 21 clusters considered (except for
{\it A1644} and {\it A3266} where the coverage extends to $\sim 0.5 R_{200}$).

To determine the galaxy stellar mass limit of our sample, we compute the mass 
of an object whose
observed magnitude is equal to the faint magnitude limit
of the survey, and whose color is the reddest color 
of a galaxy at the highest redshift considered. 

\begin{table}
\centering
\begin{tabular}{ccccc }
\hline
cluster name & z & $\sigma$  		& DM		&$R_{200}$\\
	   &	& ($km \, s^{-1}$)		&  {\it (mag)}	&{\it kpc}\\
\hline
A1069 & 0.0653 & 690$\pm$68 & 37.34& 1.65\\
A119 & 0.0444 & 862$\pm$52& 36.47	& 2.09\\
A151 & 0.0532 & 760$\pm$55 &36.87 & 1.83 \\
A500 &0.0678  & 658$\pm$48 & 37.42 & 1.58\\
A754 & 0.0547 & 1000$\pm$48& 36.94 & 2.41\\
A957x & 0.0451 &710$\pm$53& 36.50& 1.72\\
A970 & 0.0591 & 764$\pm$47&37.11& 1.84\\
A1631a & 0.0461 & 640$\pm$33 &36.55 & 1.55 \\
A1644 &0.0467 & 1080$\pm$54& 36.58& 2.61 \\
A2382 & 0.0641 & 888$\pm$54&37.30& 2.13 \\
A2399 & 0.0578 & 712$\pm$41& 37.06& 1.71\\
A2415& 0.0575 & 696$\pm$51& 37.05& 1.67\\
A3128& 0.06 & 883$\pm$41& 37.15& 2.12 \\
A3158& 0.0593 & 1086$\pm$48&37.12 & 2.61\\
A3266& 0.0593 & 1368$\pm$60&36.12 & 3.29\\
A3376 &0.0461 & 779$\pm$49&36.55 & 1.88 \\
A3395 & 0.05 & 790$\pm$42&36.73 & 1.91\\
A3490 & 0.0688 & 694$\pm$52&37.46 & 1.66 \\
A3556 & 0.0479 & 558$\pm$37&36.64  & 1.35\\
A3560 & 0.0489  & 710$\pm$41&36.68 & 1.72\\
A3809 & 0.0627 & 563$\pm$40&37.25 & 1.35\\
\hline
\end{tabular}
\caption{List of WINGS clusters analyzed in the mass-limited sample,
their redshift, velocity dispersion, distance modulus and $R_{200}$.
\label{tab:wi.cl}  }
\end{table}

The spectroscopic magnitude limit of the WINGS survey is
 V=20. Considering the distance module of the most
 distant WINGS cluster is $\sim 37.5$ , and the reddest galaxy has a
 color of $(B-V)=1.2$, the
 magnitude limit corresponds to a mass limit $M_{\ast} = 10^{9.8}
 M_{\odot}$, above which the sample is unbiased.  
Adopting this limit, the final sample 
consists of 951 early-type galaxies, of which
364 are ellipticals and  587 are S0s. 
The corresponding numbers weighted for incompleteness 
are: 1469 early-types,  557 ellipticals and
912 S0s. 
Numbers of WINGS galaxies above the EDisCS mass limit 
($M_{\ast} = 10^{10.2} M_{\odot}$, see below) are 594 
early-types (920 once weighted), 224 ellipticals (341 once weighted)
and 370 S0s (579 once weighted) (see \tab\ref{numb}).

\begin{table*}
\centering
\begin{tabular}{|l||cccccccc}
\hline
& \multicolumn{5}{|c|}{WINGS} 	&&  \multicolumn{2}{|c|}{EDisCS} \\ 
& \multicolumn{2}{|c|}{$M_{\ast} \geq 10^{9.8} M_{\odot}$} & \multicolumn{2}{|c|}{$M_{\ast} \geq 10^{10.2} M_{\odot}$} & mag && $M_{\ast} \geq 10^{10.2} M_{\odot}$ & mag \\
\hline
            	&N$_{obs}$ & N$_{w}$	&N$_{obs}$	& N$_{w}$ 	&N	&& N & \\
\hline
ellipticals 	&364	&557		&224	&341& 580 &&	145 &101 \\
S0s		&587	&912		&370	&579& 914 &&	61 &43 \\	
early-types 	&951	&1469		&594	&920& 1494 &&	206 &144 \\
\hline
\end{tabular}
\caption{Number of galaxies in the mass-limited and in the magnitude-``delimited'' samples. 
For the mass-limited sample, for WINGS both observed numbers and numbers weighted for spectroscopic incompleteness
are given. \label{numb}}
\end{table*}

\subsubsection{Magnitude-``delimited'' sample}

\begin{table}
\centering
\begin{tabular}{ccccc}
\hline
cluster name & z & $\sigma$  		& DM		&$R_{200}$\\
	   &	& ($km \, s^{-1}$)		&  {\it (mag)}	&{\it kpc}\\
\hline
 A85 & 0.0521 & 1052$\pm$68 & 36.83 & 2.54 \\ 
A133 & 0.0603 & 810$\pm$78 & 37.16 & 1.95 \\ 
A147 & 0.0447 & 666$\pm$13 & 36.48 & 1.61 \\ 
A160 & 0.0438 & 561$\pm$53 & 36.44 & 1.36 \\ 
A168 & 0.0448 & 503$\pm$43 & 36.49 & 1.22 \\ 
A193 & 0.0485 & 759$\pm$59 & 36.67 & 1.83 \\ 
A311 & 0.0657 &NULL&37.35 & 0.00 \\ 
A376 & 0.0476 & 852$\pm$49 & 36.62 & 2.06 \\ 
A548b & 0.0441 & 848$\pm$59 & 36.45 & 2.05 \\ 
A602 & 0.0621 & 720$\pm$73 & 37.22 & 1.73 \\ 
A671 & 0.0507 & 906$\pm$58 & 36.77 & 2.19 \\ 
A780 & 0.0565 & 734$\pm$10 & 37.01 & 1.26 \\ 
A1291 & 0.0509 & 429$\pm$49 & 36.77 & 1.04 \\ 
A1668 & 0.0634 & 649$\pm$57 & 37.27 & 1.56 \\ 
A1736 & 0.0461 & 853$\pm$60 & 36.55 & 2.06 \\ 
A1795 & 0.0633 & 725$\pm$53 & 37.27 & 1.74 \\ 
A1831 & 0.0634 & 543$\pm$58 & 37.27 & 1.30 \\ 
A1983 & 0.0447 & 527$\pm$38 & 36.48 & 1.28 \\ 
A1991 & 0.0584 & 599$\pm$57 & 37.08 & 1.44 \\ 
A2107 & 0.0410 & 592$\pm$62 & 36.29 & 1.44 \\ 
A2124 & 0.0666 & 801$\pm$64 & 37.38 & 1.92 \\ 
A2149 & 0.0675 & 353$\pm$53 & 37.41 & 0.85 \\ 
A2169 & 0.0578 & 509$\pm$40 & 37.06 & 1.22 \\ 
A2256 & 0.0581 & 1273$\pm$64 & 37.07 & 3.06 \\ 
A2271 & 0.0584 & 504$\pm$10 & 37.08 & 1.21 \\ 
A2457 & 0.0584 & 580$\pm$39 & 37.08 & 1.40 \\ 
A2572a & 0.0390 & 631$\pm$10 & 36.18 & 1.53 \\ 
A2589 & 0.0419 & 816$\pm$88 & 36.34 & 1.98 \\ 
A2593 & 0.0417 & 701$\pm$60 & 36.33 & 1.70 \\ 
A2622 & 0.0610 & 696$\pm$55 & 37.18 & 1.67 \\ 
A2626 & 0.0548 & 625$\pm$62 & 36.94 & 1.51 \\ 
A2657 & 0.0400 & 381$\pm$83 & 36.23 & 0.92 \\ 
A2665 & 0.0562 &NULL& 37.00 & 0.00 \\ 
A2717 & 0.0498 & 553$\pm$52 & 36.73 & 1.34 \\ 
A2734 & 0.0624 & 555 $\pm$42& 37.23 & 1.33 \\ 
A3164 & 0.0611 &NULL& 37.19 & 0.00 \\ 
A3497 & 0.0680 & 726$\pm$47 & 37.43 & 1.74 \\ 
A3528a & 0.0535 & 899$\pm$64 & 36.89 & 2.17 \\ 
A3528b & 0.0535 & 862$\pm$64 & 36.89 & 2.08 \\ 
A3530 & 0.0544 & 563$\pm$52 & 36.92 & 1.36 \\ 
A3532 & 0.0555 & 621$\pm$53 & 36.97 & 1.50 \\ 
A3558 & 0.0477 & 915$\pm$50 & 36.63 & 2.21 \\ 
A3667 & 0.0530 & 993$\pm$84 & 36.87 & 2.39 \\ 
A3716 & 0.0448 & 833$\pm$39 & 36.49 & 2.07 \\ 
A3880 & 0.0570 & 763$\pm$65 & 37.03 & 1.84 \\ 
A4059 & 0.0480 & 715$\pm$59 & 36.64 & 1.73 \\ 
IIZW108 & 0.0483 & 513$\pm$75 & 36.66 & 1.24 \\ 
MKW3s & 0.0444 & 539$\pm$37 & 36.47 & 1.30 \\ 
RX0058 & 0.0484 & 637$\pm$97 & 36.66 & 1.54 \\ 
RX1022 & 0.0548 & 577$\pm$49 & 36.94 & 1.39 \\ 
RX1740 & 0.0441 & 582$\pm$65 & 36.45 & 1.41 \\ 
Z1261 & 0.0644 &NULL& 37.31 & 0.00 \\ 
Z2844 & 0.0503 & 536$\pm$53 & 36.75 & 1.29 \\ 
Z8338 & 0.0494 & 712$\pm$60 & 36.71 & 1.72 \\ 
Z8852 & 0.0408 & 765$\pm$63 & 36.28 & 1.86 \\
\hline
\end{tabular}
\caption{List of additional WINGS clusters used for the magnitude-``delimited'' 
sample,
their redshift, velocity dispersion, distance modulus and $R_{200}$.
\label{tab:wi.cl2}  }
\end{table}

For the magnitude-``delimited'' sample we use the photometric
data for 76 WINGS clusters.\footnote{A3562 was excluded due to bad V-band
seeing.}
The clusters used are those
presented in \tab\ref{tab:wi.cl} plus those in \tab\ref{tab:wi.cl2}.

We follow the selection criteria proposed by \cite{h09}.
They selected a sample of early-type galaxies (ellipticals
and S0s) that lie on the red sequence (determined with spectroscopic
members and accepting all galaxies lying within $2\sigma$ from the
sequence -- for details see \citealt{mei09}). At low-z, they used only
spectroscopically confirmed members.  At high-z they
considered all red sequence
galaxies in the photometric catalogue except those that
are interlopers confirmed by spectroscopy.  Moreover, they selected galaxies
within a magnitude range, taking into account passive evolution:
$-19.3 >M_{B} +1.208z >-21$. Finally, they considered only galaxies
within $2R_{200}/\pi$ of the cluster center. They computed
ellipticities using the results from GALFIT \citep{peng02} and adopted
visual morphologies from the literature (\citealt{dressler80} for the
sample at low-z, \citealt{desai07} and \citealt{postman05} for the
sample at high-z).

For our WINGS magnitude-``delimited'' sample, 
to be strictly consistent with what we do for the EDisCS
data-set, we consider all galaxies in the photometric catalogue,
excluding those that are non-members based on the spectroscopy.
Contamination on the red sequence at low-z is minimal, and
we have checked that the results remain the same
using only spectroscopic members corrected for completeness.
We exclude from our analysis galaxies located outside of $R_{200}$, to
be consistent with what we do at high-z (see \S2.2.2).

We then select only 
galaxies lying within $2\sigma$ from the red sequence. 
Like \cite{mei09}, we define the red sequence using only spectroscopic
members and we build a color-magnitude diagram for each cluster. To do
this, we use the observed B and V aperture magnitude measured within a
diameter of 5 kpc around each galaxy baricenter and the total
V SExtractor AUTO magnitude, both corrected for
distance modulus and foreground Galaxy extinction, and k-corrected
using tabulated values from \cite{poggianti97}.  
For those clusters for which spectroscopy was not
available, we define the red sequence using the photometry of
morphologically selected
early-type galaxies.  
We determine the slope and the dispersion
of the red sequence in the color- magnitude diagram perfoming a
weighted least-square-fit on the data, giving less weight to the
outliers and reiterating 10 times to have a better determination of
the parameters.  

We include galaxies with $-19.3 >M_{B}
+1.208z >-21$, where $M_{B}$ is the magnitude derived from the total
(SExtractor AUTO) observed B magnitude \citep{varela09}, corrected as the
color.  In this way we automatically exclude the BCGs.  
Finally, we consider only galaxies that are elliptical and
S0 following our morphological classification.

Our final magnitude-''delimited'' 
sample consists of 580 ellipticals and 914 S0s,
 for a total of  1494 early-type galaxies (see \tab\ref{numb}).

\subsection{High-z sample: EDisCS} \label{dataediscs}
The multi-wavelength photometric and spectroscopic survey of distant clusters
named EDisCS \citep{white05} has been developed to characterize both
the clusters themselves and the galaxies within them.  It observed 20
fields containing galaxy clusters at $0.4< z <1$.

Clusters were drawn from the Las Campanas Distant Cluster Survey
(LCDCS) catalog \citep{gonzales01}. They were selected as surface
brightness peaks in smoothed images taken with a very wide optical
filter ($\sim 4500-7500$ \AA{}). 
The 20 EDisCS fields were chosen among the 30 highest surface
brightness candidates, after confirmation of the presence of an
apparent cluster and of a possible red sequence with VLT 20 min
exposures in two filters \citep{white05}. 

For all 20 fields, EDisCS has obtained deep optical multiband
photometry with FORS2/VLT
 \citep{white05} and near-IR photometry with SOFI/NTT \citep{aragon09}.
Photometric redshifts were measured 
using both optical and infrared imaging 
(see \citealt{pello09} and  \citealt{rudnick09} for details).  
They were computed for every object in the
EDisCS fields using two independent codes, a modified version
of the publicly available Hyperz code \citep{bolzonella00}
and the code of \cite{rudnick01} with the modifications
presented in \cite{rudnick03}. 
Photo-z membership (see also
\citealt{delucia04} and \citealt{delucia07} for details) 
was established using a modified version of
the technique first developed in \cite{brunner00}, in
which the probability of a galaxy to be at redshift $z$
($P(z)$) 
is integrated in a slice around the cluster redshift to give $P_{clust}$
for the two codes.
A galaxy was rejected
from the membership list if $P_{clust}$ was smaller than a 
certain probability $P_{thresh}$ for either code. 
The $P_{thresh}$ value for each cluster was calibrated from
EDisCS spectroscopic redshifts
and was chosen to maximize the efficiency with which 
spectroscopic non-members are rejected while
retaining at least $\sim 90\%$ of the confirmed cluster members,
independent of their rest-frame (B-V) color or observed
(V-I) color. In practice it was possible to choose thresholds
such that this criterion was satisfied while rejecting 45\%-70\% of
spectroscopically confirmed non-members. Applied to the entire
magnitude limited sample, these thresholds reject 75\%-93\% of all
galaxies with $I_{tot}$ $<$ 24.9. A posteriori, it was
verified that in the sample of galaxies with spectroscopic redshift
and above the mass limit described below,  20\% of those
galaxies that are photo-z cluster members are spectroscopically
interlopers and, conversely, only 6\% of those galaxies that are
spectroscopic clustermembers are rejected by the photo-z technique.

Deep spectroscopy with FORS2/VLT was obtained for 18 of the fields
\citep{halliday04, milvang08}. 
Spectroscopic targets were selected from I-band catalogs, producing
an essentially I-band selected sample with no selection bias
down to $I=22$ at $z\sim 0.4-0.6$ and $I=23$ at $z\sim 0.6-0.8$
\citep{halliday04, milvang08}.
Typically, spectra of
more than 100 galaxies per field were obtained.

ACS/HST mosaic imaging in $F814W$ of 10 of the highest redshift
clusters was also acquired \citep{desai07}, covering with four
ACS pointings a $6.5^{\prime} \times6.5^{\prime}$ field with an additional
deep pointing in the center. This field covers the $R_{200}$ of all
clusters, except for {\it 1232.5-1250} where it reaches 
$0.5 R_{200}$ \citep{poggianti06}. The $R_{200}$
values for our structures are computed from the velocity dispersions
by \cite{poggianti08}.

\subsubsection{Morphologies, ellipticity measurements and galaxy stellar masses}
Morphologies are
discussed in detail in \cite{desai07}.  The morphological
classification of galaxies is based on the visual classification of
{\it HST/ACS} F814W images sampling the rest-frame $\sim
4500-5500$ \AA{} range, similarly to WINGS.

The determination of ellipticities is presented in \cite{simard09}.
They have been estimated using the tool GIM2D (Galaxy IMage
2D) version 3.2, a fitting program
\citep{simard02} that performs a detailed surface
brighteness profile analysis of galaxies in low signal-to-noise (S/N) images 
in a fully automated way.
In this paper, we use the ellipticities derived fitting
every source in the HST/ACS images with a single Sersic fit model. 
Since the shape of the PSF on the HST/ACS images varies significantly as a 
function of position, spatially-varying PSF 
models for the EDisCS cluster images were constructed. 

In this analysis, we consider only 8 of the 10 EDisCS
clusters for which {\it HST} images are available. 
In fact, ellipticity measurements are
not available for {\it cl 1227.9-1138} and no galaxies
of {\it cl 1037.9-1243} enter our final samples (see below the selection criteria).

For EDisCS galaxies, we use stellar masses estimated using the same
relation we use for the WINGS dataset, hence again following the \cite{bj01}
method and then converting masses to a \cite{kr01} IMF.  
Total absolute magnitudes are derived from photo-z fitting
\citep{pello09}, rest-frame luminosities have been derived using
\cite{rudnick03} and \cite{rudnick06} methods and presented in \cite{rudnick09}.
Stellar masses for spectroscopic members were also estimated 
using the {\it kcorrect} tool 
\citep{blanton07},{\footnote{http://cosmo.nyu.edu/mb144/kcorrect/} that
yields masses in agreement with those used in this paper.
For a detailed discussion of our mass estimates and of
the consistency between different methods 
see \cite{morph}.

\subsubsection{Mass-limited sample}
\begin{table}
\centering
\begin{tabular}{|c|c|c|c}
\hline
cluster name & z & $\sigma$ & $R_{200}$\\
	   &	& ($km \, s^{-1}$)		& $Mpc$\\
\hline
cl 1040.7-1155 & 0.70 &418$^{+55}_{-46}$ &0.70\\
cl 1054.4-1146 & 0.70 &589$^{+78}_{-70}$ & 0.99\\ 
cl 1054.7-1245 &0.75 &504$^{+113}_{-65}$ & 0.82\\
cl 1103.7-1245    &0.62 &336$^{+36}_{-40}$& 0.41\\
cl 1138.2-1133    &0.48 &732$^{+72}_{-76}$& 1.41\\
cl 1216.8-1201 & 0.79 &1018$^{+73}_{-77}$&1.61\\
cl 1232.5-1250 & 0.54&1080$^{+119}_{-89}$& 1.99\\
cl 1354.2-1230    & 0.76 &648$^{+105}_{-110}$& 1.08\\
\hline
\end{tabular}
\caption{List of EDisCS clusters analyzed in this paper, with cluster
name, redshift, velocity dispersion and $R_{200}$ (from \citealt{halliday04,
milvang08,poggianti08}).
\label{tab:ed_cl}}
\end{table}

For the EDisCS mass-limited sample we use all photo-z members, following
the membership criteria described above. 

The choice to use the photo-z membership instead of spectroscopically
confirmed members is dictated by
the fact that otherwise the number of galaxies would be low,
not allowing a statistically meaningful analysis.  

Moreover, the spectroscopic magnitude limit ranges
between I=22 and I=23 depending on redshift, and the corresponding
spectroscopic stellar mass limit is $M =10^{10.6} M_{\odot}$
\citep{vulcani10}.  
The photo-z tecnique allows us to push the mass limit to much
lower values than the spectroscopy. 
We adopt a conservative magnitude completeness limit for the EDisCS 
photometry equal to $I \sim
24$ (though the completeness remains very high to magnitudes significantly
fainter than $I = 24$, White et al. 2005).  
We consider the most distant cluster, {\it cl 1216.8-1201}, 
that is located at $z\sim0.8$ and determine the value of the
mass of a galaxy with an absolute B magnitude corresponding to $I=24$,
and a rest-frame color $(B-V) \sim 0.9$, which is the reddest color of galaxies
in this cluster.
In this way, the EDisCS mass completeness limit based on photo-z's 
is $M_{\ast} = 10^{10.2} M_{\odot}$. This is the mass limit we adopt
for our analysis.
As we discuss in \cite{morph},
spectroscopic and photo-z techniques give very consistent results
for the galaxy mass functions in the mass range in common. 
Also comparing the ellipticity distribution determined using spectroscopic
and photo-z data down to the spectroscopic mass limit, 
we find that they are not statistically different: a Kolmogorov-Smirnov
test cannot reject the null hypothesis that the distributions are drawn
from the same parent distribution with a probability of $\sim 22\%$ 
(for details on the K-S test see \S \ref{res_mass}).
This gives additional support to our choice to use photo-z data.
 
As for the WINGS mass-limited sample, both BCGs and all galaxies at radii
greater than $r=0.6R_{200}$ have been excluded from the analysis.
Table \ref{tab:ed_cl} presents the list of clusters used and some
relevant values.

The final mass-limited EDisCS sample of galaxies with a measured
ellipticity for $M_{\ast} \geq 10^{10.2} M_{\odot}$
consists of 206 early-type galaxies, 145
of which are classified as ellipticals and 61 as S0s (see
\tab\ref{numb}).

\subsubsection{Magnitude-``delimited'' sample}

For the  magnitude-``delimited'' sample at high-z, 
to follow the same criteria of \cite{h09},
we do not consider the  photo-z membership, but
we exclude only those galaxies that have been identified
spectroscopically as non-members.
Then, we use all early-type galaxies within $2\sigma$ of the red sequence
and with $-19.3 >M_{B} +1.208z >-21$.

To determine the red sequence of each cluster, as
\cite{mei09} did, we use only spectroscopic members of our clusters 
\citep{halliday04, milvang08}.
We build color-magnitude diagrams using
the $R-I$ color (that corresponds to $\sim B-V$ in WINGS). 
Only for {\it cl 1232.5-1250} we use the $V-I$ color  
because the $R$ band is not available.  

Similarly to what we do for WINGS, we determine the red sequence by performing
a weighted least-square-fit of our data. However,  
since the resulting red sequences are not always reliable, for all
clusters but {\it cl 1232.5-1250}, for which we use a different color, we 
determine a mean slope using only 
{\it cl 1216.8-1201} and {\it cl 1054.4-1146}, two clusters located almost
at the same redshift for which 
the red sequence is well defined from the spectroscopy, and adopt this 
slope for all clusters.
The mean dispersion is determined averaging the dispersion
of all clusters except  {\it cl 1354.2-1230} that 
has been excluded because, having too few
points, it would give too small a value of the dispersion.
Then, we determine separately the red sequence of each cluster, using the slope and
the dispersion just determined and finding 
the most appropriate value of the intercept. 
Obviously, the red sequence for {\it cl1232.5-1250} 
is determined separately using its spectroscopic
data.

Subsequently, we consider only galaxies within $R_{200}$ (as for WINGS), 
instead of 2$R_{200}/\pi$ as Holden et al. (2009) did, to
improve the statistics.

Our final sample consists of 101 ellipticals and 43 S0s, 
for a total of  144 early-type galaxies (see \tab\ref{numb}).

\section{Results: the ellipticity evolution in Mass-limited samples} \label{res_mass}
\begin{figure*}
\centering
\includegraphics[scale=0.65]{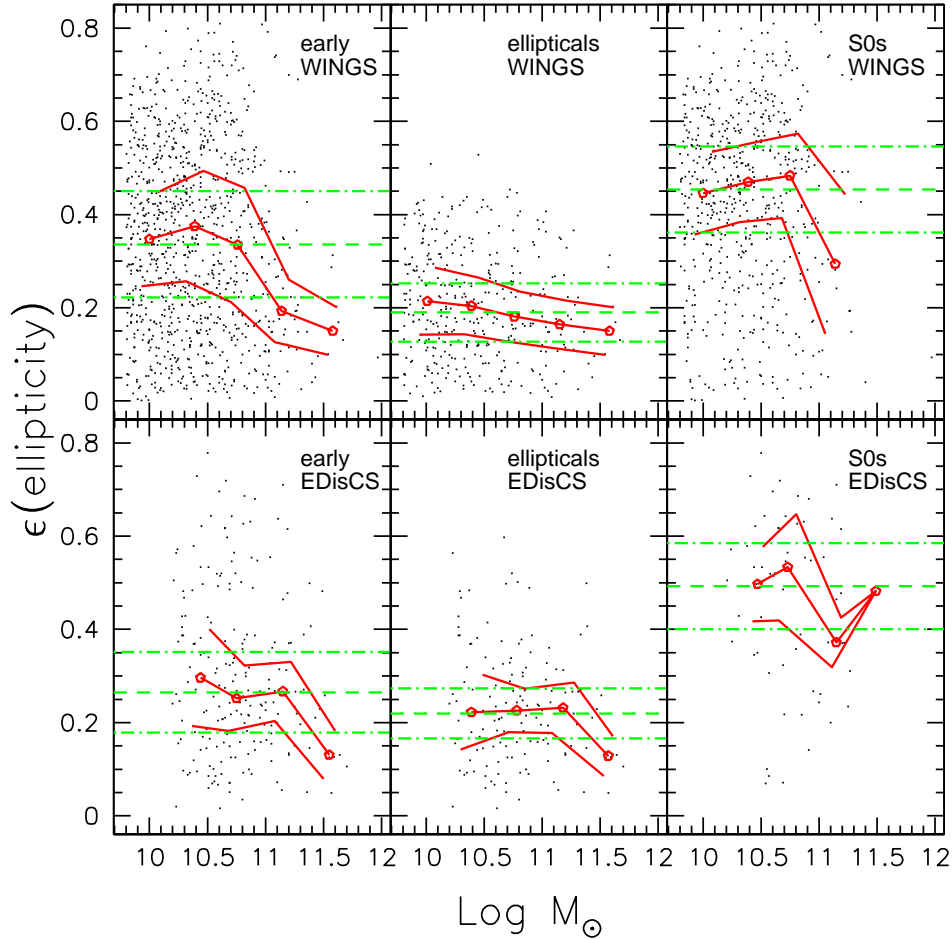}
\caption{Ellipticity vs mass in the mass-limited sample. Black points:
galaxies. Red solid lines: median and 1 $\sigma$ values, where $\sigma$ is the rms, estimated in mass bins.  
Green dashed lines: median and 1 $\sigma$ values computed
over the whole mass range.  Top panels:
WINGS data.  The WINGS median are corrected for spectroscopic incompleteness.
Left: early-type galaxies, Central: elliptical
galaxies, Right: S0 galaxies. Bottom panels:
EDisCS data (panels as for WINGS). 
\label{ellitticita}}
\end{figure*} 

In this section, we analyze the ellipticities of galaxies in our mass
limited samples.

\subsection{Ellipticity and S0/E number ratio
as a function of galaxy stellar mass}

\fig\ref{ellitticita} 
shows the trend of ellipticity as a function of galaxy stellar mass
for early-type, elliptical and S0 galaxies for the WINGS and EDisCS samples, 
above their mass completeness limits.  We compute the
median values of ellipticity both over the whole mass range (green
dashed lines) and in mass bins of 0.4 dex (red solid lines).

At both redshifts, the trend of the ellipticity 
of all early-type galaxies together clearly depends on galaxy mass.
Less massive galaxies tend to have ellipticities that extend to much higher
values compared to higher mass galaxies which populate 
only the lower end of the range.
Considering only ellipticals, the trend is much
less striking, though still present over the whole mass range
in WINGS, while a drop is observed 
only above $M_{\ast}\sim10^{11.2}M_{\odot}$ in EDisCS. 
The median ellipticity of S0s shows no clear trend with galaxy mass,
at least below $M_{\ast}\sim10^{11.1}M_{\odot}$ in WINGS. 
At high masses, both the apparent fall in WINGS and the rise in EDisCS
may be simply due to low number statistics. 

We note that 
ellipticals always have an ellipticity lower than 0.6 and
mostly below 0.4, while S0s cover a wider range of ellipticities, with
the majority being concentrated at high values of ellipticities, above
0.4.  Furthermore, ellipticals reach higher mass values than S0s (see
\cite{morph} for the mass distribution of ellipticals and S0s in these
samples, see also \S5).

Clearly, the strong trend of ellipticity
with mass observed in early-types is due
both to the trend of ellipticity of elliptical galaxies with mass,
and, mostly, to the fact the ellipticals and S0s are found in
different proportions at different masses: S0 galaxies, with their
average higher ellipticities, become more frequent going to lower
masses.  \fig\ref{frac} shows the ratio of the number of S0 to
elliptical galaxies at different masses. In WINGS (left panel), the S0/Ell ratio
strongly depends on mass: at higher masses there are proportionally
more elliptical galaxies than at lower masses. In the highest mass bin, the
ratio drops to $\sim$0, indicating that there are almost only elliptical galaxies, while
at $M_{\ast}\sim10^{10.5}M_{\odot}$ S0s are twice as numerous as
ellipticals. In contrast,  in EDisCS (right panel) we find that the trend
is almost flat up to $M_{\ast}\sim10^{11.5}M_{\odot}$ and S0s are less than
half of the ellipticals.

EDisCS clusters are seen at an epoch prior to the 
build up of the S0 cluster population and  \fig\ref{frac} clearly shows
that such build-up occurs mainly at masses below $10^{11} \, M_{\odot}$.

\begin{figure}
\centering
\includegraphics[scale=0.4]{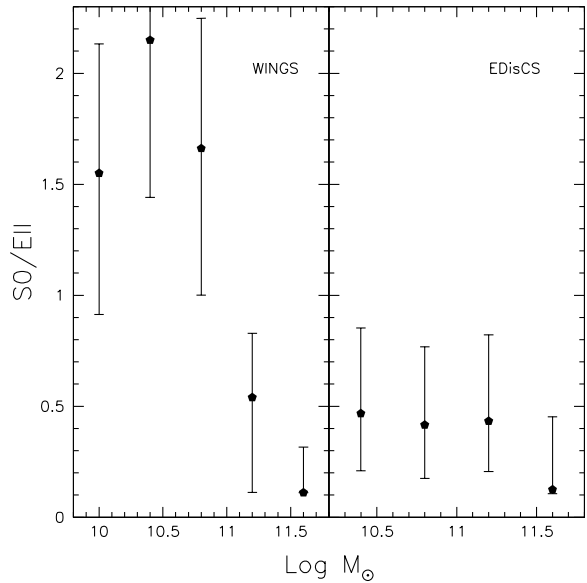}
\caption{Ratio of the number of S0/Ell galaxies
at different masses both for WINGS (left panel)
and for EDisCS (right panel) in our mass-limited sample. Errors are binomial \citep{gehrels86}. \label{frac}}
\end{figure}

\begin{table*}
\centering
\begin{tabular}{l|ll|l|}
\hline
 	& \multicolumn{2}{c}{WINGS} &\multicolumn{1}{c}{ EDisCS} \\
	&\multicolumn{1}{c}{$M_{\ast}/M_{\odot}\geq 10^{9.8}$}& \multicolumn{1}{c}{$M_{\ast}/M_{\odot}\geq10^{10.2}$}& \multicolumn{1}{c}{$M_{\ast}/M_{\odot}\geq10^{10.2}$}\\
\hline
ellipticals 	&0.190$\pm$0.011  &0.179$\pm$0.011  &0.220$\pm$0.011  \\
S0s		&0.454$\pm$0.014  &0.462$\pm$0.015 &0.493$\pm$0.032 \\
early-types	& 0.336$\pm$0.012 &0.328$\pm$0.016 &0.265$\pm$0.013 \\
\hline
\end{tabular}
\caption{Ellipticity median values for both mass-limited samples with errors defined with bootstrap resampling. For WINGS, medians are computed
taking into account the weights. For WINGS,  also 
values above the EDisCS mass limit are given.
\label{ellvalu}}
\end{table*}

\subsection{The evolution of the median ellipticity and of the ellipticity
distributions} \label{result}

\tab\ref{ellvalu} summarizes the median values of ellipticities for both
samples over the whole range of masses. 
Errors are estimated using the bootstrap resampling method. We adopt these estimates 
because we want to characterize the errors on the medians and not 
the dispersion of the points around the median value (that is the standard deviation).
In WINGS, the choice of the mass limit  ($M_{\ast} \geq 10^{9.8}M_{\odot}$ or $M_{\ast}\geq10^{10.2}M_{\odot} $) 
does not alter the final results. 

Comparing low- and high-z, the median 
ellipticity of S0s is compatible within the errors at the two redshifts, while
it slightly changes with redshift for ellipticals, and more noticeably
for early-types.
In particular,  it slightly {\it decreases} 
going to the current 
epoch for ellipticals, while it  clearly
{\it increases} for the early-types. 
This raise for early types isdue to the fact that, 
as shown in \fig\ref{frac}, the fraction of S0s increases at low-z, 
mainly in the low mass range.
Since S0s are more flattened than ellipticals, the
median ellipticity of early-types shifts to higher values at low redshift.
As for the evolution of the median of
elliptical galaxies, this will be discussed later in
this section.

We now compare the high- and low-z ellipticity distribution,

  to see if it evolves.
At both redshifts, 
we consider only galaxies above the common 
mass limit that is $M_{\ast}=10^{10.2}M_{\odot}$.

\begin{figure}
\centering
\includegraphics[scale=0.4]{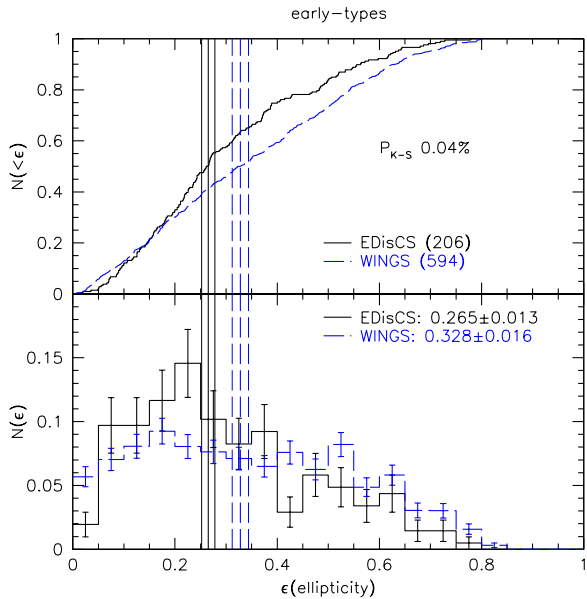}
\caption{Comparison of ellipticity distributions of early-type
galaxies in the mass-limited samples. Medians and bootstrap errors are
also indicated.  Black solid lines: EDisCS; blue dashed lines:
WINGS. Top panel: cumulative distributions of ellipticity. $P_{K-S}$
is the probability that the two distributions are drawn from the same
parent distribution. Numbers in brackets are the number of
galaxies in the considered samples.  Bottom panel: histograms in bins
of 0.05 dex, normalized to 1.\label{early_ell}}
\end{figure} 

We build both the cumulative distributions and histograms (in bins of
ellipticity equal to 0.05) for each class of galaxies analyzed. For
WINGS, both of them take into account the spectroscopic completeness
weights.

\fig\ref{early_ell} shows how the ellipticity distribution of
early-type galaxies evolves with redshift.  
As expected given the evolution of the median ellipticity, 
there are proportionally more galaxies with higher ellipticities at low-
than at high-z, indicating that low-z early-type galaxies
are on average more flattened. The overall WINGS 
distribution (blue dashed lines) is quite flat,
in particular at intermediate values of ellipticity.  Instead,
the ellipticity distribution of EDisCS early-type galaxies (black solid
lines) shows a peak around $\epsilon \sim 0.2$.  

To quantify the differences between the two distributions, we perform
a Kolmogorov-Smironov (K-S) test.\footnote{The standard K-S, in building the
cumulative distribution, assigns to each object a weight equal to
1. Instead, our WINGS data are characterized by spectroscopic
completeness weights.  So, we modified the test, to make the relative
importance of each galaxy in the cumulative distribution depend on its
weight, and not being fixed to 1. In the following, we will always use
this modified K-S test. Obviously, when using photo-z's all galaxies
have a weight equal to 1, and using the modified test is equivalent to
using the standard one.}  
Throughout this paper, we will consider significantly different two distributions
if the K-S test gives a probability $<5\%$.  
For early-type galaxies, the K-S test
allows us to exclude the similarity of the ellipticity distributions 
at the two redshifts, giving a probability 
of $\sim 0.04\%$.

\begin{figure*}
\begin{minipage}[c]{175pt}
\centering
\includegraphics[scale=0.4,clip = false, trim = 0pt 150pt 0pt 0pt]
{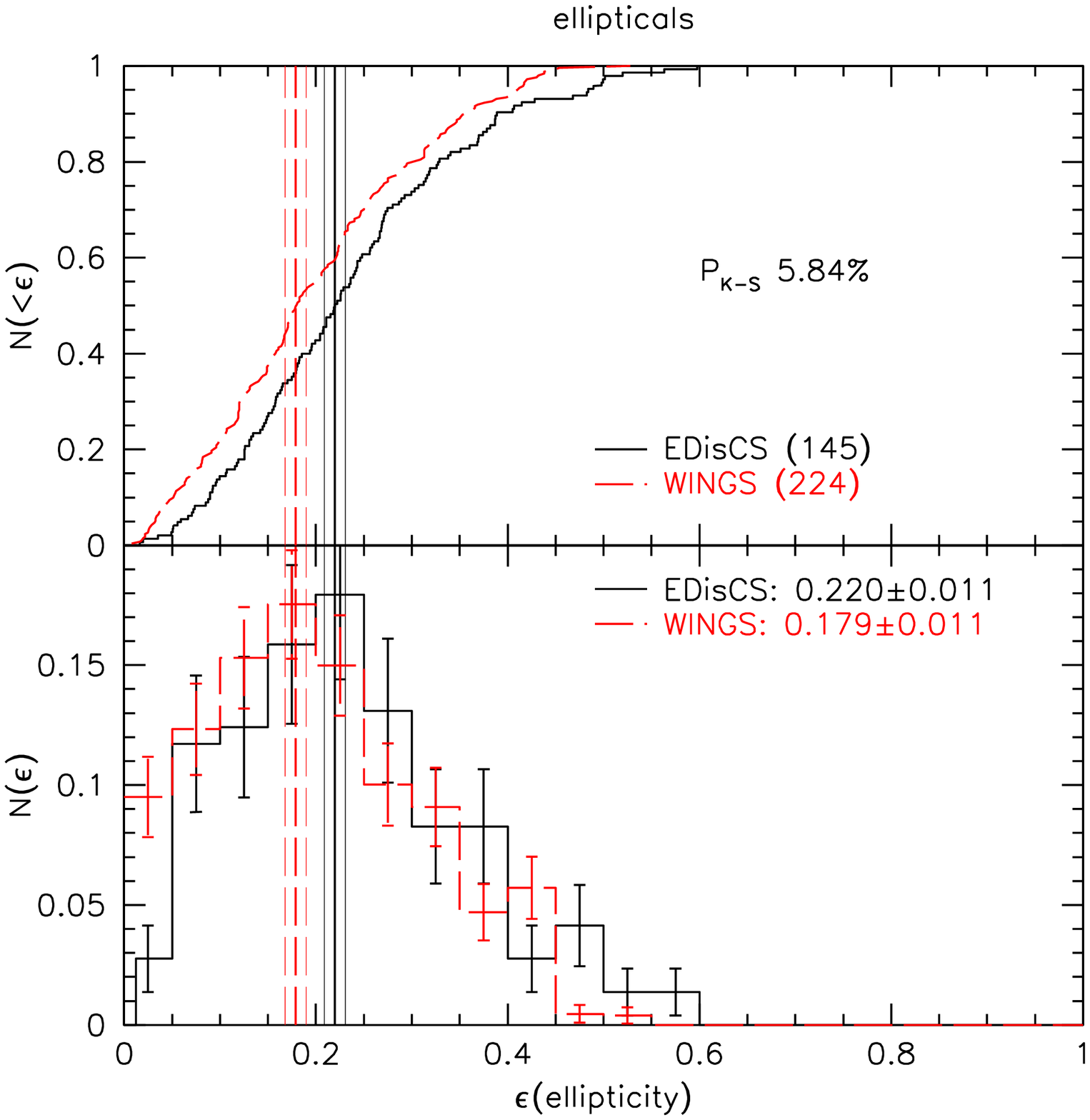}
\vspace*{2.2cm}
\end{minipage}
\hspace{2.5cm}
\begin{minipage}[c]{175pt}
\centering
\includegraphics[scale=0.4,clip = false, trim = 0pt 150pt 0pt 0pt]
{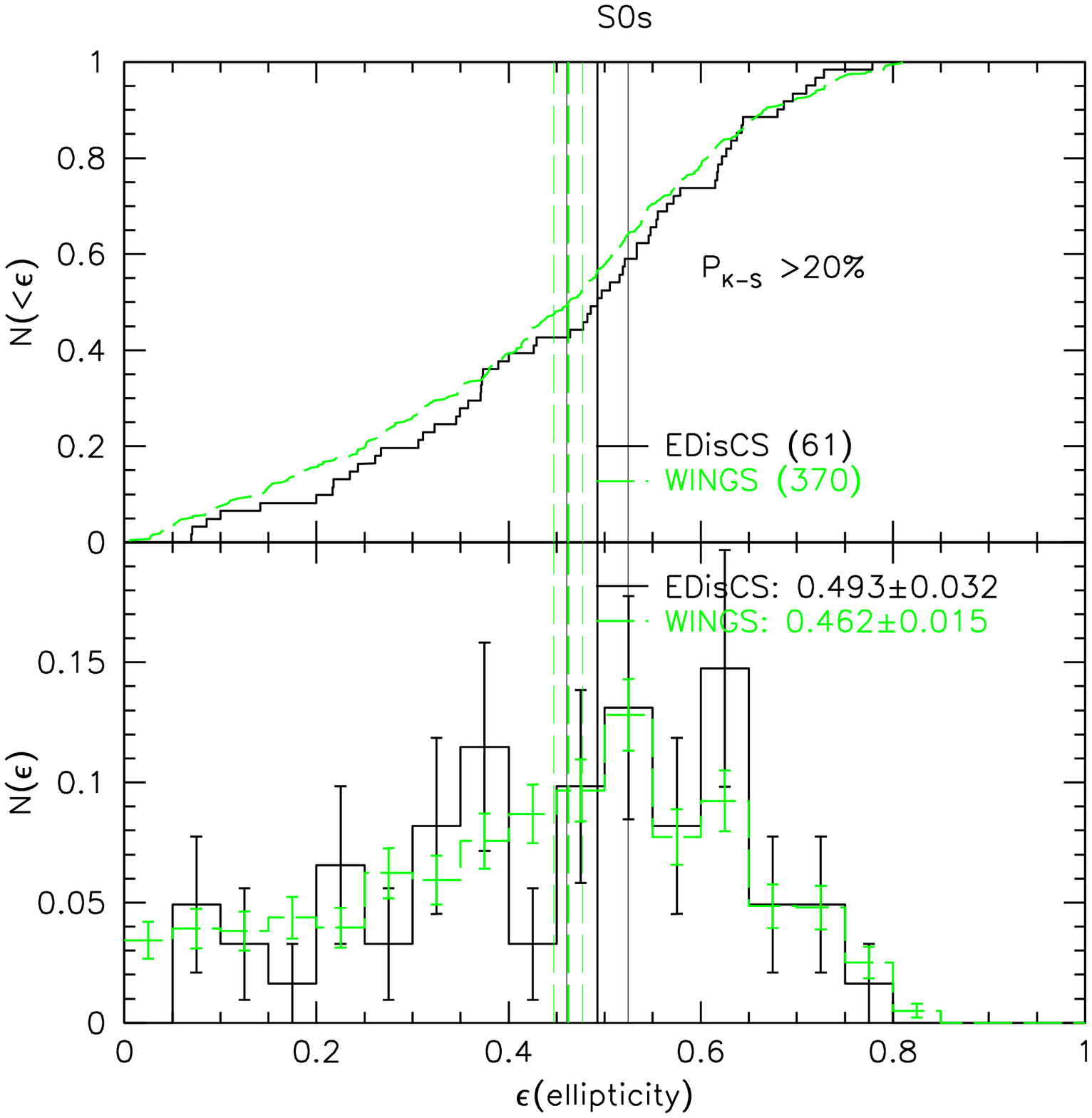}
\vspace*{2.2cm}
\end{minipage}
\hspace{1cm}
\caption{Comparison of ellipticity distributions of elliptical (left
panel) and S0 (right panel) galaxies in the mass-limited
samples. The medians and their bootstrap 
errors are also indicated.  Black solid
lines: EDisCS; red (in the left panel) and green (in the right panel)
dashed lines: WINGS. Top panels: cumulative distributions of
ellipticity. $P_{K-S}$ is the probability that the two distributions are
drawn from the same parent distribution. Numbers in the brackets are
the number of galaxies the considered samples.  Bottom panels:
histograms in bins of 0.05 dex, normalized to 1. \label{ellS0_ell}}
\end{figure*}

Next, we investigate whether the observed differences are due
to an evolution of the ellipticity distribution of ellipticals, of
S0s, or both.
\fig\ref{ellS0_ell} shows that 
for ellipticals results are ambiguous: the K-S test, giving a 
probability of 5.84\% of similarity
of the distributions, is not strictly conclusive. 
In contrast, 
for S0s 
the distributions are compatible with being similar
(P$_{K-S}\geq$20\%).
EDisCS S0s are very few and this could 
influence the results of the K-S test; 
however, the cumulative distributions appear to resemble each other, 
indicating that the result of the K-S test
should be reliable.

We wish to go deeper into our analysis, trying
to understand if the K-S results are confirmed and above all if they
are driven by a different shape of the distributions or simply by a
different location of the two populations. To do this, we perform
two other statistical non-parametric tests (i.e. they do not assume
the normal distribution) which make no assumptions about the
distributions of the populations.

In Appendix \ref{app_test} we present the detailed analysis of the 
\cite{moses63} and \cite{mann47} tests. In summary, 
although the K-S test is
inconclusive, from these additional tests it emerges that  
some mild differences
exist in the ellipticity distribution of ellipticals at high- and low-z.

\subsubsection{Round ellipticals at low-z}

Inspecting the histograms in \fig\ref{ellS0_ell}, it is clear that,
at both redshifts, there are no ellipticals with  $\epsilon \geq$
0.6 and both distributions are peaked around $\epsilon \sim 0.2-0.3$.
It seems that the greatest differences are confined to the
extremes of the distributions: 
in the highest ellipticity bins there are
proportionally more EDisCS elliptical galaxies than WINGS', and
more noticeably,
in the first bin there are proportionally more WINGS galaxies with
$\epsilon \leq0.05$ than EDisCS galaxies.
It is important to stress that this goes in the opposite
direction of what could be expected from morphological classification
biases at high-z: face-on S0s would be systematically
mistaken for ellipticals more frequently at high- than at low-z. 
Analyzing more accurately the first bin (plot not shown), we find that 
it is mainly dominated
by a second peak 
around $\epsilon\sim0.03$ that instead is not detected in the 
EDisCS distribution.
At the moment, we are not able to explain why at low redshift 
there is an exceeding population
of rounder ellipticals compared to high-z; anyway we think it is real, 
having accurately checked both morphologies 
and ellipticities for those galaxies.\footnote{We 
find a similar peak also in our low-z magnitude-``delimited'' 
sample (see \S4) and also
in the low-z sample analyzed by \cite{h09} (see \S6), so it can
neither due to the adopted selection criteria nor to a bias in our samples.} 

We have also visually inspected the ellipticity profiles of these round WINGS
ellipticals.
For a few of them, the profile is altered by crowding, for some other
the twist of the isophotes is very marked and 
the ellipticity value strongly changes
with radius.
However, 60\% of the analyzed galaxies really have a very low ellipticity 
at all radii. 
Analyzing and understanding this population goes beyond the scope of this work
and it will be discussed in a forthcoming paper. Here, we just notice 
that this population exists and that it could be the result of  
dry merger events \citep{vd99, tran05}
which might preferentially result in round galaxies building over time.

If elliptical galaxies with $\epsilon<0.05$ are excluded, the
WINGS and EDisCS ellipticity distributions for ellipticals 
are indistinguishable (the K-S test gives a probability $>20\%$
that the poopulations are drawn from the same parent distribution).

\subsection{What drives the evolution of the ellipticity distribution of
early-type galaxies: the evolution of the galaxy mass distributions,
or the evolution of the relative proportions of ellipticals and S0s?}
To summarize the most important points, in the previous sections we have found that,
for mass-limited samples,
the ellipticity distribution of early-type galaxies strongly
changes with redshift. 
The ellipticity distributions of ellipticals and S0s 
on the whole do not evolve significantly, although
for ellipticals there is a non negligible shift of
the medians of the distributions due to a relative
excess of round ellipticals at low-z compared to high-z.

We have found in \fig\ref{ellitticita} that above all for early-types 
there is a trend of
ellipticity with mass, and that these trends are different in EDisCS and in
WINGS. Moreover, as we
have discussed at length in a previous paper \citep{morph}, 
WINGS and
EDisCS have different galaxy stellar mass distributions.  We have found that
the mass distribution of {\it each} morphological type evolves with
redshift and that all types have proportionally more massive galaxies
at high- than at low-z.  As a consequence, if we want to understand
the origin of the ellipticity distribution of galaxies of
different morphological types and at different redshifts, we have to
try to disentangle the influence of the evolution of the mass distribution 
from the effects of the morphological evolution.

\subsubsection{The evolution of the galaxy mass distribution}
First, we analyze separately ellipticals and S0s.
We perform 1000 Monte Carlo simulations extracting
randomly from the WINGS sample a subsample with the same mass
distribution as the EDisCS sample, separately for the two
different morphological classes. Each time we extract from WINGS
the same number of galaxies that are in the EDisCS sample (i.e. 
145 ellipticals and 61 S0s). 
For each simulation we determine the ellipticity distribution, the
median value of the ellipticity and we perform a K-S test to compare
the result of the simulation with the EDisCS data-set.  Then, taking
into account all the simulations, we determine both the median value
of the K-S and the fraction of simulations that give a conclusive
statement (see \tab\ref{K-Stab}).

\begin{figure*}
\begin{minipage}[c]{175pt}
\centering
\includegraphics[scale=0.4,clip = false, trim = 0pt 150pt 0pt 0pt]
{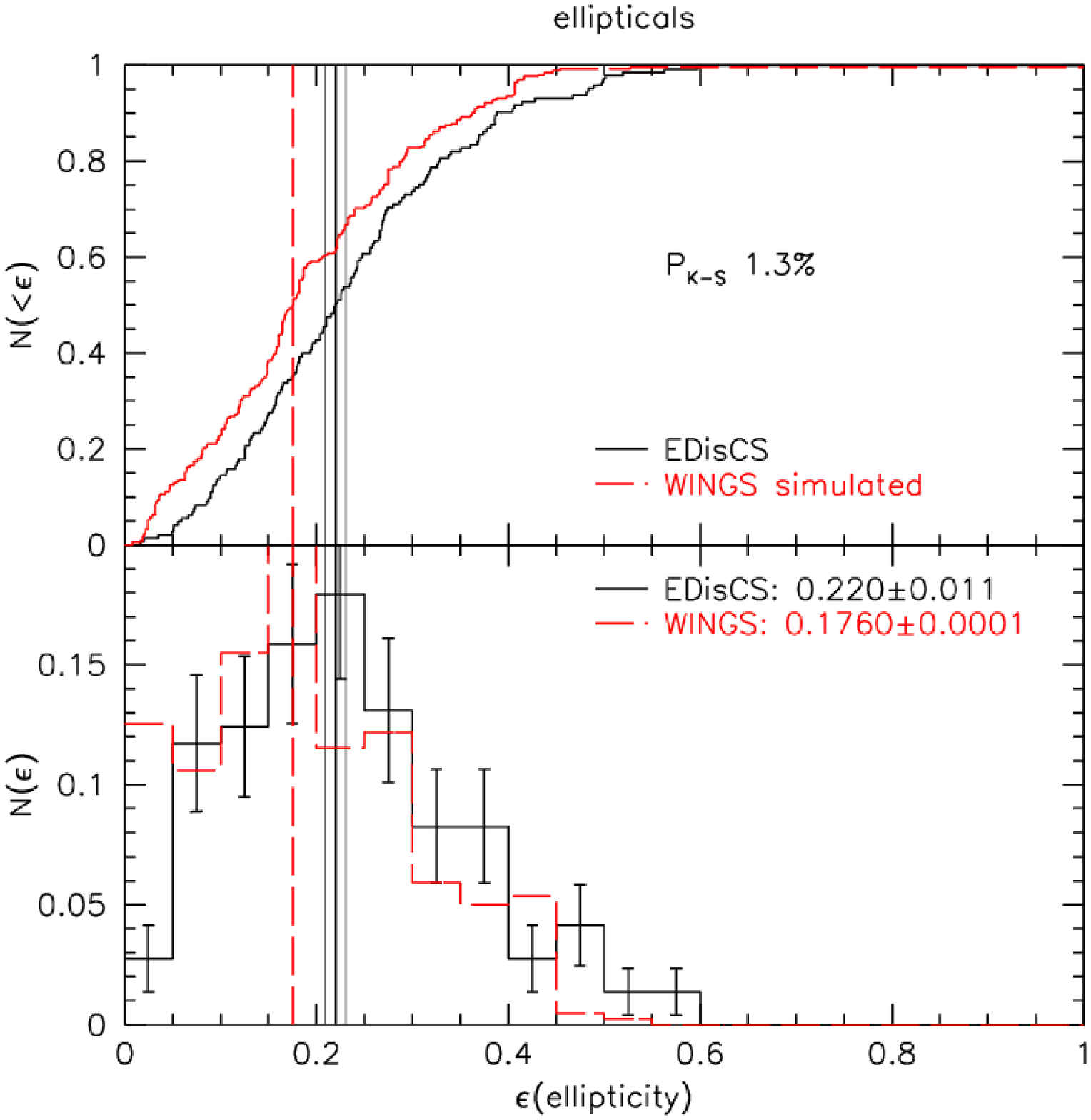}
\vspace*{2.2cm}
\end{minipage}
\hspace{2.5cm}
\begin{minipage}[c]{175pt}
\centering
\includegraphics[scale=0.4,clip = false, trim = 0pt 150pt 0pt 0pt]
{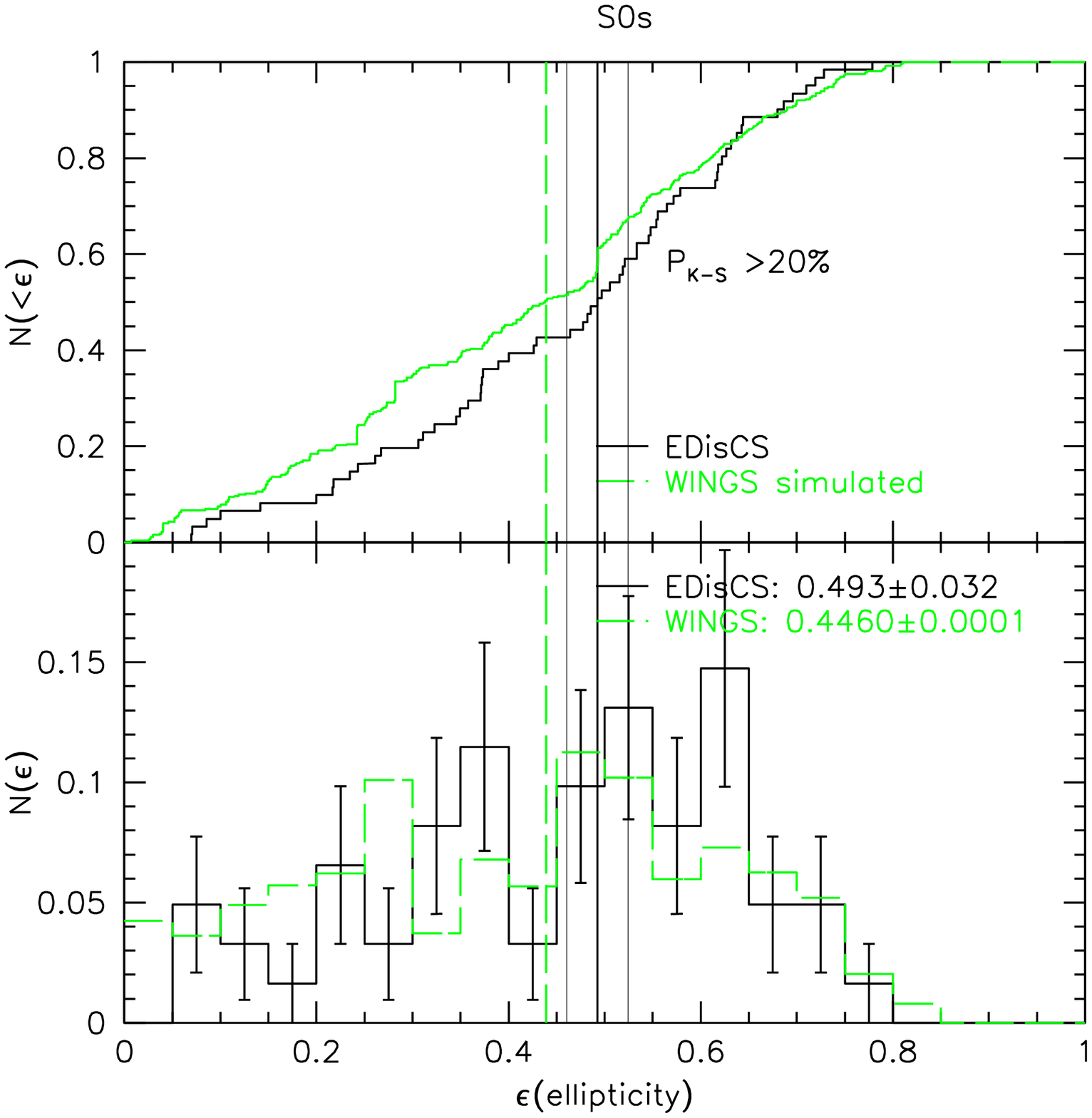}
\vspace*{2.2cm}
\end{minipage}
\hspace{1cm}
\caption{Ellipticity distribution of elliptical 
(left panel)  and S0 
(right panel) galaxies assuming for WINGS the same mass distribution  as 
EDisCS (cf. with \fig\ref{ellS0_ell}). Panels and symbols are the same as in \fig\ref{ellS0_ell}.
The plotted WINGS histogram and cumulative distributions are 
the average of the 1000 Monte Carlo simulations. The $P_{K-S}$ is the median K-S value.
\label{ellS0_sim}}
\end{figure*}

\fig\ref{ellS0_sim} shows our results for the two morphological
classes. 
\begin{table*}
\centering
\begin{tabular}{lcccc|cc}
\hline 
 		& \multicolumn{4}{c}{$P_{K-S}$}  \\
		& $<1\%$ & $1\%-5\%$ &$5\%-10\%$& $>10\%$ && median \\
\hline \hline
& \multicolumn{6}{c}{mass-matched simulations} \\
ellipticals 	&50\% 	&30\% 		&9\% 		&11\%	&&1.3\%\\
S0s		&1\% 	&6\% 		&4\% 		&89\% 	&&$>$20\%\\
early-types	&18\% 	&30\% 		&20\% 		&32\%	&&5.2\%\\
\hline 
& \multicolumn{6}{c}{morphology-matched simulations} \\
early-types	&0\% 	&3\% 		&8\% 		&89\%	&&$>$20\%\\
\hline
\end{tabular}
\caption{Results of the Kolmogorov-Smirnov test performed on the 1000
mass-matched and morphology-matched simulations (see text in \S3.3).
$P_{K-S}$ is the probability that the WINGS and EDisCS 
distributions are drawn from the same parent distribution. \label{K-Stab}}
\end{table*}
The plotted WINGS histogram and cumulative distribution 
(dashed lines) are the
average of the 1000 Monte Carlo simulations.
The K-S probabilities and WINGS medians
reported in the plots are the medians of the values of all 
the simulations.
Adopting the same mass distribution, 
for elliptical galaxies, 
the match of the mass distributions at high- and low-z makes 
the ellipticity distribution of WINGS galaxies 
(red dashed lines in the left panel of \fig\ref{ellS0_sim})
become significantly different from the EDisCS one. 
The K-S test gives conclusive results ($P_{K-S}\leq 5$\%) in 80\% of 
the simulations, with a median value of 1.3\%, 
 rejecting the null hypothesis of similarity of the two distributions. 
This seems to imply that WINGS ellipticals tend to be
on average rounder than EDisCS ellipticals of the same mass.
This effect is mitigated in the observed WINGS vs EDisCS distributions
(\fig\ref{ellS0_ell}) (yielding a non-significant K-S test) by the fact that
a) there are proportionally more less massive galaxies at low-
than at high-z (for details on the mass functions see \citealt{morph}); and
b) low-mass ellipticals are more flattened (on average) than high-mass ones.
Hence, the increase in the number of low mass ellipticals at low-z
largely compensates the existence of a significant evolution in
the ellipticity distribution of ellipticals with the same mass
distribution (\fig\ref{ellS0_sim}), and produces an ambiguous or at best weak
evidence for evolution in our analysis of \S3.2.

If we neglect galaxies with $\epsilon<0.05$,\footnote{In this way we wish to 
compare the whole
general distribution without being too much influenced by 
 galaxies located in only one bin.} for which the most oustanding differences
are detected (see \S3.2),
adopting the same mass distribution, the median value of the K-S test 
is $\sim 8\%$, indicating that the major differences between the
ellipticals at high- and low-z is indeed the enhanced population of
round ellipticals at low-z.
Turning to S0s (green dashed lines in right panel of Fig. \ref{ellS0_sim}), 
the mass matched ellipticity distributions of WINGS and EDisCS
remain statistically indistiguishable:
the K-S test cannot reject the null hypothesis, 
giving a probability $\geq 5\%$ in 93\% of the simulations and a median
value $>20\%$.

Now we wish to test if the different mass distribution at different redshifts 
alters the ellipticity 
distribution of early-types, so 
 we 
put together WINGS ellipticals and S0s,  
being sure to extract randomly galaxies in order to have the same EDisCS mass
distribution and maintaining  the WINGS morphological fractions 
(i.e. $\sim 40\%$
ellipticals and $\sim 60\%$ S0s - see \tab\ref{frac_morf}). 
In this way we test whether 
the observed evolution of the ellipticity distribution of early-type galaxies
can be entirely explained by the evolution of the mass distributions.

\begin{figure}
\centering
\includegraphics[scale=0.4]{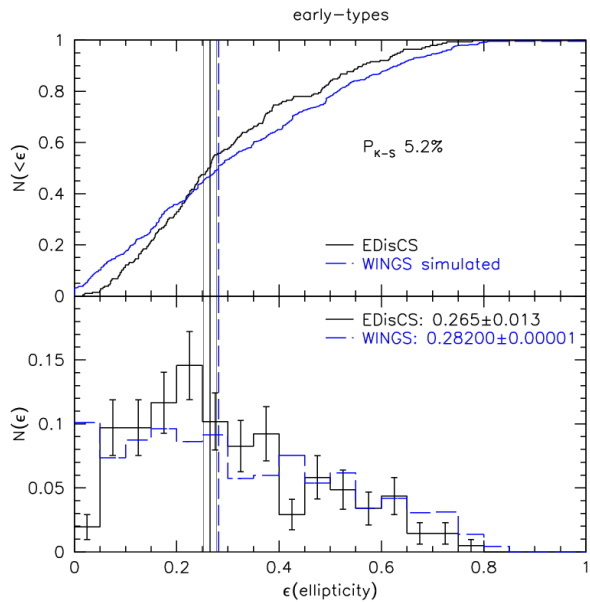}
\caption{Ellipticity distribution 
of early-type galaxies assuming for WINGS
the same mass distribution as EDisCS galaxies  
and maintaing the WINGS morphological mix (cf. with
\fig\ref{early_ell}). Panels and symbols are the same as in \fig\ref{early_ell}.
The plotted WINGS histogram and cumulative distributions are 
the average of the 1000 Monte Carlo simulations. The $P_{K-S}$ is the median
K-S value.\label{early_sim}}
\end{figure}

In \fig\ref{early_sim} 
the K-S gives 
a probability  $\leq5\%$ that the two distributions are driven 
from the same parent distribution
in 48\% of the simulations (\tab\ref{K-Stab}), 
and a median probability for
all simulations of 5.2\%. 
Excluding galaxies with $\epsilon <0.05$, the K-S results remain similar 
(median $P_{K-S} \sim 4\%$,
 $P_{K-S} <5\%$ in 55\% of the simulations, plot not shown). 
These values 
suggest that
the different mass distribution at the different redshifts 
influences at some level the evolution of the ellipticity distribution, 
even if probably it is not the main factor as it
cannot fully explain the observed evolution.
 
\begin{table*}
\centering
\begin{tabular}{|l|c|c|||c|ccc||c|c}
\hline
& \multicolumn{5}{|c|}{WINGS} 	& & \multicolumn{2}{|c|}{EDisCS} \\ 
& \multicolumn{4}{c} {Mass-limited} &Mag-delimited && Mass-limited & Mag-delimited\\
& \multicolumn{2}{|c|}{$M_{\ast} \geq 10^{9.8} M_{\odot}$} & \multicolumn{2}{|c|}{$M_{\ast} \geq 10^{10.2} M_{\odot}$}&&& $M_{\ast} \geq 10^{10.2} M_{\odot}$ &\\
\hline
            	&\%$_{obs}$     & \%$_{w}$	&\%$_{obs}$	& \%$_{w}$ 	&\%&&\%&\%		\\
\hline	
ellipticals	&38.3$\pm$1.7\% & 37.9$\pm$1.3\%& 37.7$\pm$2.1\% 	&37.1$\pm$1.7\%	&38.8$\pm$1.3\%	&&70.4$\pm$3.3\% 	&70.1$\pm$4.0\%\\
S0s		&61.7$\pm$1.7\% &62.1$\pm$1.3\% &62.3$\pm$2.1\%	&62.9$\pm$1.7\%	&61.2$\pm$1.3\%	&&29.6$\pm$3.3\%	&29.9$\pm$4.0\%\\
\hline
\end{tabular}
\caption{Relative 
morphological fractions of galaxies in both mass and
mag-(de)limited samples. For the WINGS mass-limited sample, also 
numbers above the EDisCS mass limit are given. In both cases, both observed and
completeness-weighted numbers are listed. Errors are binomial, as defined in \citet{gehrels86}. \label{frac_morf}}
\end{table*}

\subsubsection{The evolution of the morphological fractions}
We now wish to assess the role played by the evolution with redshift 
of the relative morphological fractions.
In \tab\ref{frac_morf} we show how much the morphological fractions
change with time, both for the mass-limited sample and for the magnitude-``delimited'' one (see \S\ref{sec_mag}).
We observe that while at high redshift ellipticals are more common than S0s ($\sim 70\%$ and $\sim 30\%$ respectively), in the Local Universe 
S0s dominate, representing  $\sim 62\%$ of the early-types.

 To analyze the importance of this evolution,
 we now perform a second test,
extracting randomly from the WINGS data-set a subsample of galaxies
with  approximately the same relative fraction of S0s and ellipticals as 
EDisCS (i.e. in each simulation
we extract 70 ellipticals and 30 S0s from the WINGS sample)
and paying attention to maintain the WINGS mass distribution.
In this way we wish to test whether the morphological evolution
can account for the ellipticity evolution of early-type galaxies,
letting the mass distribution to naturally evolve.
We perform 1000 such simulations.


We then compare the ellipticity distribution of the ``modified'' WINGS sample, 
to the real EDisCS one. Considering all galaxies
(also those with $\epsilon<0.05$), (plot not shown)
the K-S test is conclusive ($P_{K-S}\leq 5\%$) in 69\% of the simulations, 
while the median value of the probability is 0.2\%, indicating that even assuming the same morphological
fraction in the two samples, differences between the two ellipticities distributions are still detected.

Next, 
we exclude from our analysis those galaxies (both ellipticals and S0s)
with $\epsilon<0.05$, 
since their presence likely alters the final results. In fact, 
increasing the number of elliptical galaxies 
in WINGS from being $\sim$40\% to $\sim$70\% of the whole
population, the contribution of galaxies with $\epsilon\sim0.03$ is hugely
magnified and it would strongly influence the whole population. 
For galaxies with $\epsilon\geq 0.05$, \fig\ref{early_sim_m} 
shows that, if at both redshifts we had the same fractions of
ellipticals and S0s, the ellipticity distributions for early-type galaxies
would be indistinguishable. 
The K-S test is conclusive ($P_{K-S}\leq 5\%$) only in 3\% of the simulations, 
while the median value of the probability is $>20\%$.

Importantly, doing the same in the observed distributions and excluding
all galaxies with $\epsilon<0.05$ from Fig.~7, the low- and high-z 
early-type ellipticity distributions remain significantly different,
with a K-S test probability to be drawn from the same parent distribution
of only 0.002\%.
Except for the excess of round ellipticals at low-z, the evolution
of the ellipticity distribution of early-type galaxies can be fully  explained 
by the morphological evolution.

From this whole section we conclude that 
it is mainly the relative contribution of each morphological type 
to the total that is responsible for the evolution of
the ellipticity distribution of early-type galaxies, even if
the role of the evolution of the  mass distribution
with redshift is non-negligible.
Morphology appears to be the most decisive factor
in the evolution of the ellipticity distribution of early-type galaxies:
WINGS and EDisCS have a different morphological  mix 
and 
their ellipticity 
distribution is regulated by the different relative proportions 
of ellipticals and S0s.

\begin{figure}
\centering
\includegraphics[scale=0.4]
{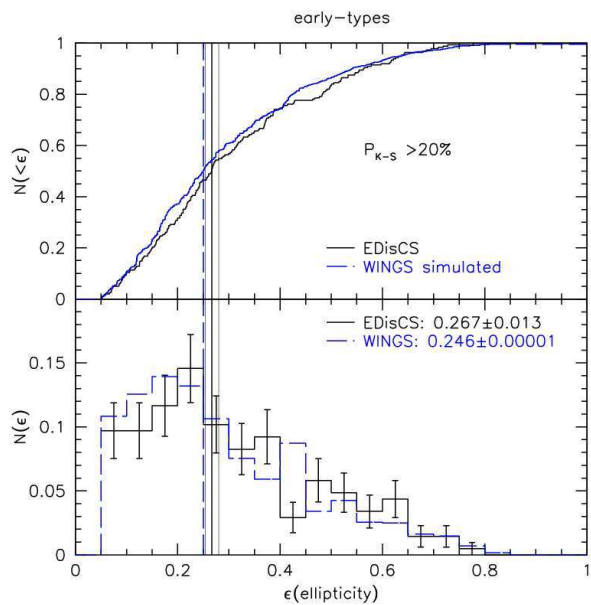}
\caption{Ellipticity distribution of EDisCS and WINGS early-type galaxies, assuming for
WINGS the  same fraction of ellipticals and S0s of EDisCS and maintaing the WINGS mass distribution (cf.
with \fig\ref{early_ell} and \fig\ref{early_sim}). 
Panels and symbols are the same as in \fig\ref{early_ell}.  Only galaxies with $\epsilon\geq0.05$
are considered.\label{early_sim_m}}
\end{figure}

\section{Results: the ellipticity distribution in the Magnitude-``delimited'' sample} \label{sec_mag}

\begin{table*}
\centering
\begin{tabular}{l|cccc|}
\hline
 		&WINGS 	&EDisCS	& \cite{h09} 	& \cite{h09}  \\
		& Mag-delimited 	&  Mag-delimited 	& low-z		&high-z\\	
\hline
ellipticals 	 &0.176$\pm$0.0073 &0.218$\pm$0.01 &0.18$\pm$0.010	&0.20 $\pm$0.010 \\
S0s		 &0.440$\pm$0.0088   &0.519$\pm$0.034&0.38$\pm$0.020  &0.47 $\pm$0.020 \\
early-types &0.300$\pm$0.0093  &0.265$\pm$0.021 	& 0.29$\pm$0.020  &0.30 $\pm$0.010 \\
\hline
\end{tabular}
\caption{Ellipticity median values for the magnitude-``delimited'' samples with errors 
estimated with bootstrap resampling. 
\citet{h09} is the sample we use as comparison for the magnitude-``delimited'' sample (for details, see text,  \S\ref{hold}).
\label{ellval}} 
\end{table*}

From EDisCS and WINGS, we have also selected a magnitude-``delimited''
sample of galaxies following
the criteria adopted by \cite{h09}, as described in \S2,
with the aim to directly compare
our finding with their results. 

First of all, we wish to check if the different selection criteria
implicate a change in our findings compared to the mass-limited sample.

In the magnitude-``delimited'' sample, we qualitatively find 
the same 
ellipticity-mass relation we found in 
\fig\ref{ellitticita} (plots not shown) for the mass-limited sample: no trend of ellipticity with mass  
for  S0s, slight trend for ellipticals, and a striking trend 
for early-type galaxies. 

\begin{figure}
\centering
\includegraphics[scale=0.4]{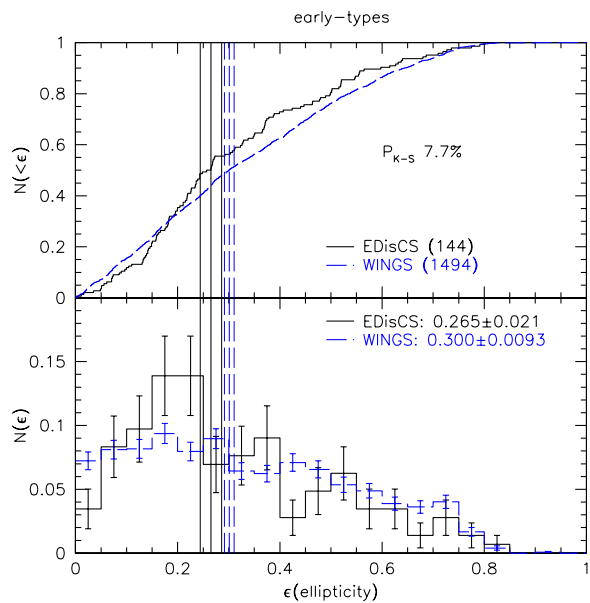}
\caption{Ellipticity distribution of early-type galaxies in the 
magnitude-``delimited'' samples. Panels and symbols are the same as in \fig\ref{early_ell}. 
Cf. with the mass-limited sample shown in \fig\ref{early_ell}.\label{early_win_ed}}
\end{figure} 

Values of the median ellipticities of the different morphological types
can be found in \tab\ref{ellval}.

Analyzing the ellipticity distributions in the magnitude-``delimited'' 
sample, for early-types 
(\fig\ref{early_win_ed}) 
the K-S test cannot detect a significant evolution 
with redshift, 
as it gives
a probability of $\sim 7.7\%$
that the two populations are drawn from the same parent distribution.  
However, the median ellipticities are different: the median
is {\it higher} at low-z (see also Table~9).

\fig\ref{ellS0_win_ed} shows separately the ellipticity distribution
for elliptical and S0 galaxies.
We find that the ellipticity of
elliptical galaxies evolves noticeably with redshift: in WINGS (red
dashed lines in the left panel of \fig\ref{ellS0_win_ed}) there are
proportionally more galaxies with low values of ellipticity, ($\epsilon<0.15$),
while in EDisCS (black solid lines) there is a noticeable 
peak around $\epsilon\sim 0.2$.
The K-S test excludes that elliptical galaxies have a common
ellipticity distribution at the different redshifts, giving a probability of
$\sim 0.41\%$ that they are drawn from the same parent distribution.
In contrast, the same test cannot distinguish any differences between the distributions of
S0s, giving a probability of $\sim 12\%$.  
However, we note that the median ellipticity of S0s is significantly
{\it lower} at low-z.\footnote{Performing the Moses test, 
we can conclude that
the early-types are unlikely 
to have the same scale parameter (with a probability of $<5\%$ in 53\% of the simulations), 
while for both ellipticals and S0s we cannot exclude a compatibility
of the scale parameter, with a probability $<5\%$ in 39\% and in 10\%
of the simulations respectively.
From the Mann- Whitney test, 
we cannot exclude 
compatible medians for early-types
(with a probability of $\sim 14\%$), while we can do it for  ellipticals and S0s (with a probability of 0.07\% and
3.4\%  respectively) (for details on the tests see Appendix \ref{app_test}).}

\begin{figure*}
\begin{minipage}[c]{175pt}
\centering
\includegraphics[scale=0.4,clip = false, trim = 0pt 150pt 0pt 0pt]
{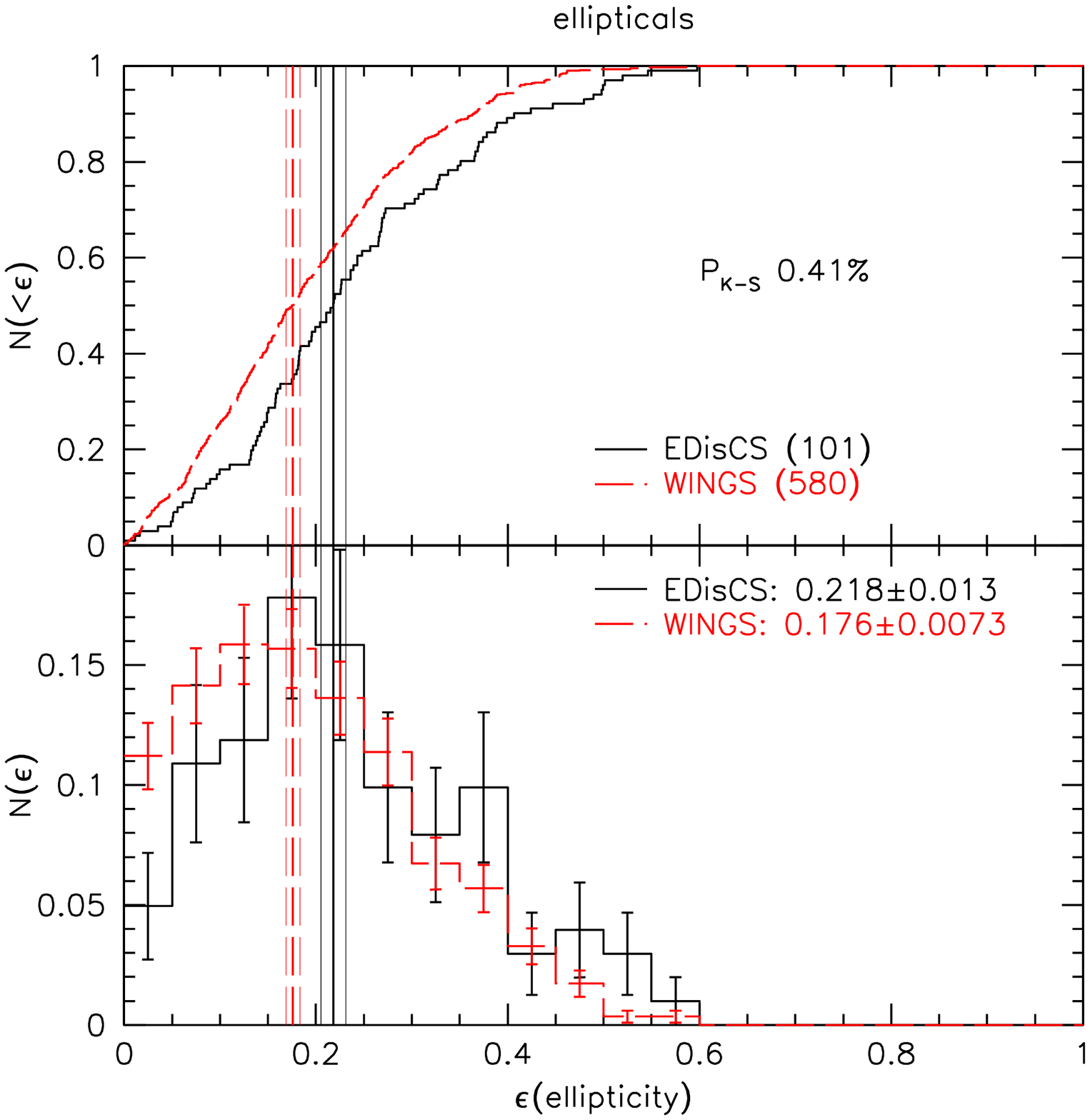}
\vspace*{2.2cm}
\end{minipage}
\hspace{2.5cm}
\begin{minipage}[c]{175pt}
\centering
\includegraphics[scale=0.4,clip = false, trim = 0pt 150pt 0pt 0pt]
{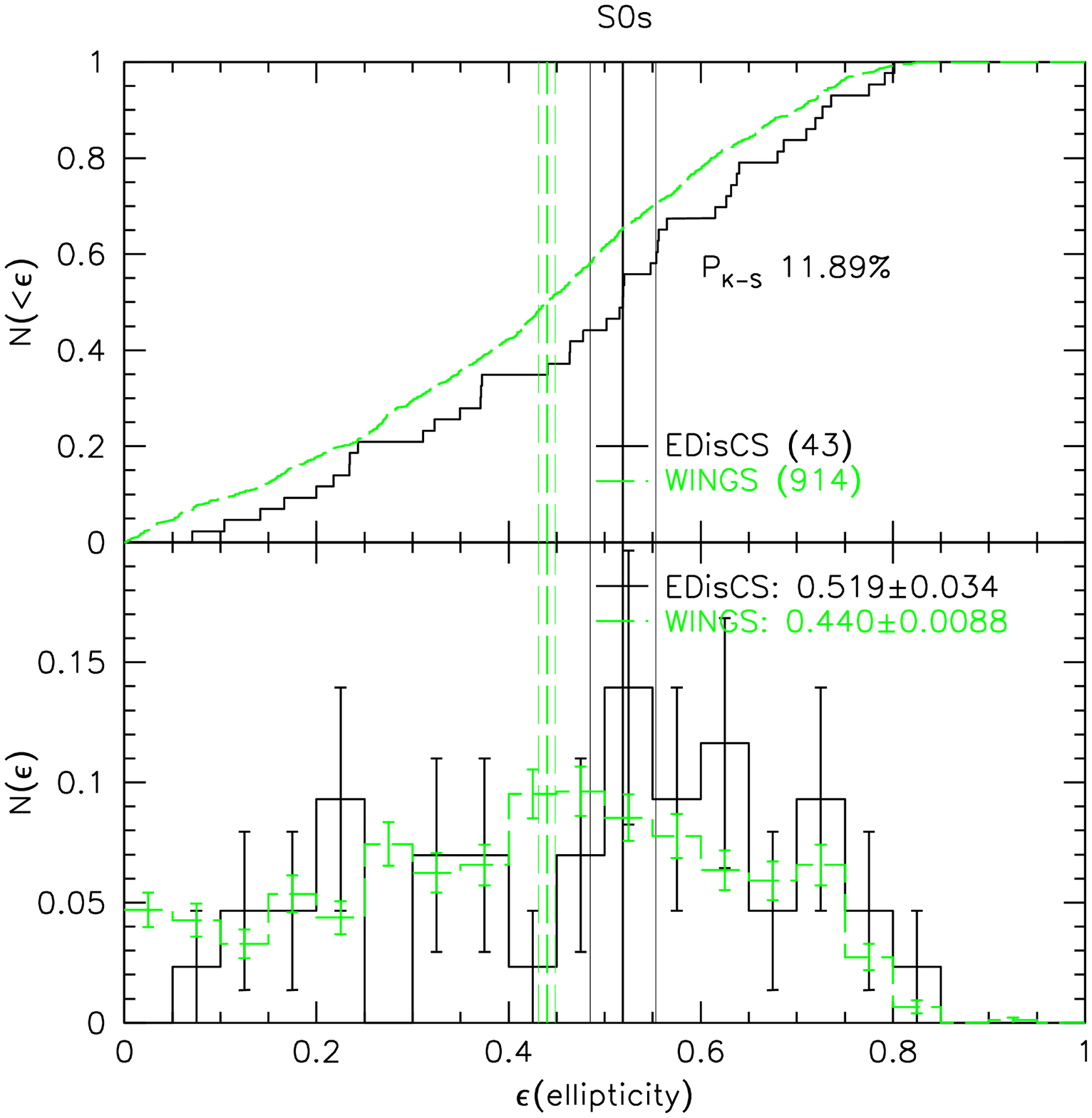}
\vspace*{2.2cm}
\end{minipage}
\hspace{1cm}
\caption{Ellipticity distribution of elliptical and S0 galaxies in our magnitude-``delimited'' samples. 
Panels and symbols are the same as in \fig\ref{ellS0_ell}.
Cf. with the mass-limited sample shown in Fig. \ref{ellS0_ell}. \label{ellS0_win_ed}}
\end{figure*}

\subsection{How can results from different samples be reconciled?}
Our mass- and magnitude- (de)limited samples give different results. 
In Appendix \ref{confr} we discuss the reasons for  the observed discrepancies, 
comparing directly the ellipticity distributions of the mass-limited and 
magnitude-``delimited'' samples for the same type of galaxies at the
same redshift. 
Briefly, the origin of the observed differences  in the distributions
lies in the fact that
galaxies in the two samples are characterized by different
properties; in particular the magnitude-``delimited'' samples are biased
and so they are not 
representative of the overall population. In fact, selecting
early-type galaxies on the red sequence only in the magnitude range $-19.3 >M_{B} +1.208z >-21$, 
we are loosing (the few) most massive galaxies, and a large fraction
of the less massive galaxies.  

\section{Comparison with literature results}\label{hold}
We now show that, even adopting the same selection criteria, 
our results are not in agreement with those reached by
\cite{h09}, who investigated the evolution in the ellipticity
distribution using two magnitude-``delimited'' samples of
cluster early-type galaxies in two redshift ranges.  Their sample
in the local Universe ($z=0.02-0.05$) consisted of 10 clusters (for a total of 210 galaxies),
 while the
sample in the distant Universe ($z=0.33-1.26$) consisted 
of 17 clusters (for a total of 487 galaxies) of
galaxies with {\it HST} images.

With this selection, their main conclusion is that there is no evolution
neither in the median ellipticity nor in the shape of the ellipticity
distribution of cluster early-type galaxies with redshift from $z>1$
to $z\sim0$. Their median ellipticity at $z>0.3$ 
is statistically identical with that at $z<0.05$ and the shapes of the distributions
broadly agree.
Moreover, they find a statistically significant evolution in
the S0s ellipticity distribution, while they do not detect
evolution for the ellipticals. 

In \tab\ref{ellval} we compare our median ellipticities with theirs.

Summarizing, comparing the two magnitude-``delimited'' samples
selected in the same way, even if we both do not detect an overall evolution
in the ellipticity distribution for early-types galaxies, we do find
an evolution in the median which they do not. Moreover, we find a significant
evolution for ellipticals and no evolution for S0s 
(though again we do find an evolution in the median), while they find no
evolution for ellipticals and strong evolution for S0s. 
This leads us to believe that the agreement for the early-types is not real, 
but simply due to a particular combination of the distributions of 
ellipticals and S0s.

We note that both at low and at high redshift, there are some clusters that 
are in common between our and their samples: at low redshift there are 4: 
{\it A119, A168, A957x, A1983}, at high redshift there are 5: 
{\it cl 1040.7-1155, cl 1054.4-1146, cl 1054.7-1245, cl 1216.8-1201} and 
{\it cl 1232.5-1250}.
For these 5 EDisCS clusters in common at high-z, also the morphological 
classifications \citep{desai07} and the images used to measure
ellipticities are the same.

So we can use this information to try to understand the reason for the discrepancy of the results.

\subsection{Origin of the discrepancies}

We proceed by comparing separately our WINGS and EDisCS samples with 
the \cite{h09} samples, 
trying to identify the reason of the observed differences.

\subsubsection{WINGS and Holden et al. (2009)}

\begin{figure*}
\centering
\includegraphics[scale=0.27,clip = false, trim = 0pt 150pt 0pt 0pt]
{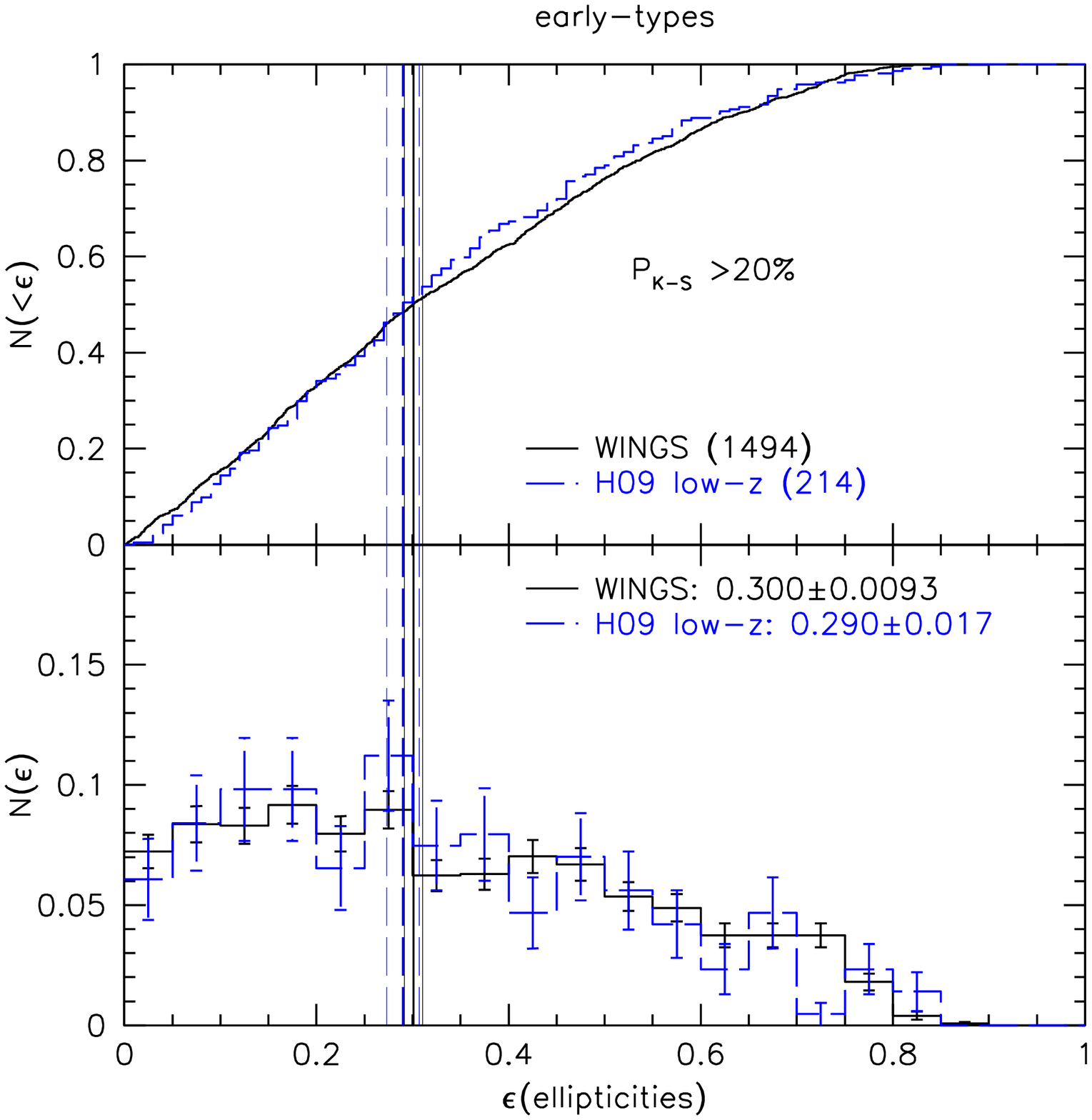}
\includegraphics[scale=0.27,clip = false, trim = 0pt 150pt 0pt 0pt]
{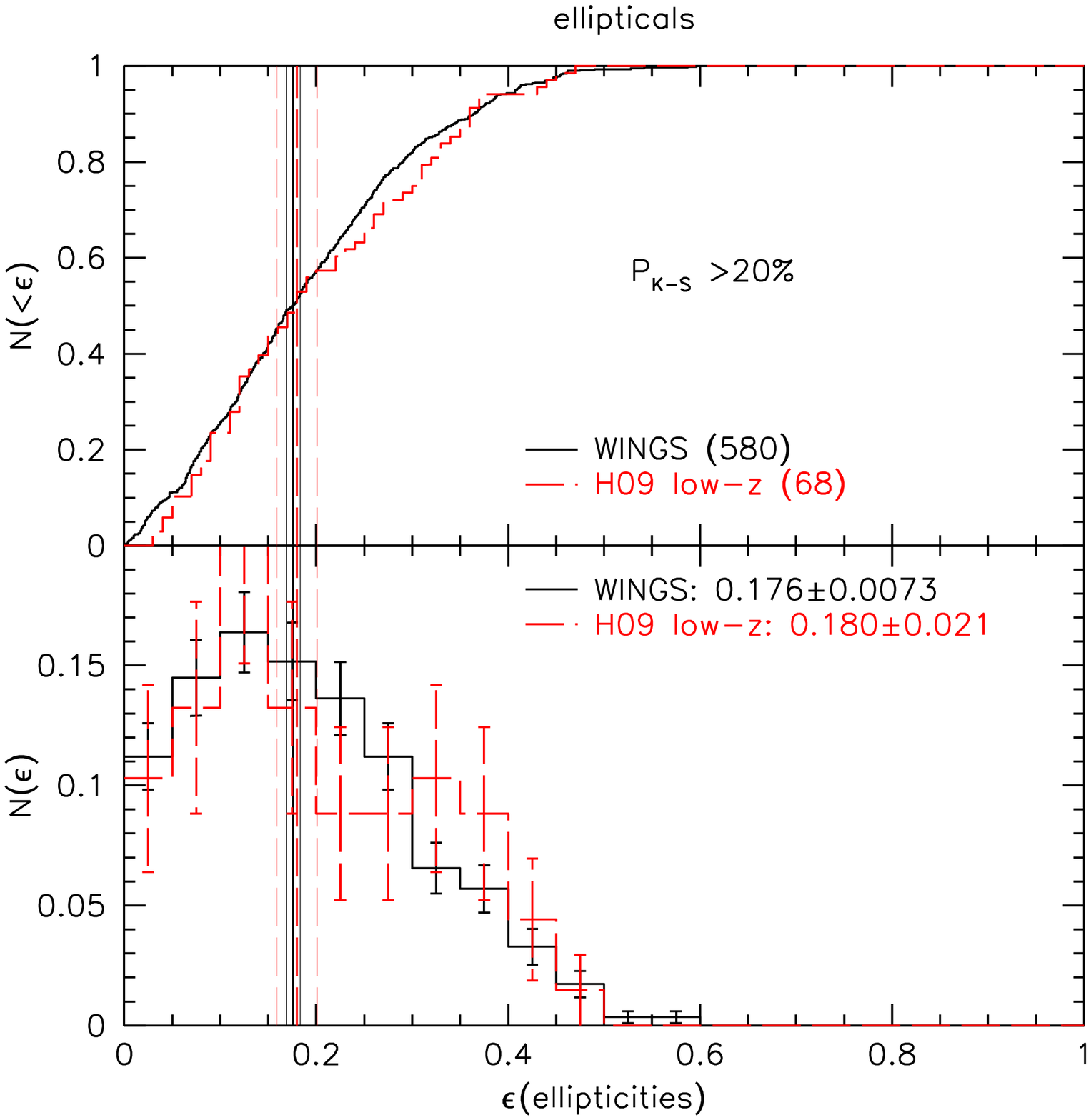}
\includegraphics[scale=0.27,clip = false, trim = 0pt 150pt 0pt 0pt]
{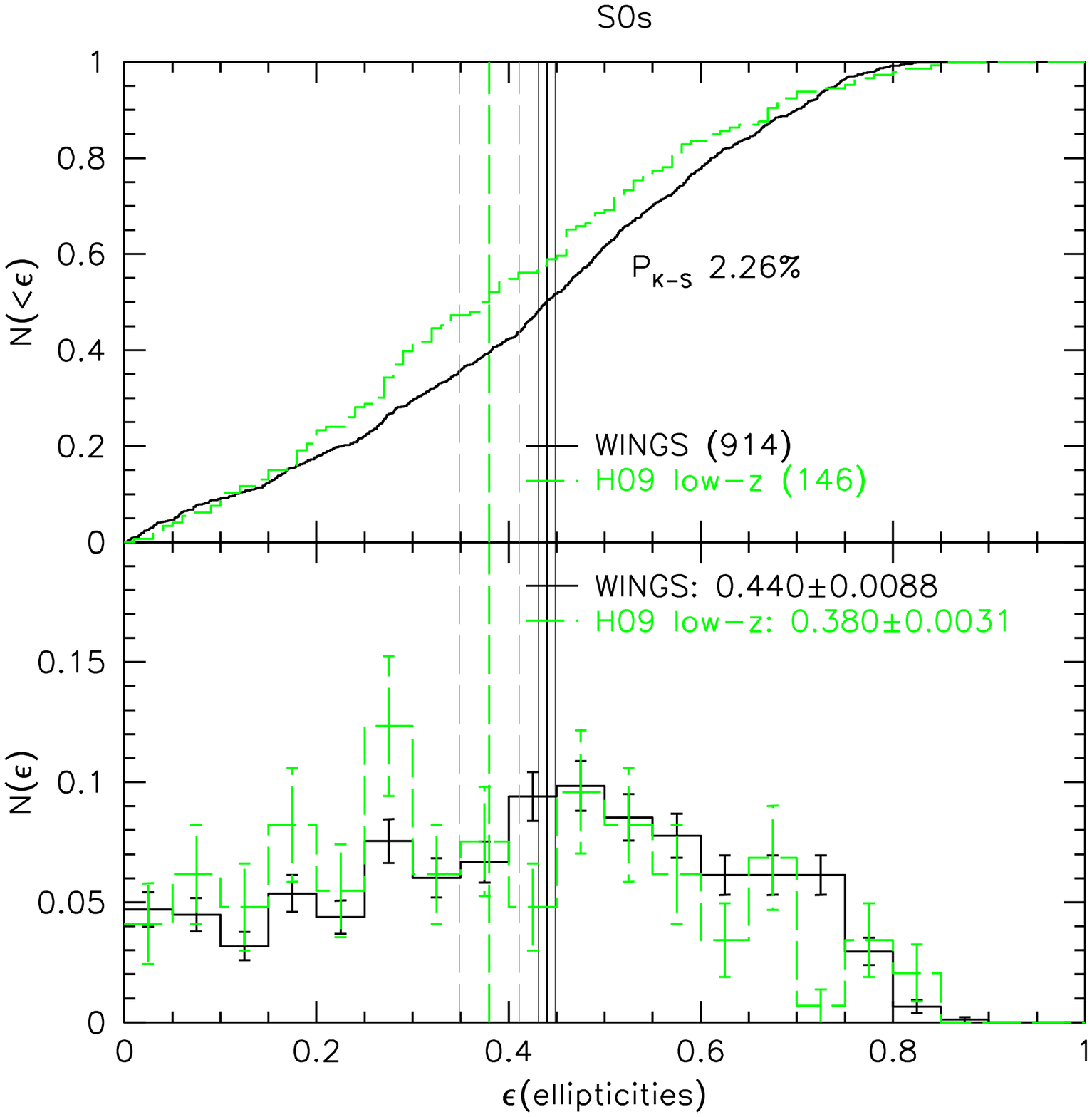}
\vspace*{1.2cm}
\hspace{1cm}
\caption{Comparison between our results and those of \citet{h09} at low-z.
Top panels: cumulative distributions; bottom panels: histograms normalized to 1. 
$P_{K-S}$ is the probability of the two distributions are drawn from 
the same parent distribution.
Left panel: comparison between WINGS (black solid lines) 
and  \citet{h09} (blue dashed lines) early-types.
Central panel:  comparison between WINGS (black solid lines) 
and  \citet{h09} (red dashed lines) ellipticals.
Right panel: comparison between WINGS (black solid lines) 
and  \citet{h09} (green dashed lines) S0s. \label{win_hol} }
\end{figure*}

In \fig\ref{win_hol} we compare our results at $0.04<z<0.07$
with those of \cite{h09} at $z<0.05$. While the ellipticity
distributions of early-type galaxies
and ellipticals are consistent with being similar in the two samples 
(in both cases the K-S test cannot reject the null hypothesis of similarity 
of the distributions, giving a probability of
 $>20\%$\footnote{This could be a problem linked to the poor statistic: if
we double the number of \cite{h09} galaxies the test 
becomes conclusive.}), 
those
of S0s is remarkably different: 
WINGS S0s (black solid lines in the right panel
 of \fig\ref{win_hol}) peak around $\epsilon\sim$ 0.5, 
while S0s of \cite{h09} (green dashed lines
 in the same plot) peak around $\epsilon\sim$ 0.3-0.35, 
indicating that in the \cite{h09} sample
galaxies
are on average rounder than in WINGS. The K-S test
finds that the distributions can be drawn from the same
parent distribution with a probability of only $\sim 2.3\%$.

Since we adopted \cite{h09} selection criteria, 
if there are
differences, they  can be due either to differences in the morphological 
classification, differences in the measurement of the ellipticities, or to 
variations of some other galaxy properties.
Here we try to analyze each one of these factors.

Since \cite{h09} draw morphologies from \cite{dressler80} 
and in their sample at low redshift 
there are some clusters in common with WINGS, 
we select from the catalog of \cite{dressler80}
galaxies in common with WINGS and see if there 
are some noticeable differences among them.
On the whole, there are 18 clusters in the \cite{dressler80}
catalog that belong also to WINGS\footnote{For this comparison we include
in the analysis also those clusters from \cite{dressler80} that do not enter the \cite{h09}
sample, in order to improve the statistics.}.

First of all, we select those galaxies that are early-types
according to the \cite{dressler80} classification
and we assign to them our measurements of ellipticity. 
In this way, we can compare directly our values of 
ellipticity with those calculated by \cite{h09}.
\fig\ref{ellithol} shows the ellipticity distribution
of the two datasets, for early-type, elliptical and S0 galaxies: 
the K-S probability is always inconclusive ($P_{K-S} \geq 20\%, \sim8.5\%, \geq 20\%$
respectively). 
\begin{figure*}
\centering
\includegraphics[scale=0.27,clip = false, trim = 0pt 150pt 0pt 0pt]
{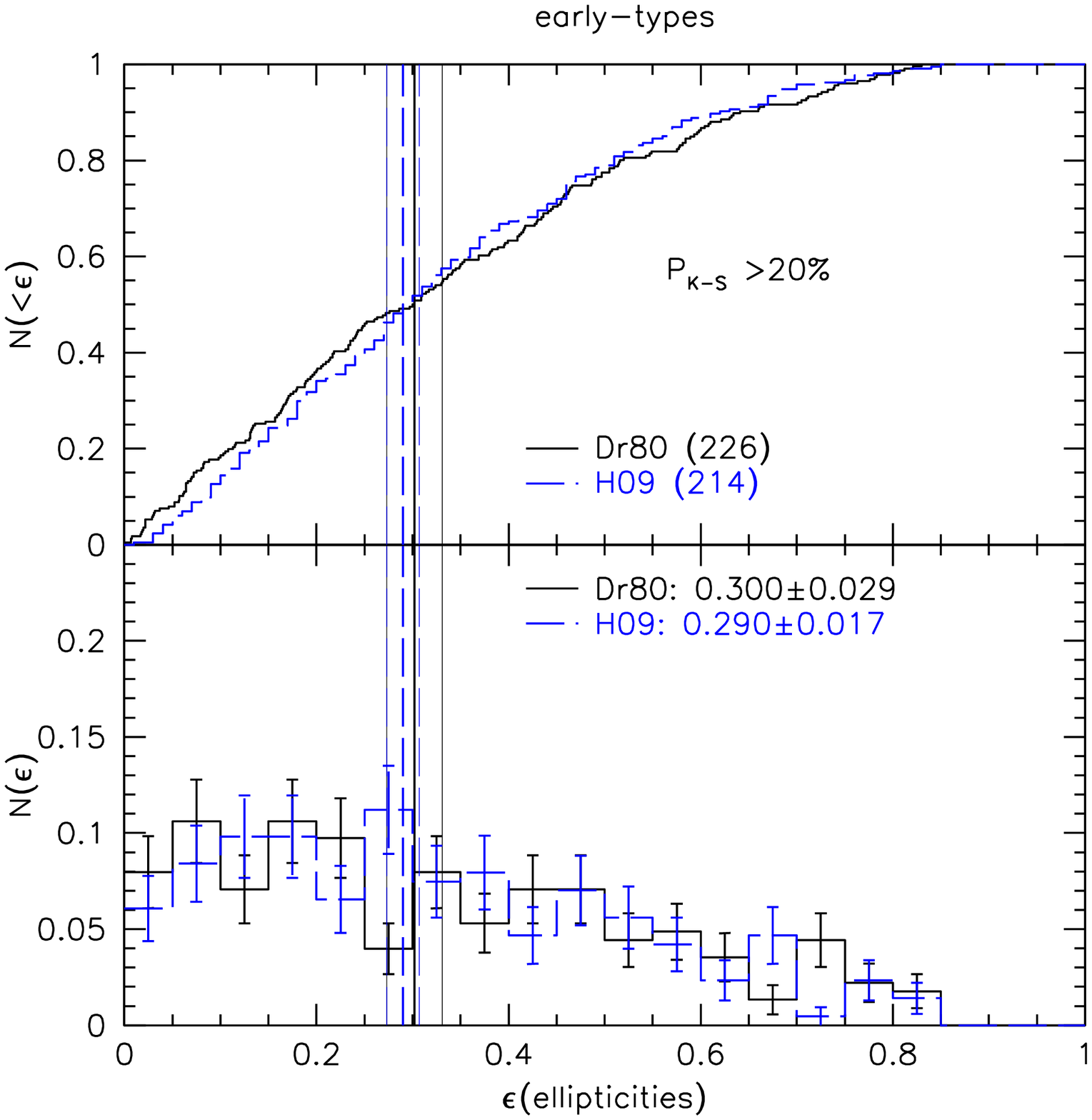}
\includegraphics[scale=0.27,clip = false, trim = 0pt 150pt 0pt 0pt]
{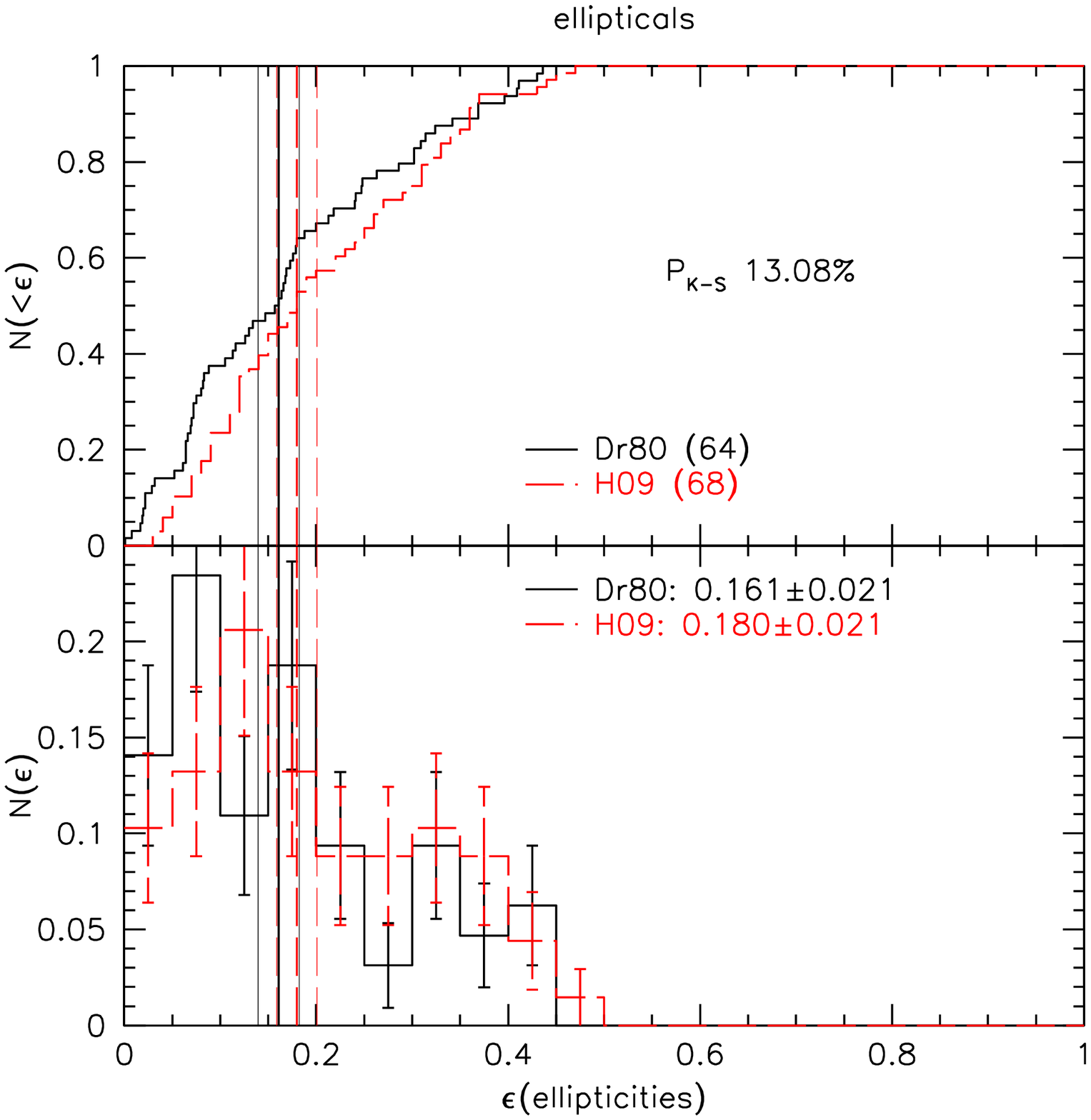}
\includegraphics[scale=0.27,clip = false, trim = 0pt 150pt 0pt 0pt]
{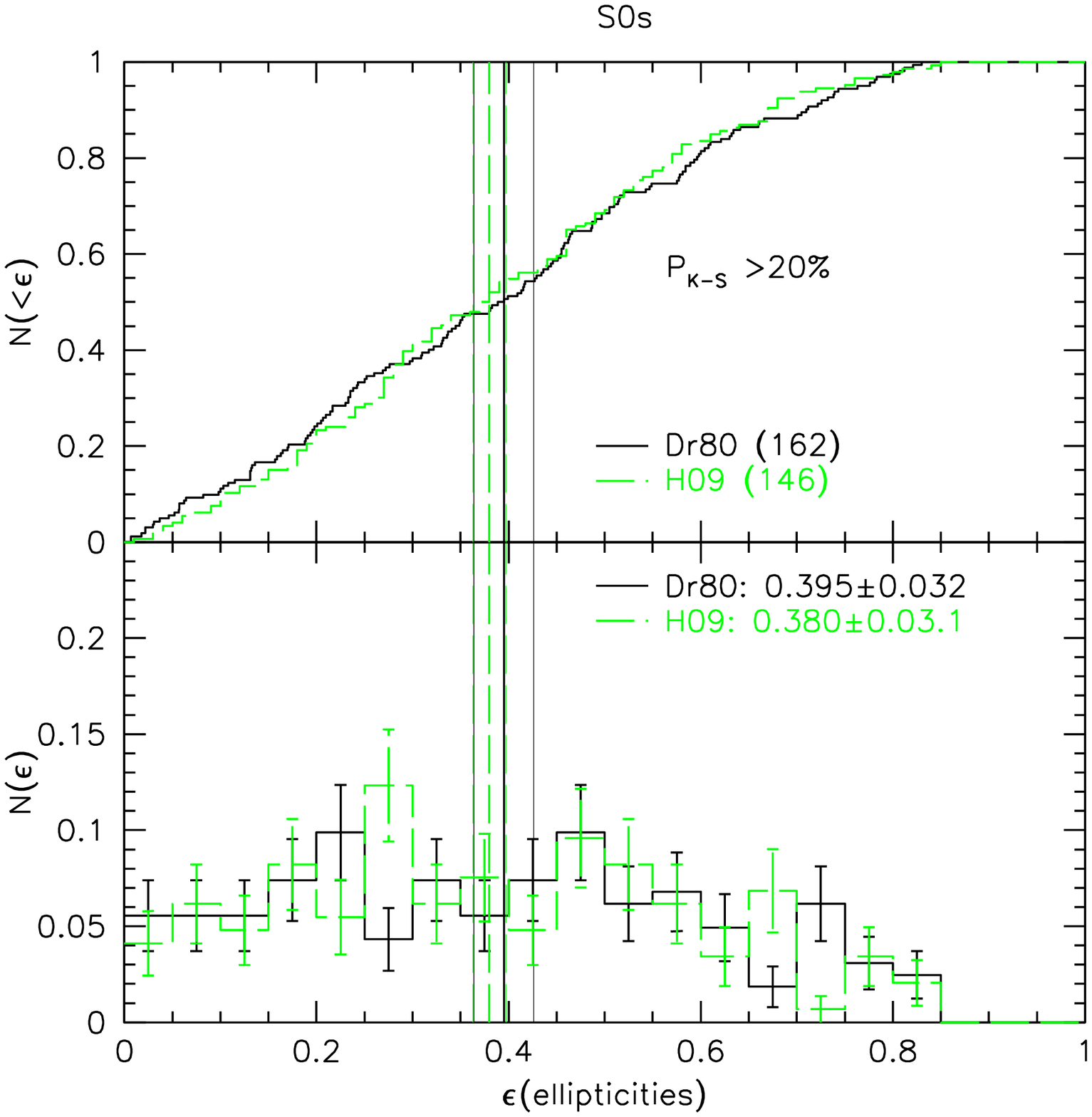}
\vspace*{1.2cm}
\hspace{1cm}
\caption{Comparison between the distribution of WINGS ellipticities
for galaxies that are early-types for \citet{dressler80}  (black solid lines)  and the low-z sample of \citet{h09}.
Top panels: cumulative distributions; bottom panels: histograms normalized to 1.
$P_{K-S}$ is the probability of the two distributions are drawn from 
the same parent distribution.
Left panel: comparison between WINGS  and  \citet{h09} (blue dashed lines) early-types.
Central panel:  comparison between WINGS and  \citet{h09} (red dashed lines) ellipticals.
Right panel: comparison between WINGS and  \citet{h09} (green dashed lines) S0s.\label{ellithol}}
\end{figure*}
This is consistent with the assumption that 
\cite{h09} estimates of galaxy 
ellipticity are compatible with ours. 

Second, we check that using only WINGS clusters classified also
by Dressler is equivalent to using the whole WINGS dataset:
 in fact the ellipticity distributions for
WINGS clusters in common with \cite{dressler80}  
(18 clusters) are in agreement with that
of our whole sample (76 clusters) (plots not shown, K-S probabilities $>>20\%$ for all
the morphological types).
The 3 distributions (early-types, ellipticals and S0s) are very similar,
indicating that the WINGS clusters in common with \cite{dressler80}
are not a biassed subsample of the whole WINGS dataset.

Since ellipticity measurements do not seem to be responsible
for the differences between the two datasets, 
we focus our attention on the other possible 
sources of differences, starting with the morphological classifications.

We find that $\sim$27\% (79/296) of the galaxies 
have been classified differently from us and \cite{dressler80}. 
This corresponds to the typical agreement between independent
classifiers, see also \S\ref{samplew}.
However, we have checked that they are too few to influence
the overall ellipticity distribution. Even 
re-classifing them and moving them to the other morphological class, 
they do not alter the ellipticity distribution
of the class in which they have been inserted. 

Moreover, comparing the ellipticity distribution
of galaxies  belonging to the same morphological class for us
and for \cite{dressler80}, once again we find no significant differences 
(the K-S test is always largely inconclusive) (plots not shown). 

So, the inconsistency is not even linked to the different morphological 
classification, 
and we have to  focus on possible biases due to other factors.

Third, we investigate the magnitude distributions of \cite{h09} and WINGS samples, to
be sure that all samples are equally deep. In \fig\ref{mb} we compare
the magnitude distribution of the analyzed samples and we find that
the magnitude distribution of the subsample of
galaxies  with \cite{dressler80}
morphologies (therefore those used by \citealt{h09}) 
(red filled histogram) is very different from that of our galaxies,
both if we consider only the clusters that are in common (green filled
histogram) and if we consider the whole WINGS sample (blue filled
histogram). 
Performing a K-S test on the magnitude distributions, we find that the whole WINGS sample and the subsample of
galaxies  with \cite{dressler80}
morphologies are drawn from different parent distributions ($P_{K-S}=2\%$).
 Since we are following exactly
the same selection criteria, the magnitude distributions should have been
similar.  It seems that the \cite{dressler80}, and hence the \cite{h09}, 
low-z sample misses galaxies
at fainter magnitudes. Probably they are too faint to have been
morphologically classified by \cite{dressler80} who used photographic
plates for
the classification. 

\begin{figure*}
\centering
\includegraphics[scale=0.8]
{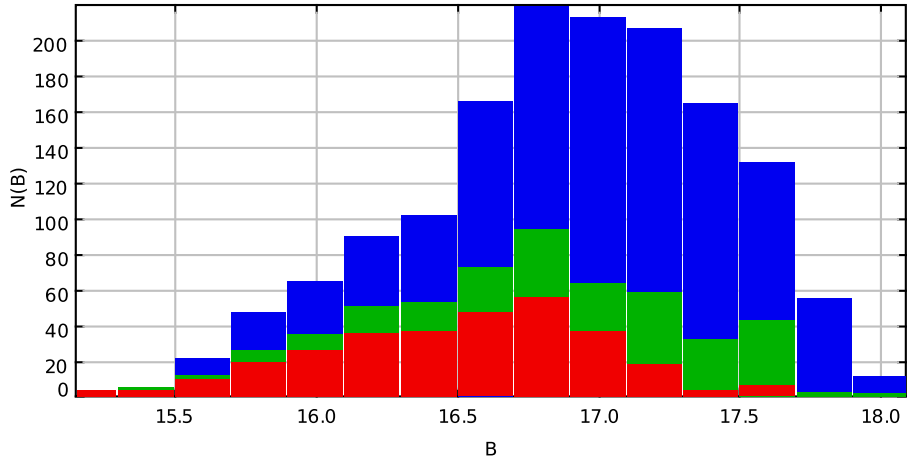}
\caption{Magnitude distribution of galaxies in different samples. Blue
histogram: early-type galaxies in the whole WINGS sample (1494 galaxies).  Green histogram:
early-type galaxies in the WINGS subsample of clusters that are in common with
\citet{dressler80} (547 galaxies).  Red histogram: galaxies early-types in the WINGS
sample that are in common with \citet{dressler80} (296 galaxies). 
\label{mb}}
\end{figure*}

To test whether this bias considerably alters the ellipticity
distribution, we create magnitude matched samples, selecting from our
WINGS sample a subsample of galaxies with the same magnitude (within
$\pm$0.05 mag), the same morphology and in the same
cluster as the sample of \cite{dressler80} used by \cite{h09}. As
shown in \fig\ref{Mb_confr}, comparing the WINGS magnitude matched
simulated sample with
the Dressler (hence \citealt{h09}) sample
the ellipticity distributions are
compatible, both for early-types, for ellipticals and for S0s (the
K-S test can not reject the null hypothesis of common origin of the
distribution, giving a probability always $>20\%$).

\begin{figure*}
\centering
\includegraphics[scale=0.27,clip = false, trim = 0pt 150pt 0pt 0pt]
{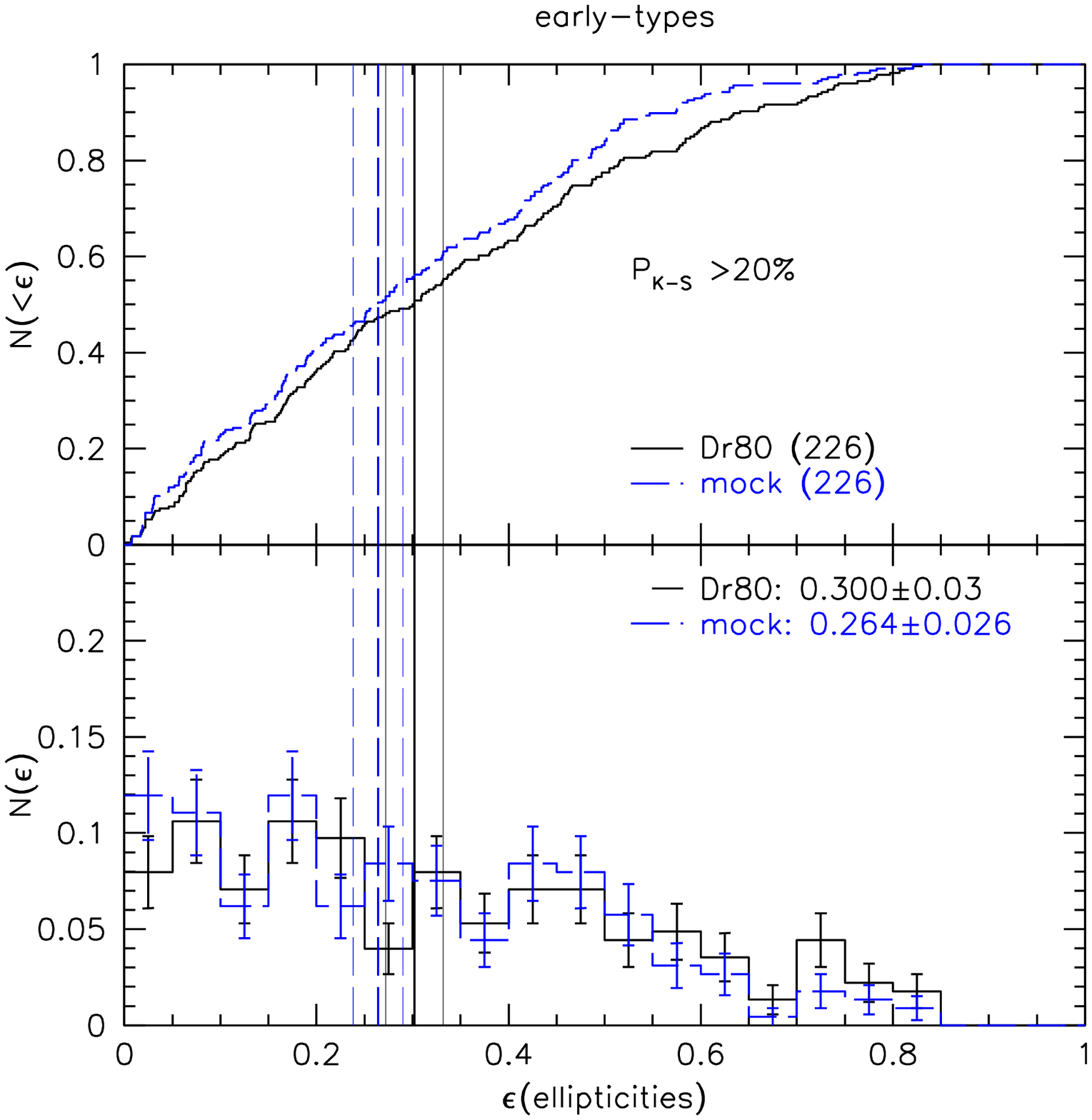}
\includegraphics[scale=0.27,clip = false, trim = 0pt 150pt 0pt 0pt]
{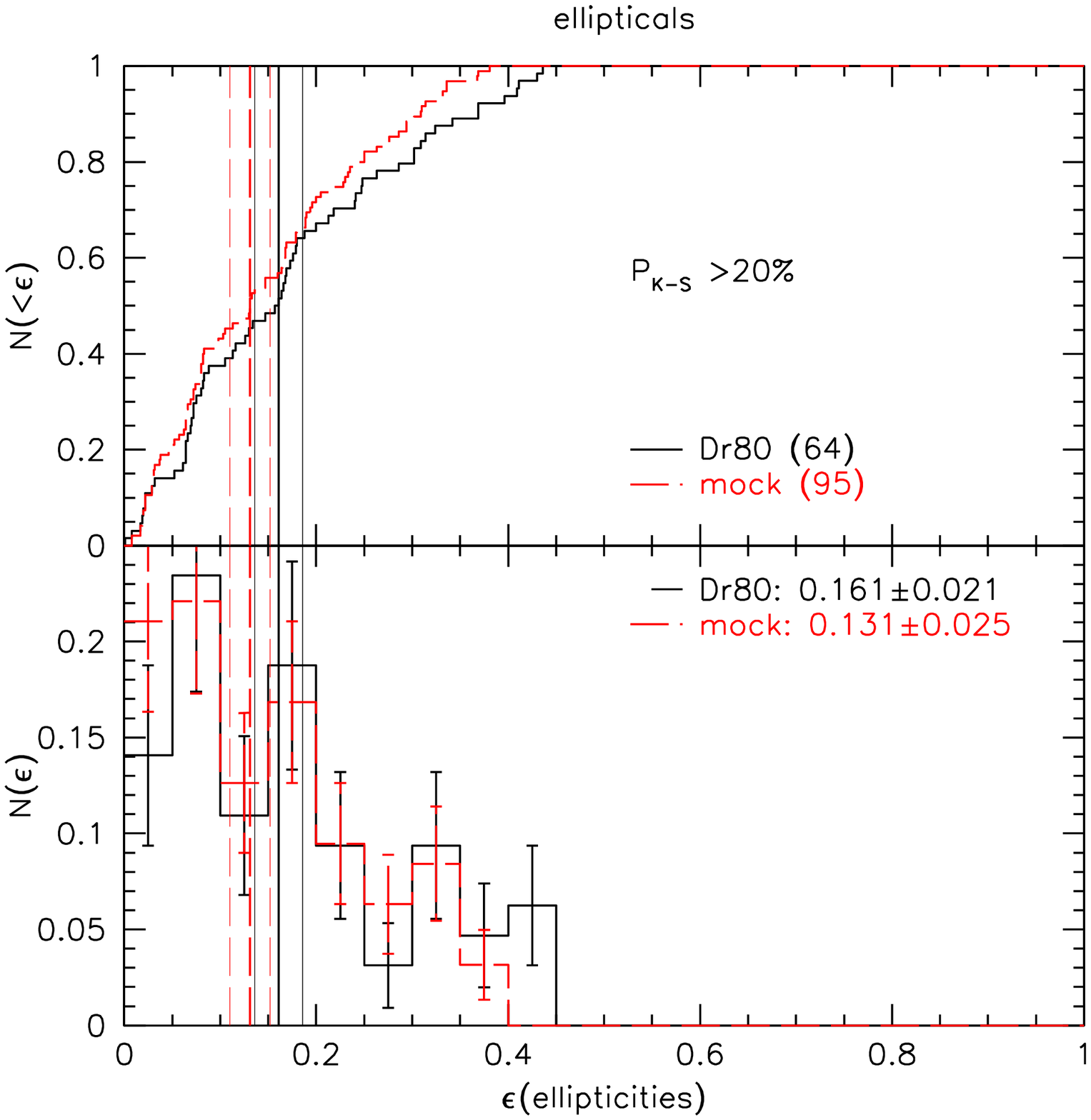}
\includegraphics[scale=0.27,clip = false, trim = 0pt 150pt 0pt 0pt]
{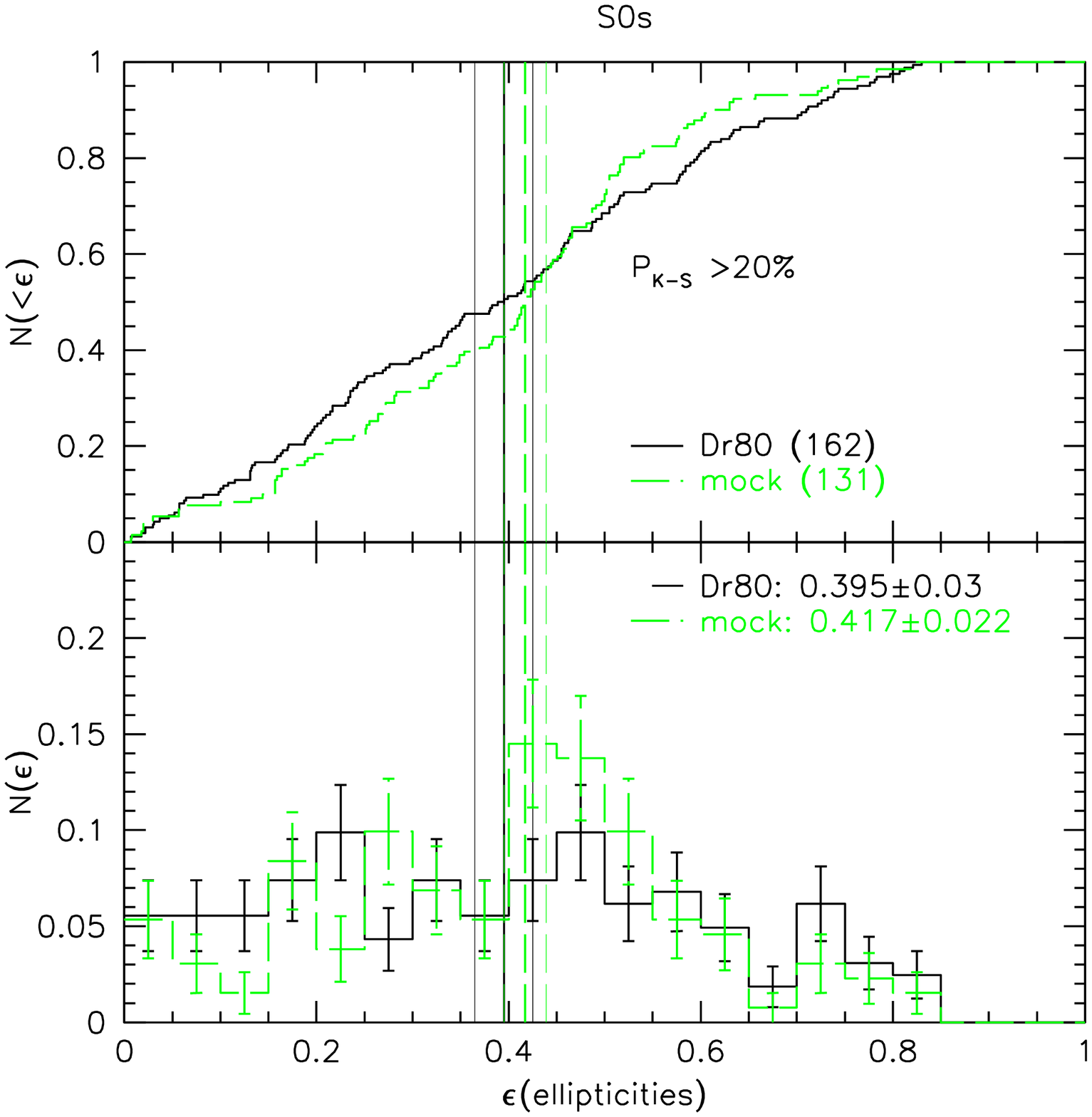}
\vspace*{1.2cm}

\hspace{1cm}
\caption{Ellipticity distribution of magnitude-matched samples (see text for details) 
for early-types (left panel), ellipticals (central panel) and S0s (right panel). Panels and symbols 
are as usual. Black solid lines
represent WINGS galaxies present in \citet{dressler80} catalog; 
coloured dashed lines represent WINGS mock distributions.
\label{Mb_confr}}
\end{figure*}

We conclude that the differences observed between the WINGS and
the Holden
samples in the Local Universe are due to the fact that the latter
includes only those galaxies that were morphologically classified
by \citet{dressler80} and does not correspond to a complete
sample within the adopted magnitude limits. Therefore, when comparing
WINGS and the low-z \cite{h09} sample we are
not comparing samples with the same properties, i.e. with the
same magnitude distribution.

\subsubsection{EDisCS and Holden et al. (2009)}
Now we wish to check if there are some differences also between EDisCS and 
the \cite{h09} sample.

Having the ellipticities of both samples, 
 we can compare directly
the ellipticity distributions
of the two samples at high redshift. 
\fig\ref{ed_hol} shows the comparison between the EDisCS and Holden high-z
ellipticity distributions for the different morphological types.  
We find that there are no significant
differences ($P_{K-S}$ $\sim11\%$, $\geq 20\%$ and $\geq 20\%$ for early-types, 
ellipticals and S0s respectively).
Also
comparing the ellipticity distribution of galaxies 
belonging only to the clusters in common (plots not shown), 
we can state that there are no discrepancies between the samples, 
as the K-S is always largely inconclusive ($P_{K-S}$ always
$>>20\%$).

\begin{figure*}
\centering
\includegraphics[scale=0.27,clip = false, trim = 0pt 150pt 0pt 0pt]
{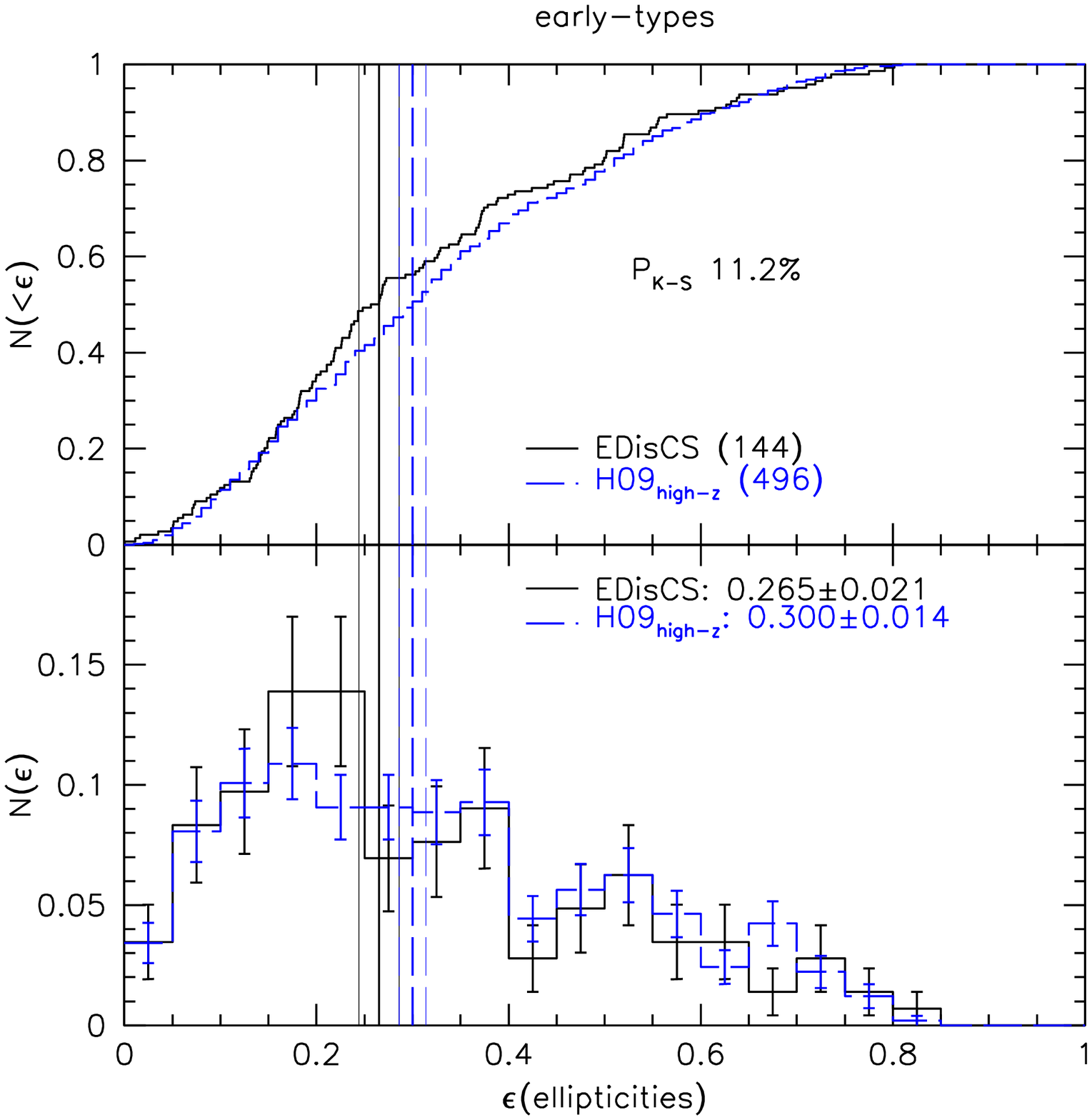}
\includegraphics[scale=0.27,clip = false, trim = 0pt 150pt 0pt 0pt]
{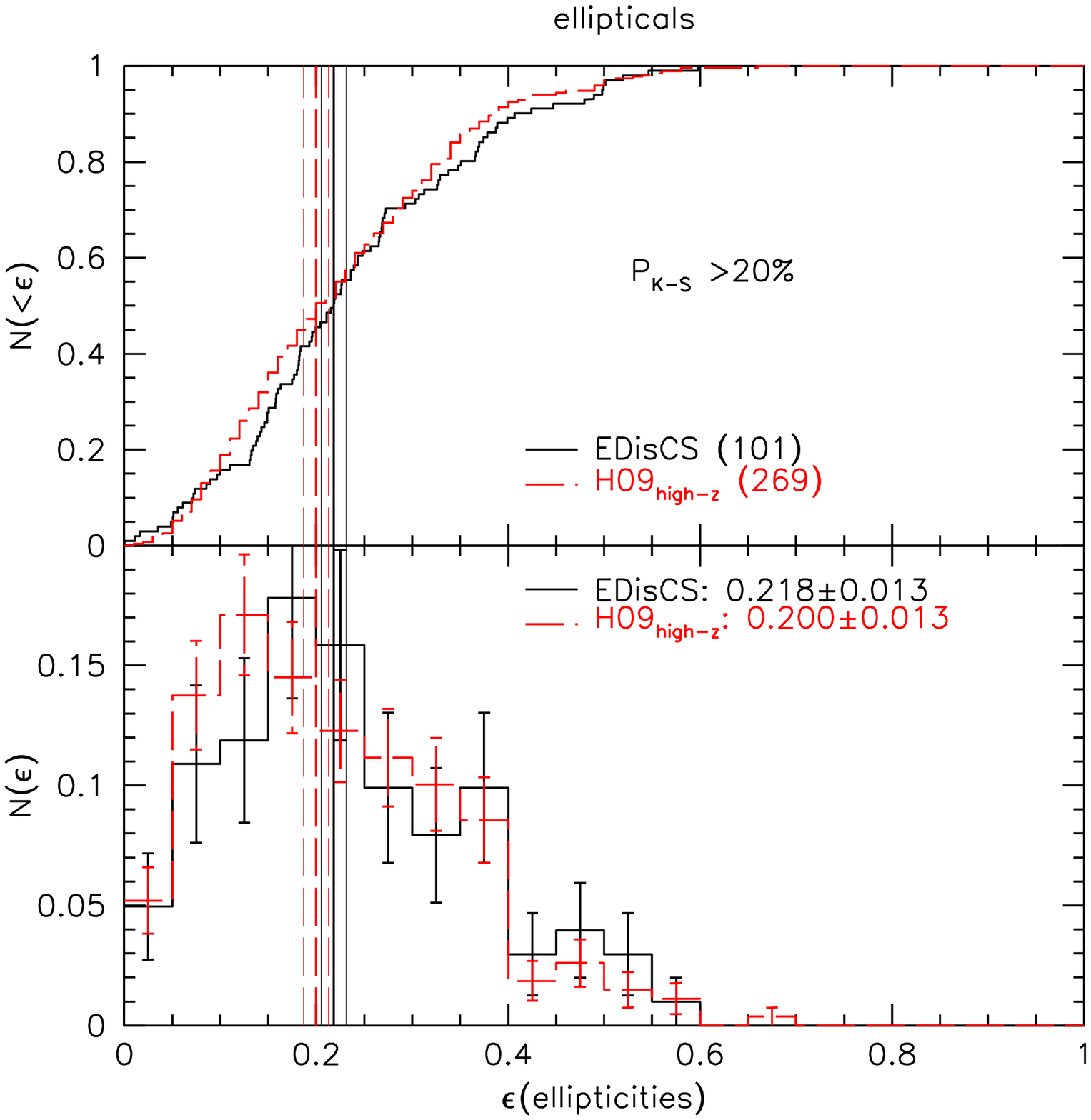}
\includegraphics[scale=0.27,clip = false, trim = 0pt 150pt 0pt 0pt]
{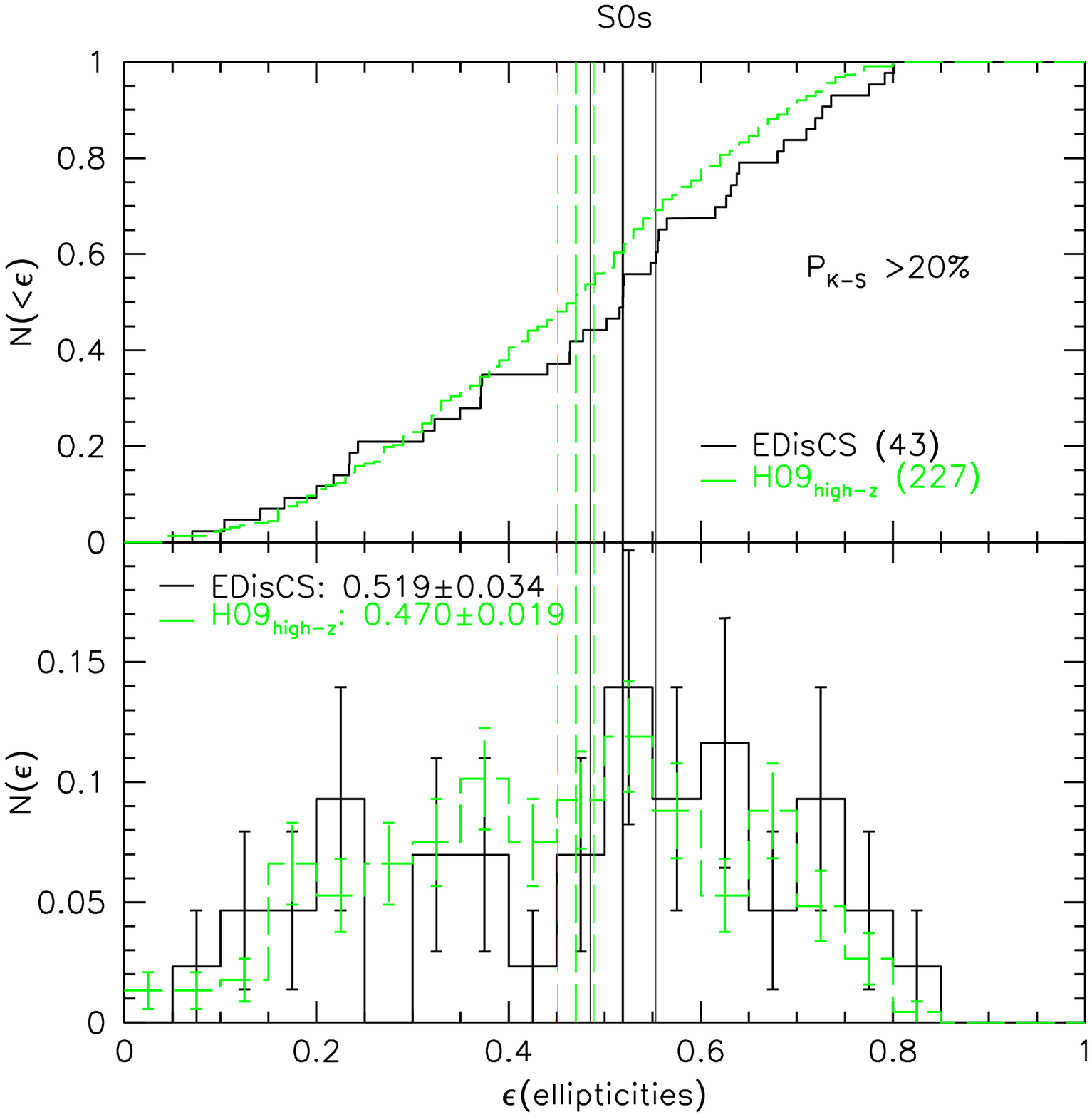}
\vspace*{1.2cm}
\hspace{1cm}
\caption{Comparison between our results and those of \citet{h09} at high-z. 
All clusters are used in both samples.
Top panels: cumulative distributions; bottom panels: histograms normalized to 1. 
$P_{K-S}$ is the probability of the two distributions are drawn from 
the same parent distribution.
Left panel: comparison between EDisCS (black solid lines) 
and  \citet{h09} (blue dashed lines) early-types.
Central panel:  comparison between EDisCS (black solid lines) 
and  \citet{h09} (red dashed lines) ellipticals.
Right panel: comparison between EDisCS (black solid lines) 
and  \citet{h09} (green dashed lines) S0s. \label{ed_hol}}
\end{figure*}
  
\subsection{Conclusions}
To conclude, the different results drawn analyzing our samples and
that of \cite{h09} mainly arise from the fact that, despite 
the fact that in principle galaxies are 
selected 
following the same criteria, actually at low redshift they have a different 
magnitude distribution.
The \cite{h09} low-z sample suffers from incompleteness 
at faint magnitudes, likely due to the lack of availability of Dressler (1980)
morphologies at faint magnitudes.
No differences have been detected instead at high redshift between
Holden's sample and ours.

\section{summary and conclusions}
In this paper we have analyzed the ellipticity distribution
of early-types galaxies, and of ellipticals and S0s separately,
in clusters at $z=0.04-0.07$ and  $z=0.4-0.8$. 
We have taken into account both a  mass-limited sample and a 
magnitude-``delimited'' sample of galaxies. 

\begin{itemize}
\item
In our mass-limited samples, above the common mass limit 
($M_{\ast}\geq 10^{10.2}M_{\odot}$) the ellipticity distribution of early-type 
galaxies strongly varies with redshift. This is due to a change  both of
 the median and of the shape of the distributions with redshift.
 For ellipticals,  no statistically significant
differences are observed in the high- and low-z distribution, even if 
an evolution of the medians is detected and we observe an excess population
of round ellipticals at low-z compared to high-z. 
Finally, no evolution is observed for S0s.
The evolution of early-type galaxies is not simply 
related to the different mass
distributions at high- and low-z. In fact, removing
the influence of the mass, the results remain inconsistent.
Instead, it is mainly related to the evolution of the morphological
 mix with redshift and hence to the relative
contribution of ellipticals and S0s at the two epochs. 

\item
As mentioned in the previous point,
in our low-z sample, we find a population of very round ($\epsilon\leq0.05$)
elliptical galaxies that is less conspicuous at high-z. 
This population seems 
real and not due to selection effects or measurement problems.
 
\item
In our  magnitude-``delimited'' sample, 
for early-types  and S0s the evolution is not evident
(though the medians of both early types and S0s change with z), 
while for ellipticals
we have found a change of the distribution with redshift.  

\item
The observed differences between the mass-limited sample
and the  magnitude-``delimited'' one can be due to  the different mass 
distribution of the two samples:
in fact in the magnitude-``delimited'' samples we 
are loosing some galaxies that enter the mass-limited one, both at high
 but especially and more importantly at low masses.

\item
Our magnitude-''delimited'' results are not in agreement with those 
of \cite{h09}, who also analyzed a magnitude-``delimited''
sample of early-types belonging to the red sequence. The main reason of
the observed discrepancy  is that, despite galaxies 
being selected following in principle
the same criteria, in practice the two low-z
samples have a different magnitude distribution because
the \cite{h09} sample suffers from incompleteness 
at faint magnitudes.
\end{itemize}

\section*{Acknowledgments}

We want to thank B. P. Holden for providing us the lists of ellipticities from
his paper. We thank G. Rudnick for his useful comments that improved our paper. 
BV and BMP acknowledge financial support from ASI contract I/016/07/0.

\appendix
\section{Additional statistical tests} \label{app_test}
In \S~\ref{result} we characterize the evolution of the ellipticity 
distribution performing the K-S test. Anyway, since this test is as general 
as possible,  
we wish to go deeper into our analysis, trying
to understand if the K-S results are confirmed and above all if they
are driven by a different shape of the distributions or simply by a
different location of the two populations. To do this, we perform
two other statistical non-parametric tests (i.e. they do not assume
the normal distribution) which make no assumptions about the
distributions of the populations.

We use the \cite{moses63} test to check the equality of the scale
parameter, taking into account that each population has a different
median.  This test is very useful to compare the shape of two
distributions and to evaluate their dispersion.
Following this
procedure, we subdivide each population into a certain number of groups,
each one containing 10 observations.  For each group, we compute
its average and the sum of the residuals. Then we put all together the
residuals of the two populations, paying no attention  which one each value belongs to,
and we sort them. Afterwards, we sum the rank of each population separately and we
compare the sums. If they are very different, the probability that the
populations are drawn from the same parent distribution is very
small.
Since this test requires to consider randomly a subsample of 
the observations of the populations, we repeat the test 1000 
times.\footnote{Since this test is based on random  samples, we do not take into account the WINGS's weights.}
It emerges that early-type galaxies at different redshifts are unlikely to
have the same scale parameter (with a probability $<5\%$ in 85\% of the simulations), 
while both ellipticals and S0s show a high compatibility of it 
(with a probability $<5\%$ only in 14\%  of the simulations in both cases), 
suggesting that the shapes of the distributions are similar at different redshifts.

Then, to test if there could be a shift in location between the populations, we adopt 
the U-statistic proposed by \cite{mann47}. This allows us to assess if there are
differences in the median values, regardless of the choice of the errors adopted to characterize the medians.
This procedure requires to rank all the values, without regard to
which population each value belongs to.\footnote{To take into account the WINGS incompleteness, here
we consider rounded WINGS' weights, so that WINGS galaxies can weigh 1, 2 or 3, according to their real weight.} 
Similarly to what we did for the Moses test, we sum the ranks of each 
population and we compare the sums.
Again, if they are very different, the hypothesis that the two 
populations are drawn 
from the same parent distribution is ruled out.
This test strongly
supports the hypothesis that early-types and ellipticals 
have a different 
median at different redshifts, while for S0s it cannnot exclude the similarity 
of them (giving, respectively, a probability of 0.06\%, 1.17\% and 17.12\%).
We note that these results are fully in agreement with the bootstrap errors (see \tab\ref{ellvalu}).

A detailed summary and comparison of the results of the different tests is shown in
\tab\ref{test}, both for the mass-limited sample and the magnitude-``delimited'' ones (see \S \ref{sec_mag}).

\begin{table*}
\centering
\begin{tabular} {|c|llc|llc|llc|}
\hline
{\bf test} 		& \multicolumn{3}{c|}{\bf early-types} 									&\multicolumn{3}{c|}{\bf ellipticals} 									&\multicolumn{3}{c|}{\bf S0s}\\
\hline
\hline	
KOLMOGOROV-		& MASS:				& 0.04\%		 	& $\neq$				&MASS: &5.84\%&  = 								&MASS:& $>$20\% &  =  \\
SMIRNOV			& MAG:				& 7.7\% 			&  =					&MAG: &0.41\% &  $\neq$								&MAG: &11.9\% &  = \\
\hline
\hline
			& \multirow{4}{*} {MASS:}& $<1\%$ in 60\%	&\multirow{4}{*} {  $\neq$}	& \multirow{4}{*} {MASS:}& $<1\%$ in 3\%	&\multirow{4}{*} {  =}	&\multirow{4}{*} {MASS:}& $<1\%$ in 2\%	&\multirow{4}{*} {  =}\\	
			&			&1-5\% in 25\%		&					&			&1-5\% in 11\%		&					&			&1-5\% in 12\%		&				\\	
			&			&5-10\% in 7\%		&					&			&5-10\% in 12\%		&					&			&5-10\% in 10\%		&				\\	
{MOSES}		&			&$>10\%$ in 8\%	&					&			&$>10\%$ in 74\%	&					&			&$>10\%$ in 76\%	&				\\	
\cline{2-10}
{simulations}			&\multirow{4}{*} {MAG:}&$<1\%$ in 28\%		&\multirow{4}{*} {  $\neq$}	&\multirow{4}{*} {MAG:}&$<1\%$ in 16\%	&\multirow{4}{*} {  =}		&\multirow{4}{*} {MAG:}&$<1\%$ in 1\%		&\multirow{4}{*} {  =}\\
			&			&1-5\% in 25\%		&					&				&1-5\% in 23\%		&					&			&1-5\% in 9\%		&				\\
			&			&5-10\% in 17\%		&					&				&5-10\% in 10\%		&					&			&5-10\% in 13\%		&				\\
			&			&$>10\%$ in 30\%		&					&				&$>10\%$ in 51\%		&					&			&$>10\%$ in 77\%	&				\\
\hline				
\hline
\multirow{2}{*} {MANN-WHITNEY}	& MASS:				& 0.06\%		 	&  $\neq$				&MASS: &1.17\%&   $\neq$									&MASS:& 17.12\% &  =  \\
				& MAG:				& 13.18\% 		&  =					&MAG: &0.07\% &  $\neq$											&MAG: & 3.4\% &  $\neq$ \\
\hline
\end{tabular}
\caption{Summary of the results of different tests applied both to our mass
 and magnitude-``delimited'' samples (see \S \ref{sec_mag}). The K-S, being as general as possible, gives 
an indication on whether two distributions can derive from the same parent 
distribution; the Moses tests the equality of scale parameters when the 
assumption of common medians is not reasonable (hence it tests the {\it shape}); 
the Mann-Whitney tests if there is a shift in median of the two population 
(hence it tests the compatibility of the {\it medians}). 
For the mass-limited sample, only galaxies above $\log M/M_{\odot} \geq 10.2$ are considered.
Moreover, for WINGS we also take into account the completeness weights (see text for details). 
The symbol $\neq$ means that the considered test can state that the two populations are drawn
from different parent distribution, while the symbol $=$ means that the considered test is not
conclusive. \label{test}}
\end{table*}

\section{how can results from different samples be reconciled?} \label{confr}
From \S~\ref{res_mass} and \S~\ref{sec_mag} we draw different results.
Summarizing, we have found that in our mass-limited sample
there is no clear trend between 
ellipticity and mass for S0s. For ellipticals, this trend is only hinted, with
more massive galaxies having slightly lower values of ellipticity, while
for early-types it is quite strong and  
mostly due to the fact that ellipticals and S0s are found in 
different proportions at different masses.

Comparing the ellipticity distributions at the two redshifts, we have
found an evolution for the early-types, with WINGS galaxies being
proportionally more flattened than EDisCS galaxies.  No strong evolution has
been detected for ellipticals and S0s separately, except for the likely 
presence of an enhanced population of round ellipticals at low-z.
Note that this trend for round ellipticals is {\it opposite} to the trend
for all early-type galaxies (rounder vs flatter at low-z, respectively),
therefore we must be observing two distinct evolutionary effects.

In contrast, from the analysis of the magnitude-``delimited'' sample, 
we cannot exclude that, both in the case of early-type galaxies and of
S0s, the galaxy samples at high- and low-z
 are drawn from the same parent distribution, 
although the change in the median ellipticity values with time seems to
indicate an evolution instead.
Moreover, we have found
a significant evolution (2$\sigma$ error) of the ellipticity distribution of elliptical galaxies,
due mainly to a different median of the distributions.

To understand the reasons for  the observed discrepancies, 
we have compared directly the ellipticity distributions of the mass-limited and 
magnitude-``delimited'' samples for the same type of galaxies at the
same redshift (plots not shown). 

The K-S suggests different distributions ($P_{K-S}\sim 0\%$) 
for WINGS early-types and S0s, while it is inconclusive in all other cases  
(i.e. WINGS ellipticals; EDisCS early-types, ellipticals and S0s, $P_{K-S}>>20\%$).
Going deeper into the analysis, 
WINGS early-types show
incompatible values both of median (the Mann-Whitney test gives a probability of 0.90\%) 
and of scale parameter (the Moses test gives a probability $<5\%$ in 80\%
of the simulations respectively), while WINGS S0s have 
different scale parameters (the Moses test gives a probability $<5\%$ in 
83\% of the simulations respectively).

The origin of the observed differences  in the distributions
probably lies in the fact that
galaxies in the two samples are characterized by different
properties; in particular in the magnitude-``delimited'' samples, selecting
galaxies only in the magnitude range $-19.3 >M_{B} +1.208z >-21$, we
are loosing (the few) most massive galaxies, and a large fraction
of the less massive galaxies.

\begin{figure*}
\centering
\includegraphics[scale=0.27,clip = false, trim = 0pt 150pt 0pt 0pt]
{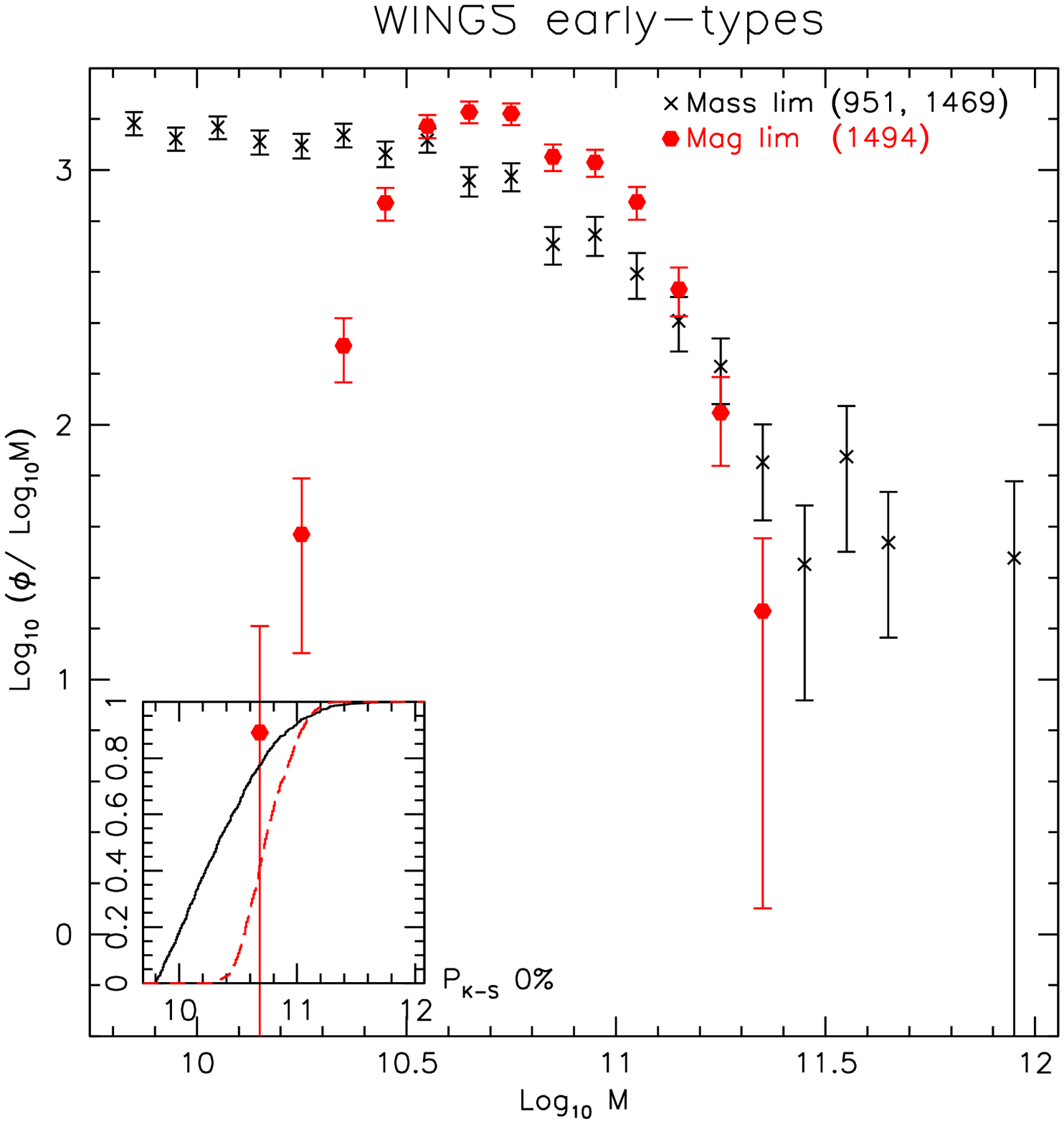}
\includegraphics[scale=0.27,clip = false, trim = 0pt 150pt 0pt 0pt]
{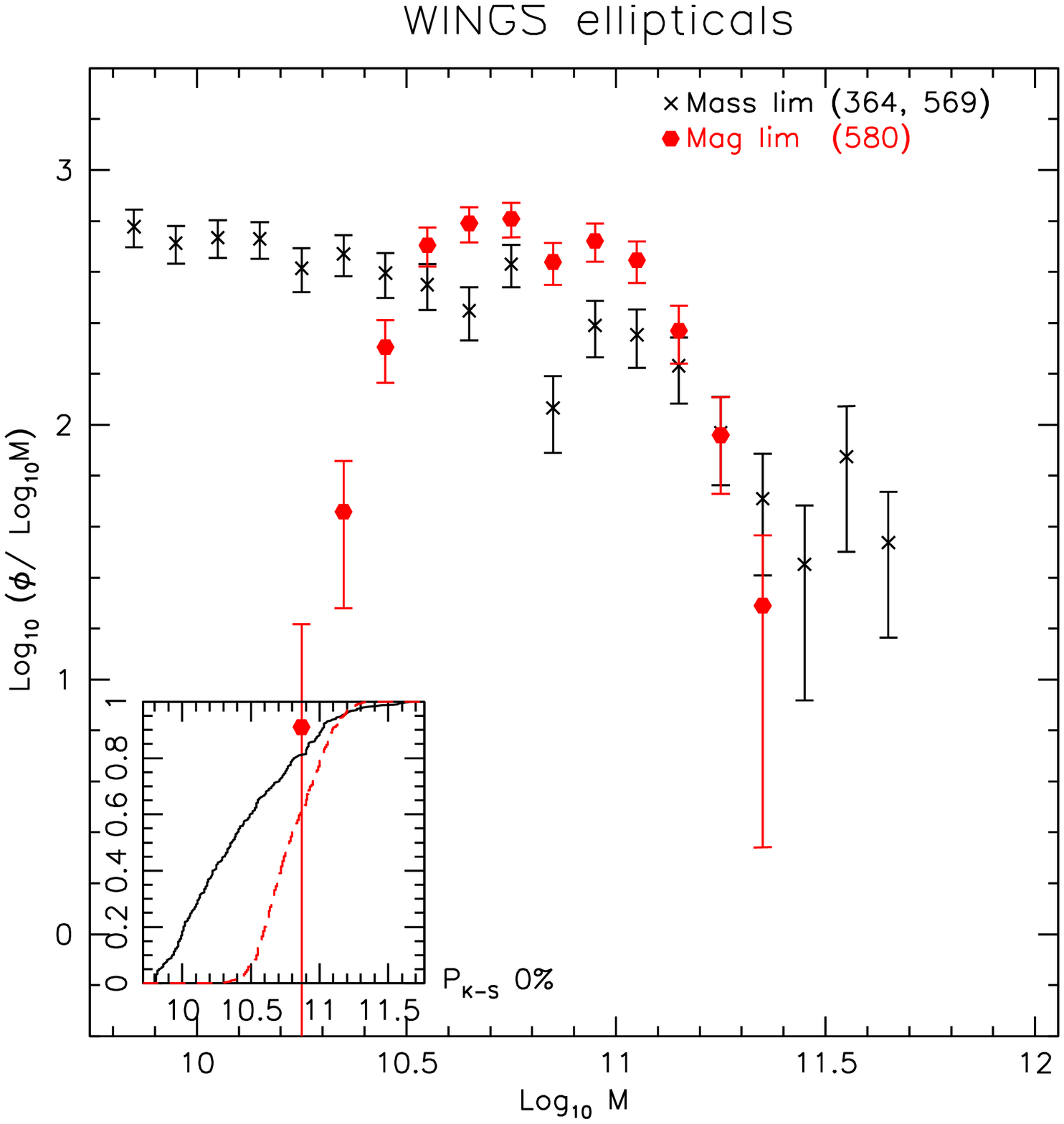}
\includegraphics[scale=0.27,clip = false, trim = 0pt 150pt 0pt 0pt]
{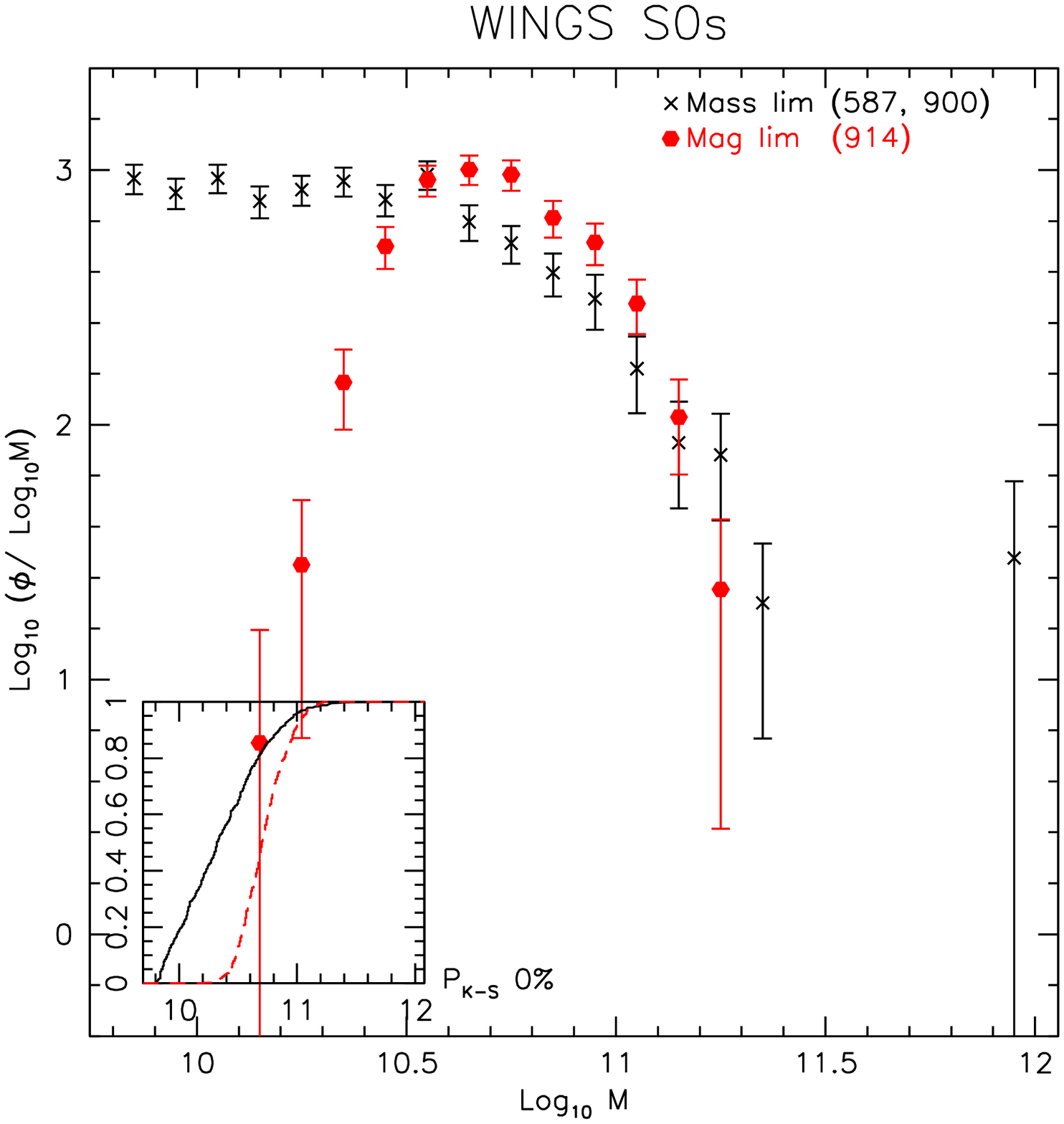}
\vspace*{1.2cm}
\hspace{1cm}
\includegraphics[scale=0.27,clip = false, trim = 0pt 150pt 0pt 0pt]
{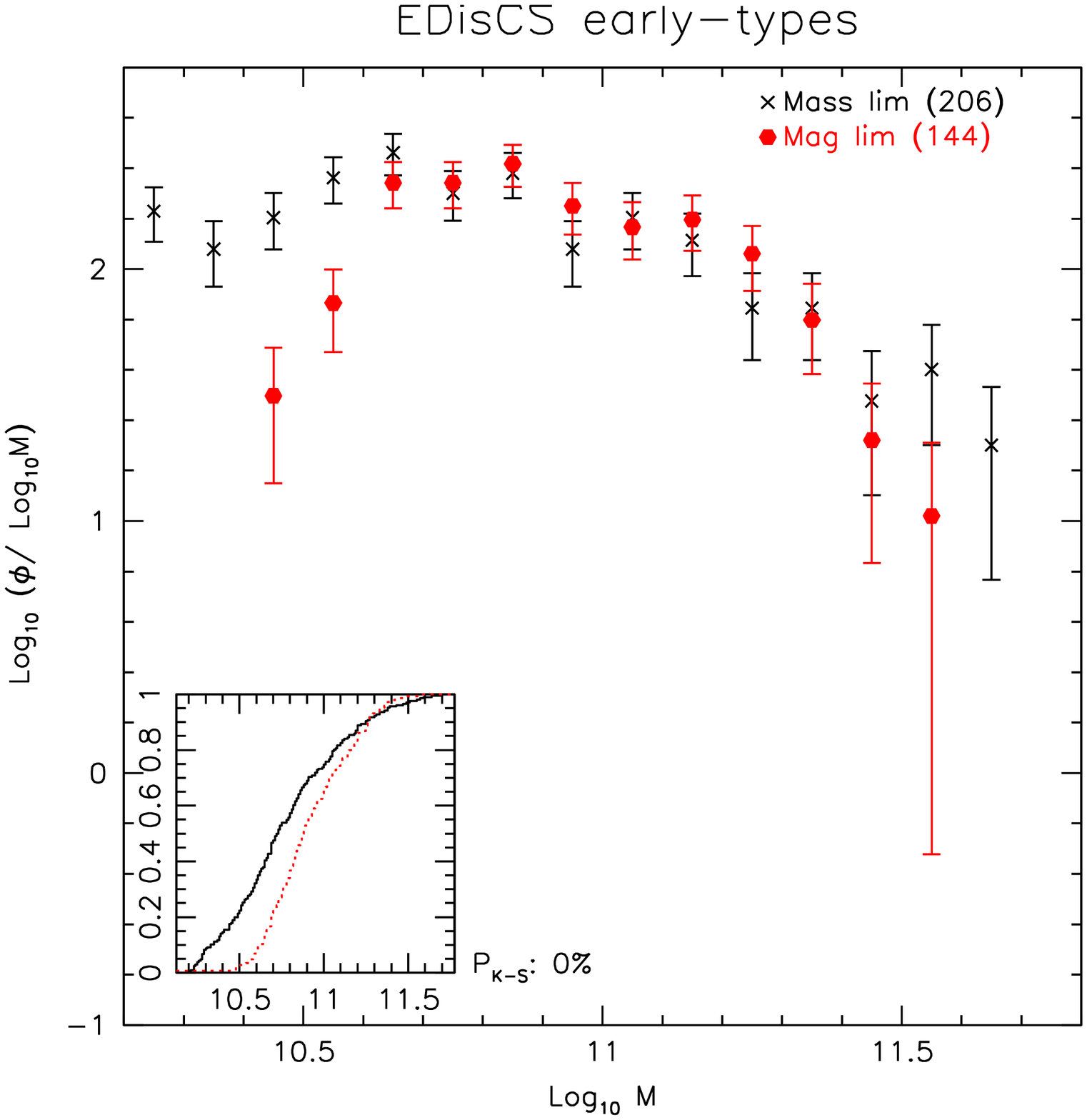}
\includegraphics[scale=0.27,clip = false, trim = 0pt 150pt 0pt 0pt]
{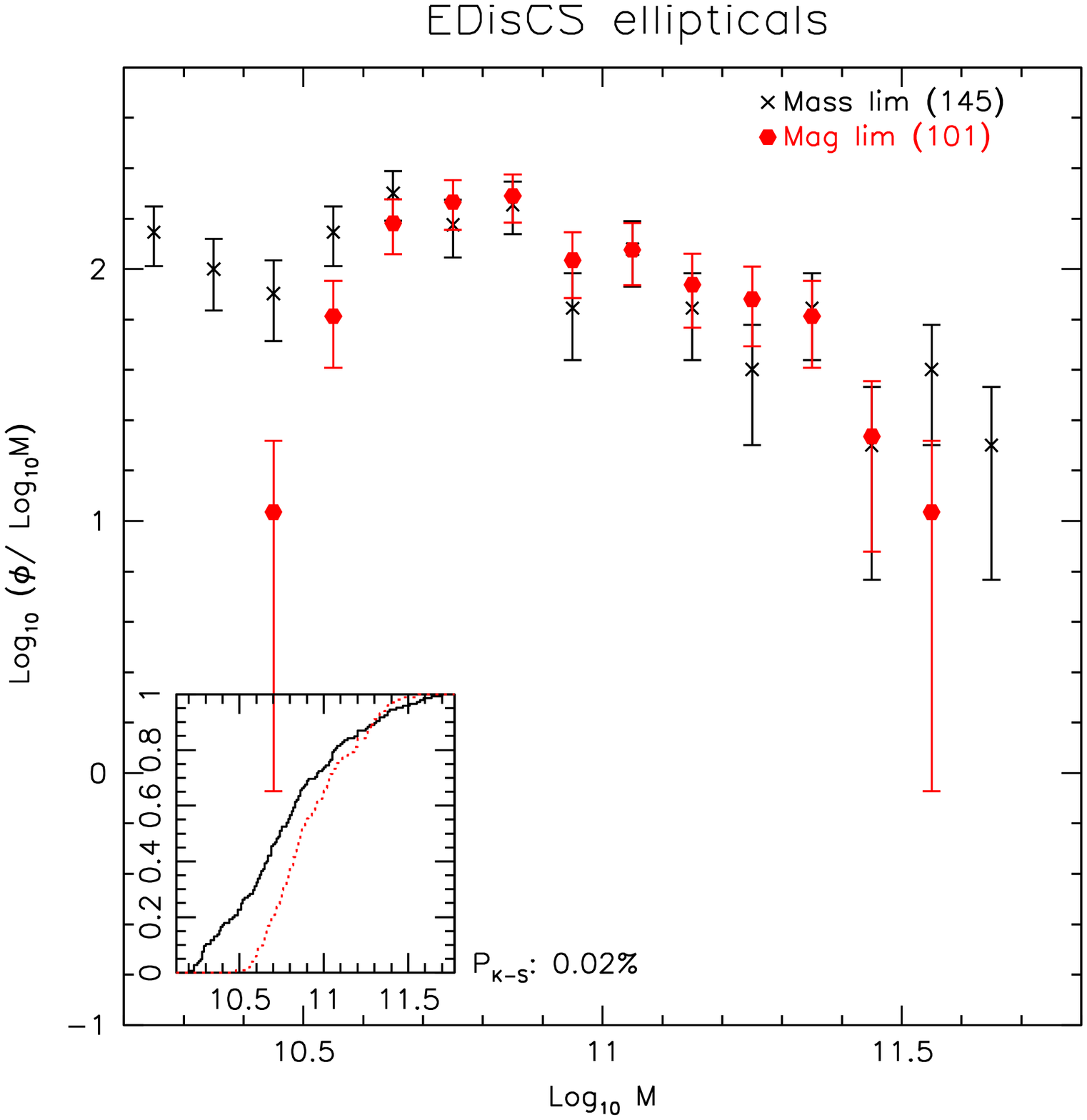}
\includegraphics[scale=0.27,clip = false, trim = 0pt 150pt 0pt 0pt]
{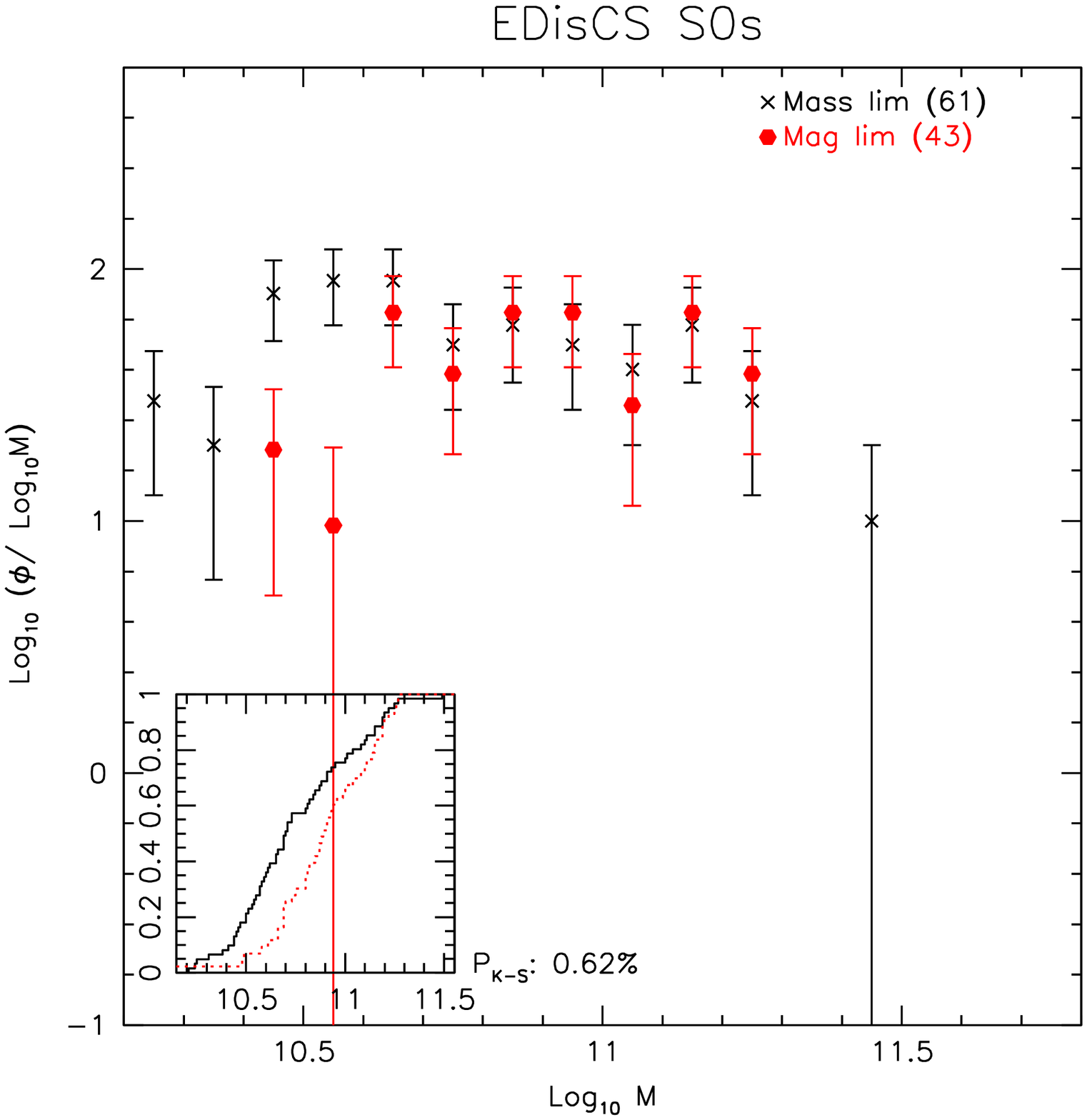}
\vspace*{1.5cm}
\hspace{1cm}
\caption{Comparison between the mass functions of the mass- (black crosses) and the magnitude- 
(red points) 
(de)limited samples for WINGS (upper panels) and EDisCS (bottom panels) early-types, ellipticals and S0s. 
\label{MF}}
\end{figure*}

This is evident in \fig\ref{MF}, where we compare the galaxy stellar mass
functions of our mass- and magnitude (de)limited samples, as
derived in \cite{morph}.
It is clear that the magnitude-``delimited'' sample is incomplete
at low and at high masses, and much more so for S0s at low 
masses in the local Universe than at high-z.

The net effect of the differential incompleteness
in mass at high- and low-z in magnitude selected samples is to
artificially deprive the low-z distribution preferentially
of high ellipticity (S0) galaxies.
The loss of low mass S0s (more flattened) at low-z
greatly reduces the differences between the high- and low-z
ellipticity distribution of early-type galaxies, bringing
their medians to be almost consistent and the K-S test to be inconclusive.

For ellipticals, the net effect of the incompleteness of the magnitude-``delimited'' 
sample is to exacerbate the differences with redshift, again
subtracting low-mass (hence higher ellipticity) ellipticals at low-z.

The incompleteness in the mass distributions of the
magnitude-``delimited'' sample therefore
seems to be consistent with the differences we observe between the
ellipticity distributions of the
mass-limited and of the magnitude-``delimited'' sample.
The magnitude-``delimited'' sample is 
biassed; in particular, early-type galaxies on the red sequence
and with $-19.3 >M_{B} +1.208z >-21$ are not 
representative of the overall population.

\label{lastpage}

\end{document}